\documentclass[twocolumn,
               preprintnumbers,
               superscriptaddress,
               numbers,
               floatfix,
               nofootinbib,
               showpacs,
               colorlinks,
               linkcolor=blue,   
               citecolor=blue]{revtex4-2}

\usepackage{amssymb}
\usepackage{amsmath}

\usepackage{booktabs}
\usepackage{hyperref}
\hypersetup{
    colorlinks,
    urlcolor={blue!80!black}
}

\usepackage{tabularx}
\usepackage{aas_macros}

\usepackage{orcidlink}

\usepackage[T1]{fontenc}

\usepackage{multirow}

\begin{document}

\defcitealias{Moline21}{M23}
\defcitealias{Molin_2017}{M17}
\defcitealias{Coronado_Bl_zquez_2022}{CB22}

\title{Impact of baryons on the population of Galactic subhalos and implications for dark matter searches} 

\author{Sara Porras-Bedmar \orcidlink{0009-0001-3512-2628}}
\email{sara.porras.bedmar@uni-hamburg.de}
\affiliation{Institute of Experimental Physics, University of Hamburg, Luruper Chaussee 149, D-22761 Hamburg, Germany}

\author{Miguel Á. Sánchez-Conde \orcidlink{0000-0002-3849-9164}}
\affiliation{Instituto de F\'isica Te\'orica UAM-CSIC, Universidad Aut\'onoma de Madrid, C/ Nicol\'as Cabrera, 13-15, Madrid, 28049, Spain}
\affiliation{Departamento de F\'isica Te\'orica, M-15, Universidad Aut\'onoma de Madrid, Madrid, E-28049, Spain}

\author{Alejandra Aguirre-Santaella \orcidlink{0000-0002-9581-7288}}
\affiliation{Departamento de Astronomía y Astrofísica, Universidad de Valencia,
            C/ Dr. Moliner 50, E-46100 Burjassot, 
            Valencia,
            Spain}
\affiliation{Institute for Computational Cosmology, Department of Physics, Durham University,
            South Road, 
            Durham,
            DH1 3LE,
            UK}

\date{\today}

\begin{abstract}
We have used Auriga --a set of state-of-the-art cosmological hydrodynamical simulations of Milky Way-size systems-- to study the impact of baryons on the Galactic subhalo population.
A DM-only run counterpart of Auriga allows us to compare results with and without baryons. We repopulate the original suites with low-mass subhalos orders of magnitude lighter than the mass resolution limit, starting from a detailed characterization of Auriga data in the well-resolved subhalo mass range.  
The survival of low-mass subhalos to tidal forces is unclear and under debate nowadays, thus in our study we stay agnostic and consider two different levels of subhalo resilience to tidal stripping ('fragile' and 'resilient' subhalos). 
We find baryons to alter the Galactic substructure significantly, by decreasing its overall abundance by a factor $\sim2.4$ (fragile) and $\sim1.9$ (resilient) and subhalo concentration --here defined in terms of maximum circular velocity-- by $\sim1.5$ with respect to the DM-only scenario. This has important consequences for indirect searches of DM. As an example, we investigated the case of using unidentified gamma-ray sources to set constraints on the DM particle properties, assuming some of them may be dark satellites.
Our results show the importance of including baryons to properly characterize the Galactic subhalo population, as well as to propose the most optimal subhalo search strategies, not only via its potential DM annihilation products but also through their gravitational signatures (e.g.~stellar streams, strong lensing).
\end{abstract}

\maketitle

\section{Introduction} \label{sect:intro}
    
    Our current standard model of cosmology, $\Lambda$ Cold Dark Matter ($\Lambda$CDM), is built upon several components for the energy-matter content of the Universe. Among them, a non-baryonic form of dark matter (DM) is thought to constitute roughly an 85\% of the whole matter content~\cite{Planck2018results}. DM has not been directly detected, yet its gravitational interaction has been detected on several different astrophysical and cosmological scales \cite{JUNGMAN1996195,garret2011,2018B_history,2018RPPh...81f6201R, 2018Natur.562...51B}.
    
    One of the most popular and accepted particle physics models for the nature of DM particles is the Weakly Interacting Massive Particle (WIMP) \cite[see e.g.][]{2005PhR...405..279B}. This particle seldom interacts with any other particle except itself. 
    Majorana WIMPs can self-annihilate into Standard Model particles, which in turn may produce photons through decay or radiative processes \cite{1997Bregstroem}.
    The predicted mass of WIMP particles is $\mathcal{O}$(GeV-TeV), so this self-annihilation produces a gamma-ray flux, the most energetic kind of photons. The measurement of gamma-ray fluxes is a technique widely used in current indirect searches for DM, using space-borne or terrestrial detectors such as the Fermi-LAT \cite{Atwood_2009} or current imaging atmospheric Cherenkov telescopes like MAGIC \cite{2005NIMPA.553..274F}, HESS \cite{2004NewAR..48..331H}, VERITAS \cite{WEEKES2002221}, LHAASO \cite{LHAASOsciencebook} or the HAWC observatory \cite{2012Hawc}. This is not the only existing detection method; direct detection in laboratories and particle production in colliders are complementary methods widely pursued. For a review, see \cite{2005PhR...405..279B}.

    A key aspect for astrophysical DM searches is to know in detail the distribution of DM in the Universe.
    The postulate of Cold Dark Matter (CDM) defines DM as non-relativistic, which causes a bottom-up hierarchical building of structures \cite{201200212FrenkWhite}.
    The precise minimum mass of the first halos depends on the exact nature of the DM particle as well as its decoupling temperature, being of the order of the Earth mass or below for typical CDM particle models \cite{Bringmann_2009}. These halos then grow in size by accreting surrounding DM or merging with other clumps. The hierarchical merging of halos leads to the creation of subhalos, that is, former field halos that have been accreted by a host and orbit within its potential well. Baryonic matter falls into halos, the latter thus becoming the seeds for galaxy formation. It is believed that halos with a mass $M\gtrsim 10^8 \, \mathrm{M_{\odot}}$ are capable of attracting enough baryonic material to have a visible counterpart \cite{2015MNRAS.448.2941S,Sawala_2015}.

    Numerical simulations have been developed to study structure formation of MW-like systems.
    Historically, researchers have relied on Dark Matter Only (DMO) or purely N-body simulations, which model one type of particle and its gravitational interaction. In the past decade, (magneto)hydrodynamical suites ((M)HD) have emerged, incorporating baryonic physics such as gas dynamics, star formation, feedback processes, and (in some cases) magnetic fields, alongside the DM content. They create a more realistic representation of astrophysical systems compared to DMO runs.
    Some examples of recent state-of-the-art (M)HD simulations of MW-like systems are 
    Auriga \cite{2024Grand_releasedataauriga}, Phat ELVIS \cite{10.1093PhatELVIS}, APOSTLE \cite{10.1093/mnras/stv2970} and FIRE \cite{10.1093/mnras/fire2017}; and DMO ones like Uchuu \cite{10.1093Uchuu_release}, VL-II \cite{Diemand_2008} and Aquarius \cite{Springel_2008}. For a more comprehensive list of suites, see \cite{2020NatRP...2...42V}.

    All simulations are inherently constrained by numerical resolution, regardless of the processes they simulate.
    For instance, algorithms used to find and follow the evolution of subhalos employ a minimum particle threshold to identify gravitationally bound structures, in such a way that those that contain fewer particles than this threshold are not considered bound within the simulation framework.
    For state-of-the-art high resolution runs, this threshold creates an artificial minimum subhalo mass around one million solar masses \cite{2024Grand_releasedataauriga, Diemand_2008, Springel_2008, 10.1093Uchuu_release}. 
    Also, both (M)HD and DMO suites consistently show an absence of subhalos near the GC \cite{Grand_2021, 10.1093PhatELVIS,2020MNRAS.492.5780R,2023MNRASFire,Moline21, santaella23}.
    However, comparison between simulation runs at different resolution levels demonstrates that part of the substructure disruption can be attributed to the limitations imposed by particle mass and grid resolution \cite{Grand_2021}.
    An open question in the community is whether these numerical issues solely account for the ``observed'' subhalo disruption, or whether truly physical processes could cause full disruption as well. Subhalos are subject to tidal forces while traversing their hosts, which causes mass to be stripped from the subhalo \cite{Hayashi_2003,2019MNRAS.490.2091G, 2020MNRAS.491.4591E}.
    Several studies suggest that the central regions of subhalos should survive tidal stripping, even when in some cases only a small fraction of their initial mass is retained~\cite{2018MNRAS.474.3043V, van_den_Bosch_2018, Stref_2019, 2020MNRAS.491.4591E, 2021arXiv211101148A, Aguirre-Santaella_sheddinglight}.
    Conversely, other studies indicate that tidal forces can be sufficiently strong to strip material to the full disruption of subhalos. Furthermore, the inclusion of baryonic processes in MHD runs results in a deeper gravitational potential and introduces dynamical processes, which can enhance subhalo disruption relative to the DMO runs \cite{10.1093/mnras/fire2017,10.1093PhatELVIS,Grand_2020,Grand_2021}.

    In this work, we evaluate the impact of baryons on the subhalo population in MW-like galaxy simulations, using the Auriga suite \cite{2024Grand_releasedataauriga}.
    Auriga has both MHD and DMO runs starting from the same initial conditions, allowing us to have a one-to-one comparison between them. We first use the information on existing DM subhalos at redshift $z=0$, and characterize their population. Being aware that resolution limits\footnote{As an example, in the ``Level 3'' of Auriga runs, DM particles have a mass $m_\mathrm{DM}=5\times 10^4\,\mathrm{M}_\odot$ and baryonic particles $m_\mathrm{baryon}=6\times 10^3\,\mathrm{M}_\odot$ \cite{2024Grand_releasedataauriga}.}
    prevent a precise description of subhalos in this galactic system, we then recreate a consistent, more complete population of subhalos using the repopulation algorithm first employed on VL-II data~\cite{santaella23}.
    Once the characterization of the resolved Auriga subhalo population has been performed, we repopulate the galaxy with low-mass subhalos below the original resolution limits.
    In our work, we will remain agnostic about the nature of heavy mass loss (or full disruption) of subhalos, by studying two distinct cases in which i) the mass loss is real and assumed to be properly described by the Auriga data; and ii) it is numerical in origin and thus spurious. For the latter case, we will add low-mass subhalos in the simulation according to semi-analytical recipes, that will be more resilient to tidal forces. These two cases will allow us to investigate how potential numerical resolution issues might affect predictions for those indirect DM searches focused on subhalos, while also helping to assess the level of associated systematic uncertainties.
    Finally, and to illustrate the relevance that our results may have for current gamma-ray DM searches, we will revisit WIMP cross-section constraints that were derived from so-called {\it dark satellites}--subhalos with no baryonic counterpart-- as characterized in DMO runs~\cite{Coronado_Bl_zquez_2022}, and will investigate how such constraints are modified by the inclusion of baryons.

    This work is structured as follows. First, in Section~\ref{sect:subs_as_gamma_targets} we introduce DM subhalos as optimal gamma-ray targets. Then, in Section~\ref{sect:auriga_characterization}, we characterize the subhalo population of the Auriga simulations. Section~\ref{sect:repopulation_theory} is devoted to the subhalo repopulation algorithm, which is later used in Section~\ref{sect:repopulation_results} to calculate expected subhalo DM annihilation fluxes. This section also includes an illustrative example to show how previously computed DM constraints can get modified by the inclusion of baryons. Finally, in Section~\ref{sect:discussion_conclusions}, we provide our conclusions and outlook.

\section{Subhalos as gamma-ray DM targets} \label{sect:subs_as_gamma_targets}
        
    If DM is made of WIMPs, it can self-annihilate producing Standard Model particles, which in turn generate gamma-rays \cite{1997Bregstroem}. The expected DM annihilation flux $\phi$ is given by \cite{BERGSTROM1998137,PhysRevDEvans2004}
    \begin{equation} \label{Flux_eq}
    \begin{split}
            \phi = &\underbrace{\frac{1}{D_\mathrm{Earth}^2} \int_{\Delta\Omega}d\Omega \int_{l.o.s.}dr\, r^2 \rho_\mathrm{DM}^2}_\mathrm{J-factor}\\
            &\times \underbrace{\frac{1}{4 \pi} \frac{\langle\sigma v\rangle}{2 m_{\chi}^2}\sum_fB_f\int\frac{dN_f}{dE}dE}_{f_\mathrm{PP}}
        \end{split}
    \end{equation}
    where we have the product of two terms: the so-called ``J-factor'', and the particle physics factor $f_\mathrm{PP}$. The J-factor is related to the geometry and distribution of DM in the system, codified via the density of DM, $\rho_\mathrm{DM}$, along the line of sight (l.o.s.) and the distance between the DM target and Earth $D_\mathrm{Earth}$; whereas the $f_\mathrm{PP}$ takes into account the velocity averaged annihilation cross-section of such particle $\langle \sigma v\rangle$, its mass $m_\chi$, and its possible annihilation channels, \textit{f}.

    Subhalos can be excellent targets for indirect DM searches, given both their expected abundance and proximity in the Galaxy according to $\Lambda$CDM cosmology~\cite{PhysRevD.82.063501,10.1111/j.1745-3933.2012.01287.x,Ackermann_2012,HannesSZechlin_2012,PhysRevLett.115.231301,Bertoni_2016, galaxies7040090,2019JCAP...11..045C, Coronado_Bl_zquez_2019,Coronado_Bl_zquez_2021,Coronado_Bl_zquez_2022}.
    In this work, we calculate the analytical expression of the J-factor assuming a Navarro-Frenk-White (NFW) \cite{1997ApJ490493N} DM density profile for subhalos, integrated up to their scale radius, $r_\mathrm{s}$. The latter is defined as the radius where 
    \begin{equation}
    \frac{\partial(\log(\rho_\mathrm{DM}))}{\partial(\log(r))}\Bigg\vert_\mathrm{r_\mathrm{s}} = -2,
\end{equation}
    roughly separating the high (inner) and low (outskirts) density regions.
    In field halos, almost 90\% of the J-factor is contained within $r_\mathrm{s}$, and we expect this percentage to be similar or higher in subhalos, which lose their outer parts due to tidal forces \cite{Coronado_Bl_zquez_2022}. With these considerations, the J-factor corresponding to the annihilation within $r_\mathrm{s}$, $J_\mathrm{S}$, can be expressed as
    \begin{equation}
        J_\mathrm{S}= 3.91\times10^{21}\,\frac{\mathrm{GeV}^2}{\mathrm{cm}^5}
        \left(\frac{D_\mathrm{Earth}}{1\,\mathrm{kpc}}\right)^{-2}
        \left(\frac{c_\mathrm{V}}{10^5}\right)^{0.5}
        \left(\frac{V_\mathrm{max}}{10\,\mathrm{km}/\mathrm{s}}\right)^{3}.
        \label{eq:Js}
    \end{equation}
    We detail this derivation in Appendix~\ref{appendix:J}. 
    
    Two other parameters appear in Eq.~\eqref{eq:Js} that refer to the internal structure of the subhalo. 
    The term $V_\mathrm{max}$ is defined as the maximum circular velocity that particles reach within a subhalo, and the {\it velocity concentration} $c_\mathrm{V}$ codifies the inner structure of a subhalo, with higher concentrations given by more compact subhalos.
    These terms are further developed below in Section~\ref{sect:auriga_characterization}.
    In our work, we make use of Auriga data to derive and characterize these quantities for the population of Galactic subhalos.

\section{Characterization of the Auriga subhalo population} \label{sect:auriga_characterization}
    
    \begin{figure}
        \centering
        \includegraphics[width=\linewidth]{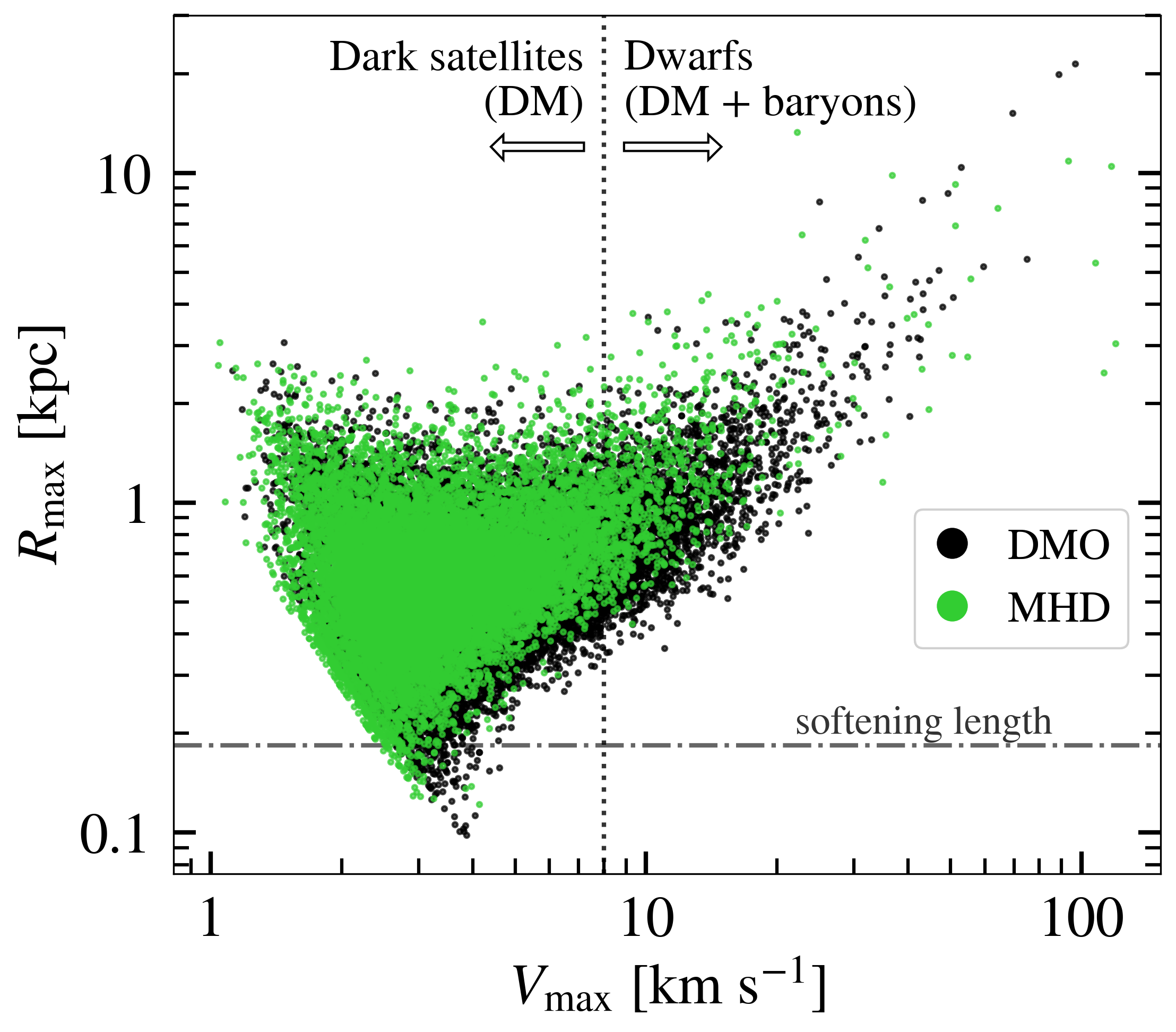}
        \caption{Pair of $V_\mathrm{max}$ and $R_\mathrm{max}$ values of subhalos in the six Auriga MW-like halos analyzed in this work. Data from both DMO (black points) and MHD (green) runs are shown. The dotted vertical line depicts the separation we adopt between those subhalos expected to host dwarf galaxies ('dwarfs') and those that remain dark ('dark satellites'). Full details are given in Section~\ref{sect:auriga_characterization}. The horizontal line at the bottom refers to the Auriga softening length for this level of resolution, i.e.~188~pc.}
        \label{fig:RmaxVmax}
    \end{figure}

    The Auriga cosmological suite is a set of zoom-in magnetohydrodynamical simulations with DMO counterparts \cite{Grand_2017, 2024Grand_releasedataauriga}.
    The DMO runs allow for one-to-one comparisons since they have the same set of initial conditions.
    Among available simulations that include hydrodynamics and magnetic fields, Auriga is distinguished by its high spatial and mass resolution. 
    We use a subset of six MW-like Auriga systems that best reproduce some observational properties like the MW mass~\cite{Grand_2018, Grand_2020}, which have been resimulated at a higher resolution. More precisely, we use data from resolution ``Level 3'', as described in the public data release.\footnote{\url{https://wwwmpa.mpa-garching.mpg.de/auriga/index.html}} The hosts have masses, $M_\mathrm{halo}$, ranging between $1-2 \times 10^{12}\, \mathrm{M}_\odot$, and virial radii, $R_\mathrm{vir}$, between $210-260\,$kpc \cite{2024Grand_releasedataauriga}.

    Due to tidal stripping, subhalos experience continuous mass loss as they orbit their host, and mostly from their outskirts. This ongoing stripping renders their $R_\mathrm{vir}$, and consequently their virial mass, ill-defined. We require alternative parameters to characterize the internal structure of subhalos. We adopt $V_\mathrm{max}$ and $R_\mathrm{max}$ (the latter simply being the radial distance from the subhalo center at which $V_\mathrm{max}$ occurs).
    Indeed, these parameters allow us to describe the subhalo population independently of a particular choice of the subhalo DM density profile \cite{Molin_2017}, and can be directly extracted from the simulations. 
    
    In Fig.~\ref{fig:RmaxVmax}, we show $V_\mathrm{max}$ and $R_\mathrm{max}$ values of all subhalos at redshift $z=0$ in the six Auriga MW-like systems analyzed.
    Black markers correspond to subhalos in the DMO runs (26,717 subhalos, 3,$409-5$,235 in each galaxy), while green markers refer to subhalos in MHD runs (13,855 subhalos, 1,$982-2$,609 per galaxy).
    This indicates that each DMO simulation hosts approximately a factor two more subhalos than its MHD counterparts, implying a higher survival rate. This is expected, as baryons  strengthen the gravitational potential of the host, amplifying the tidal forces that affect subhalos, particularly in the innermost regions~\cite{2017MNRAS.471.1709G,10.1093PhatELVIS,Grand_2020,Grand_2021,2022MNRAS.509.2624G,Aguirre-Santaella_sheddinglight}.
    The Auriga softening length at this resolution level is $l_\mathrm{soft} = 188 \,$pc at $z=0$, and it is shown in the figure as a dotted dashed horizontal line. We do not include data points below this value in our analysis in order to avoid any issues with numerical resolution. 
   In the same figure, we also show as a vertical dotted line the boundary we adopt in our work between those subhalos that could potentially host baryons (depicted as ``dwarfs'' in the plot) and those which should not (``dark satellites'') and should thus remain completely dark. We set this boundary to $M_\mathrm{sub} = 10^7\mathrm{M_\odot}$ or, equivalently, $V_\mathrm{max} = 8\,\mathrm{km/s}$. We discuss this particular choice later in Section~\ref{sect_sub:results_darkSatellites}.

    With all this wealth of subhalo simulated data at our disposal, we now proceed to fully characterize the Auriga subhalo population, both for DMO and MHD runs.     
    From a statistical point of view, we describe the subhalo population with three independent components. The probability of encountering N subhalos at a particular distance from the host halo center, with a particular $V_\mathrm{max}$, and a given concentration, is
    \begin{equation}
    \begin{split}
        \frac{dN}{dV_\mathrm{max} \, dD_\mathrm{GC}\, dc_\mathrm{V}} \propto &\frac{dN}{dV_\mathrm{max}}(V_\mathrm{max})
        \times \frac{dN}{dD_\mathrm{GC}}(D_\mathrm{GC}) \\
        &\times \frac{dN}{dc_\mathrm{V}}(V_\mathrm{max}, R_\mathrm{max}).
    \end{split}
        \label{eq:distrSubhalos}
    \end{equation}
    From left to right in Eq.~(\ref{eq:distrSubhalos}), we find the Subhalo Velocity Function (SHVF), the Subhalo Radial Distribution (SRD), and the dependency with the subhalo velocity concentration $c_\mathrm{V}$. In the following subsections, we discuss every component in detail.

    \subsection{Subhalo Velocity Function (SHVF)} \label{subsect:shvf}
    
        \begin{figure}
            \centering
            \includegraphics[width=\linewidth]{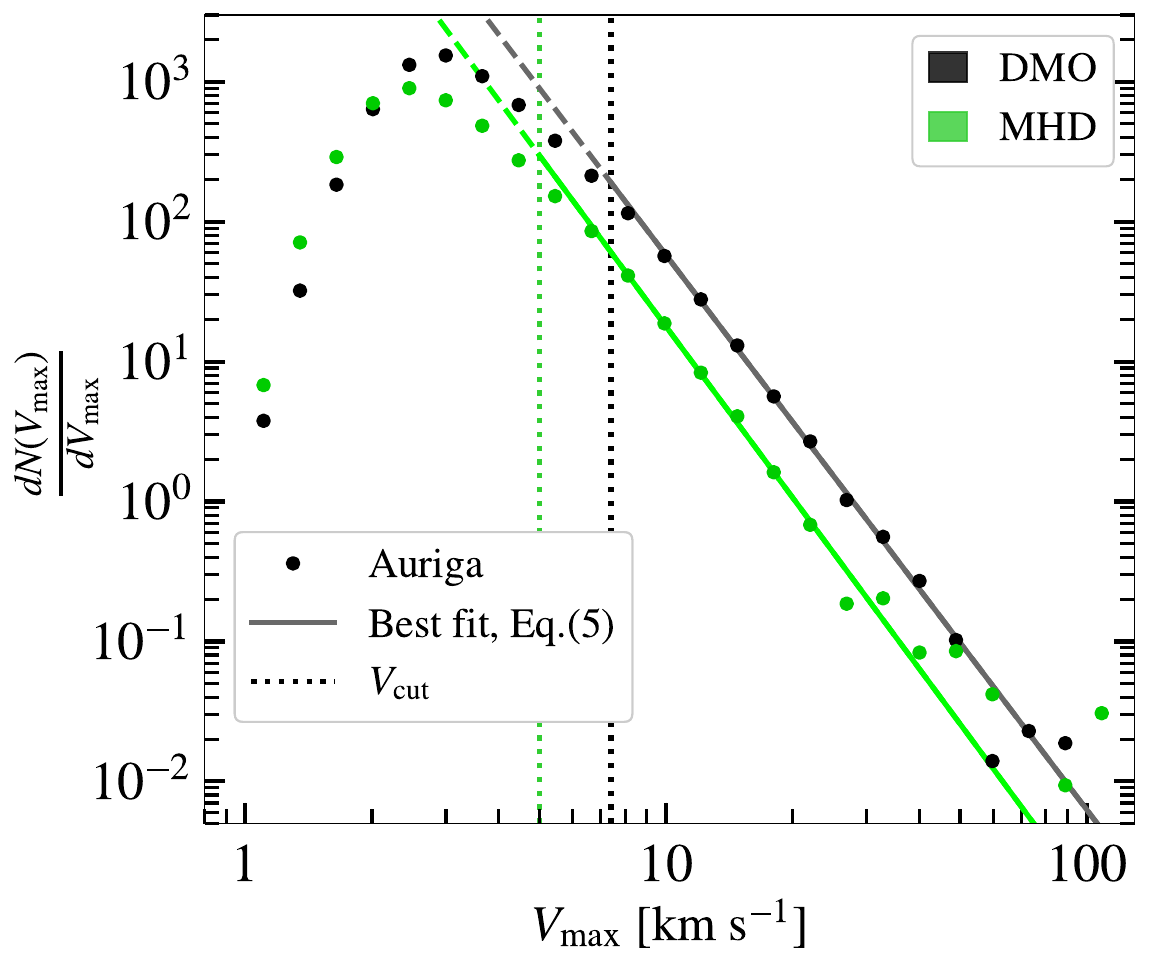}
            \caption{Average SHVF from Auriga data as derived from the six MW-size halos we use in this work, both for DMO (black markers) and MHD (green markers) runs. Power law fits to Eq.~\eqref{eq:shvf} applying bootstrapping techniques are shown as solid lines, with best-fit parameters listed in Table~\ref{tab:bestParams}. The vertical dotted lines refer to the corresponding $V_\mathrm{cut}$ values, the simulation lacking subhalos due to numerical resolution to the left of these lines.}
            \label{fig:SHVF_Grand}
        \end{figure}
        
        The SHVF determines the number of subhalos as a function of their maximum circular velocity (a proxy of the mass), regardless of their position within the host.
        To build the Auriga SHVF, we consider the $V_\mathrm{max}$ values of all subhalos in the six MW-like halos.
        We discretize this $V_\mathrm{max}$ distribution into 24 bins, equally spaced in logarithmic scale.
        We also obtain the differential SHVF, for which we quantify the number of subhalos within each bin and normalize by the bin width. We normalize by the total number of halos (six) to derive an average SHVF per host. We do so both for DMO and MHD. These SHVFs are the ones shown as markers in Fig.~\ref{fig:SHVF_Grand} for the DMO (black) and MHD (green) runs.
        The data follow a power law distribution, departing from it at low $V_\mathrm{max}$ values due to numerical resolution, which translates into a lack of small subhalos in the simulations, see e.g.~\cite{Springel_2008}.
        We define this $V_\mathrm{max}$ from which we start losing small subhalos due to numerical resolution as $V_\mathrm{cut}$, shown as vertical dotted lines in Fig.~\ref{fig:SHVF_Grand} for both DMO and MHD.
        We then fit a power law to the well-behaved part of the data as
        \begin{equation} \label{eq:shvf}
            \frac{dN}{dV_\mathrm{max}} = 10^{V_0} \, V_\mathrm{max}^{\alpha},
        \end{equation}
        where we set the normalization $V_0$ and slope $\alpha$ to be free parameters.
        These parameters are highly dependent on the fitting interval, thus we perform a bootstrapping technique to ensure their statistical robustness~\cite{santaella23}.\footnote{We perform 1000 realizations of the fitting procedure. In each realization, we draw a random sample of the subhalo data. This sample is fitted within a random $V_\mathrm{max}$ interval, always staying above $V_\mathrm{cut}$. All bins included in the fit are required to contain a minimum number of 10 subhalos, which ensures a suitable statistical sample, particularly at the high end of $V_\mathrm{max}$ distribution. Then, we calculate the average and standard deviation of each fitted parameter along all the realizations.}
        The best-fit parameters are listed in Table~\ref{tab:bestParams}, the result being shown as straight lines in Fig.~\ref{fig:SHVF_Grand} for both DMO and MHD, with their extrapolations to lower $V_\mathrm{max}$ indicated as dashed lines.
        The DMO-to-MHD normalizations ratio is $\sim2.5$, this being constant through the well-behaved $V_\mathrm{max}$ range. 
        
        Our MHD results in Fig.~\ref{fig:SHVF_Grand} are consistent to the results in~\cite{Grand_2021}, where the cumulative SHVF of the Auriga MW-like halos is calculated. They are also consistent with previous ones from different simulations, e.g.~\cite{santaella23,Springel_2008,zavala2019dark,Diemand_2008}.

    \subsection{Subhalo Radial Distribution (SRD)} \label{subsect:srd}
        
        \begin{figure*}
            \centering
            \includegraphics[width=\linewidth]{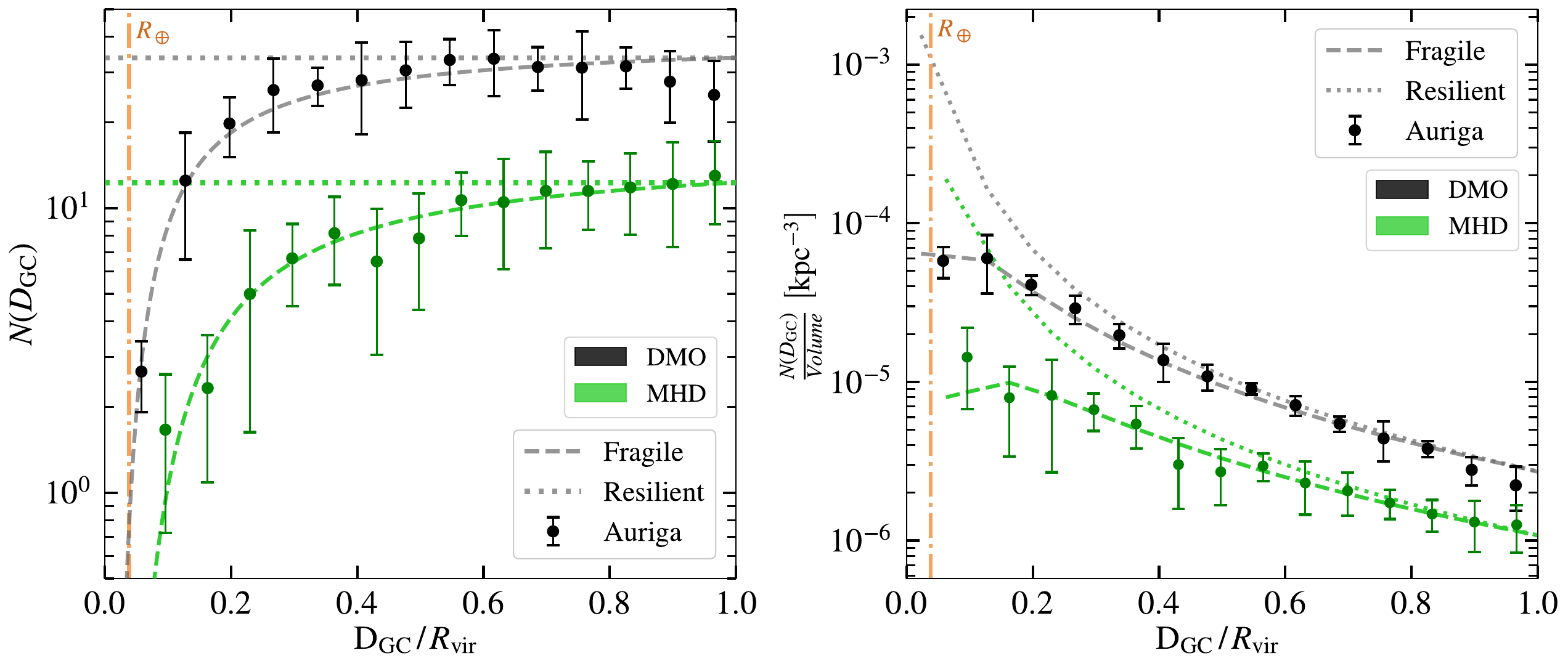}
            \caption{\textbf{Left: }Auriga SRD (data points), derived for subhalos with $V_\mathrm{max} > V_\mathrm{uni}$; see Table~\ref{tab:bestParams} and Section~\ref{subsect:srd} for details.
            Dashed lines are the best fits to the SRD data as given by Eq.~\eqref{eq:srd_original} for the fragile scenario. Also shown are the SRDs for the resilient scenario, Eq.~\eqref{eq:srd_resilient} (dotted lines), both for the DMO (black) and MHD (green) populations. Best-fit values for all these equations are given in Table~\ref{tab:bestParams}. The galactocentric distance of the Earth, $R_\oplus=8.5\,$kpc, is also displayed as a vertical orange dash-dotted line (where we have assumed the virial radius of the MW from observations, $R_\mathrm{vir} = 220\,$kpc \cite{Klypin_2002}, to transform its position into the plot units).
            \textbf{Right:} Same elements than in the left translated to the density space, dividing the subhalo number by the volume associated with the bins. The obtained curves do not have analytical formulas in the density space.
            }
            \label{fig:srd_vol}
        \end{figure*}
        
        The SRD describes how the subhalos are spatially distributed inside the host. The SRD has been proven in most simulations to be universal, i.e.~independent of the subhalo mass or $V_\mathrm{max}$ \cite{ Moline21, Coco16, 2016MNRAS.457.1208H, santaella23, Springel_2008,Diemand_2008}.
        
        The closest subhalo to the GC in Auriga is located at a distance $D_\mathrm{GC} = 5.57\,$kpc among the DMO runs, and $D_\mathrm{GC} = 15.71\,$kpc in MHD.
        Therefore, either there are no bound structures near the GC due to lack of resolution, or there is a very efficient subhalo disruption in this region, enhanced in the second case by the presence of baryons.
        
        On the left panel of Fig.~\ref{fig:srd_vol} we show the number of Auriga subhalos as a function of the $D_\mathrm{GC}$, averaged over the six MW-like hosts under study, as black (green) markers for DMO (MHD).
        We only show subhalos with $V_\mathrm{max} > V_\mathrm{uni}$. Below this value, we checked that the SRD loses its universality, i.e.~it depends on the considered $V_\mathrm{max}$ interval, which would add a further layer of complexity that will be explored elsewhere. The values of $V_\mathrm{uni}$ are given in Table~\ref{tab:bestParams}.
        Note that, since the hosts have slightly different sizes, we normalize the $D_\mathrm{GC}$ of each subhalo by the $R_\mathrm{vir}$ of their host.
        The data show that the inclusion of baryons consistently reduces the number of subhalos at all $D_\mathrm{GC}$ but does not significantly alter the SRD shape.

        We perform fits to the measured SRD adopting two scenarios of subhalo survival as discussed in Ref.~\cite{Stref_2019}. All these fits are performed over the number of subhalos, which are the data points on the left panel of Fig.~\ref{fig:srd_vol}. Our aim is to remain agnostic about the very nature of the subhalo disruption itself, by considering two opposite yet realistic subhalo survival scenarios. The first one, ``fragile'', assumes that subhalos actually do disrupt as much as the simulations suggest. The second scenario assumes that this disruption is an artifact caused by numerical resolution issues. In such scenario, the subhalo number in the innermost region of the host is not expected to decrease. We refer to this alternative scenario as ``resilient''.

        Assuming a fragile scenario, we fit the Auriga data using an exponential cutoff as described in Ref.~\cite{santaella23},
        \begin{equation} \label{eq:srd_original}N_\mathrm{fragile}\left(D_\mathrm{GC}\right) = a_1 \, \exp \left[a_0 \frac{R_\mathrm{vir}}{D_\mathrm{GC}}\right],
        \end{equation}
        where the parameters $a_0$ and $a_1$ have been fitted to our data. The best-fit parameters we found are listed in Table~\ref{tab:bestParams}.
        
        In contrast, the resilient scenario allows for substructure to exist near the GC, even if it cannot be resolved in our simulations. The level of resilience of these low-mass subhalos in the center of their hosts is a free parameter that can lead to large variations in the number of overlooked subhalos~\citep{Stref_2019}. Here, we aim for a conservative approach compared to other, more extreme (yet possible) scenarios as the ones also explored in~\citep{Stref_2019}. Indeed, we decide to adopt a constant subhalo number distribution, independent of the $D_\mathrm{GC}$ (dotted lines on the left panel of Fig.~\ref{fig:srd_vol})
        \begin{equation}
            N_\mathrm{resilient} = N_\mathrm{fragile}\left(D_\mathrm{GC}=R_\mathrm{vir}\right). \label{eq:srd_resilient}
        \end{equation}
        
        The right panel of Fig.~\ref{fig:srd_vol} shows these data points and fits translated to subhalo number density, i.e., number of subhalos divided by the volume of each spherical shell.
        
        The resilient SRD boosts the probability of a subhalo being located within 10\,kpc or closer to the GC by a factor $\sim90$ in the DMO scenario, and $\sim1570$ in MHD. These factors are calculated as the ratio between the normalized cumulative SRDs of the resilient and fragile scenarios, and do not account for the intrinsic difference in subhalo abundance that exists between the fragile and resilient scenarios. We discuss the boost that accounts for subhalo abundance below, in Section~\ref{sect:subsect_extrapolations}.

    \subsection{Velocity concentration (\texorpdfstring{$c_\mathrm{V}$}{Cv})} \label{subsect:cv}
    
        \begin{figure*}
            \centering
            \includegraphics[width=\linewidth]{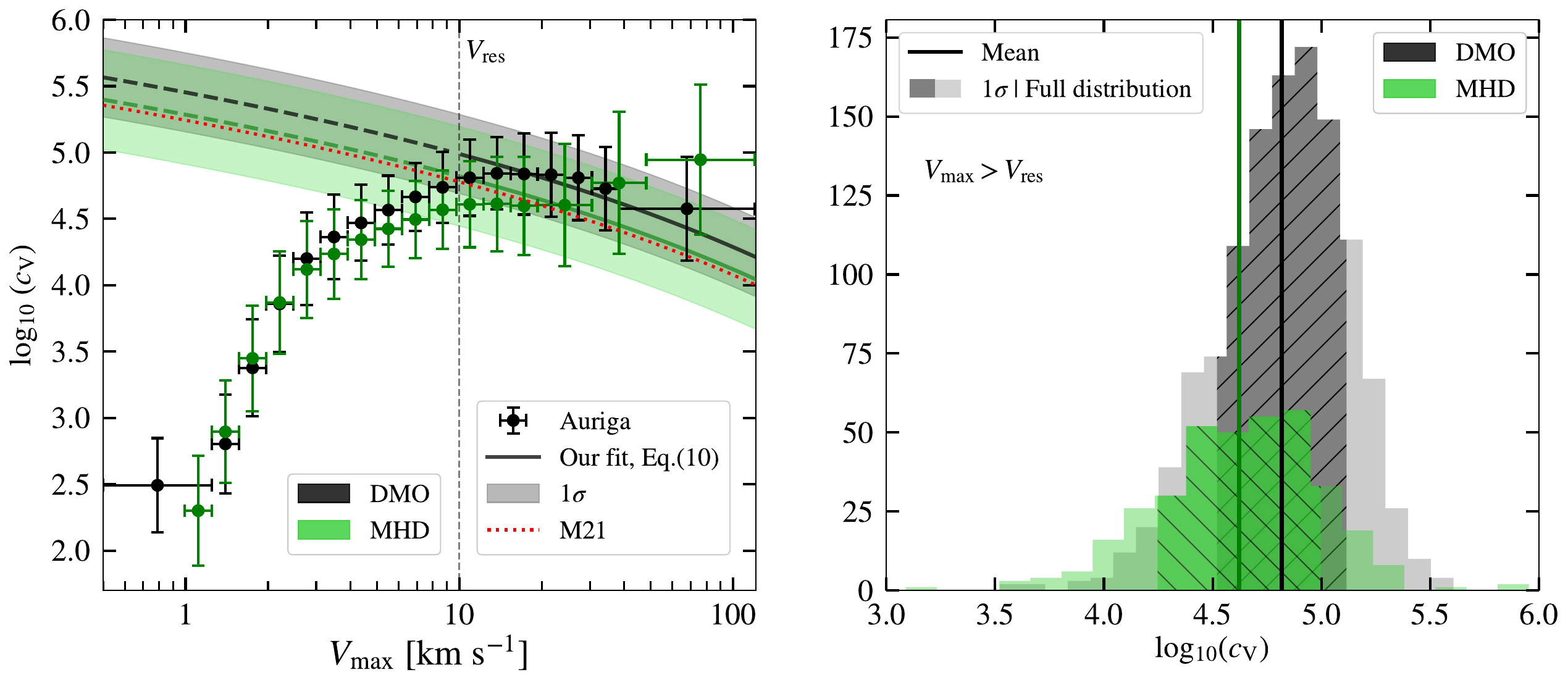}
            \caption{\textbf{Left: }
            Geometric mean values of $c_\mathrm{V}$ as a function of $V_\mathrm{max}$ as found in Auriga for both DMO and MHD scenarios (black and green markers, respectively). 
            The velocity concentration model for DMO by \citetalias{Moline21} at $z=0$ is shown as a dotted red line, with the expression and parameters given in Eq.~\eqref{eq:cv}.
            We have fitted Auriga data to the same parametric form at $V_\mathrm{max} \geq V_\mathrm{res}$, the latter shown as a dashed vertical line.
            Our fits to actual data are shown as straight lines while their extrapolations to smaller $V_\mathrm{max}$ appear as dashed lines. The best-fit values are provided in Table~\ref{tab:bestParams}.
            Shadowed regions depict 1$\sigma$ scatter from the mean values.
            \textbf{Right: }Histogram of individual $c_\mathrm{V}$ values for those data over the resolution limit in the left figure. Opaque stripped regions contain the 1$\sigma$ intervals of the respective distributions. The vertical lines are the geometric mean values of the corresponding DMO (black) and MHD (green) distributions.
            Further explanations are found in the text.}
                \label{fig:cv}
        \end{figure*}

        The concentration of field halos is usually defined as (see, e.g., \cite{2001MNRAS.321..559B,2008MNRAS.391.1940M,2013MNRAS.432.1103L,2012MNRAS.423.3018P,2017MNRAS.472.4918P} and references therein):
        \begin{equation}
            c_{200} = \frac{R_\mathrm{vir}}{r_\mathrm{s}}.
        \end{equation}
        However, as previously mentioned, $R_\mathrm{vir}$ is ill-defined for subhalos, since tidal forces strip mass from the outskirts, indeed removing in most cases all the material at the ``original'' virial radius of the infalling subhalo as it was defined before accretion time. Instead, for subhalos, it is now customary to adopt a definition of the concentration that solely depends on $V_\mathrm{max}$ and $R_\mathrm{max}$, thus independent of any assumptions on the functional form used for the DM density profile nor on $R_\mathrm{vir}$~\cite{2007ApJ...667..859D,Molin_2017,santaella23}. This so-called {\it velocity concentration} is given by \cite{Diemand_2008}:
            \begin{equation}
                c_\mathrm{V} = 2 \left(\frac{V_\mathrm{max}}{H_0\,R_\mathrm{max}}\right)^2.
                \label{eq:cv_definition}
            \end{equation}
        where $H_0$ is the Hubble constant at present time, which we take as $H_0 =67.7 \, \mathrm{km}/\mathrm{s}/\mathrm{Mpc}$ to be consistent with the Auriga simulations \cite{2024Grand_releasedataauriga}. Fig.~\ref{fig:cv} shows mean $c_\mathrm{V}$ and corresponding standard deviation values as a function of $V_\mathrm{max}$, as found in Auriga data. These data were first discretized using 24 $V_\mathrm{max}$ bins equally spaced in logarithmic scale. However, since some of these bins contained less than ten subhalos, especially in the high end of $V_\mathrm{max}$, bins were merged with neighboring ones until each bin contained at least ten subhalos so as to reach statistical robustness.
        
        In $\Lambda$CDM, a flattening of the $c_\mathrm{V}$ curve is expected towards low $V_\mathrm{max}$ \cite{Molin_2017, Moline21,S_nchez_Conde_2014}. Instead, our results in Fig.~\ref{fig:cv} show a decline in Auriga data at the lowest measured $V_\mathrm{max}$ values. This is simply a consequence of lack of numerical resolution in the simulations, thus we restrict ourselves to those $V_\mathrm{max}$ values  safely located to the right of such decline in concentration, that we set at $V_\mathrm{res} = 10\,$km/s as a compromise of statistics and data behavior. It is in this safe range of $V_\mathrm{max}$ values above $V_\mathrm{res}$ where we proceed and perform fits to the $c_\mathrm{V}-V_\mathrm{max}$ relation. Complex models for such relation $c_\mathrm{V} \left(V_\mathrm{max}\right)$ have been explored \cite{Molin_2017, Moline21,S_nchez_Conde_2014}. We adopt the one in \citet{Moline21}, from now on M21, which reads as 
        \begin{equation} \label{eq:cv}
            c_\mathrm{V} \left(V_\mathrm{max}, z=0\right) = c_0 \left[ 1 + \sum_{i=1}^3a_{i} \left[ \mathrm{log}_{10}\left(\frac{V_\mathrm{max}}{\mathrm{km}/\mathrm{s}}\right)\right]^i \right],
        \end{equation}
        where $c_0=1.75\times 10^5$ and $a_i = \left(-0.90368, 0.2749, -0.028\right)$ are the best-fit parameters found in that work. We show this $c_\mathrm{V}-V_\mathrm{max}$ parametrization as a dotted red line on the left panel of Fig.~\ref{fig:cv}.
        Prior to this work, Eq.(\ref{eq:cv}) had only been applied to describe DMO runs. In our study, we extend its application by assuming that it can also properly characterize the subhalo population in our MHD runs. The main challenge in fitting the data to this expression is the limited number of data points in the well-sampled region. The original formulation includes four free parameters, which introduces excessive degrees of freedom and leads to overfitting. We explored several fitting configurations with varying numbers of parameters, and found that allowing two to four parameters to vary consistently resulted in poor generalization. To address this, we adopted a compromise: we fixed the parameters $a_i$ to the best-fit values reported by \citetalias{Moline21}, and only allowed the normalization parameter $c_0$ to vary. The resulting best-fit value is presented in Table~\ref{tab:bestParams}. The corresponding fits for both DMO and MHD are shown as solid lines on the left panel of Fig.\ref{fig:cv}, with their extrapolations to lower $V_\mathrm{max}$ indicated as dashed lines. 
        Interestingly, we find the DMO normalization to be $\sim1.5$ times higher than the MHD one, which indicates that the inclusion of baryons in the simulations results in less concentrated subhalos.
        We can also compare our best-fit DMO and MHD normalizations with the original one in \citetalias{Moline21}.
        The three $c_\mathrm{0}$ values overlap when taking into account the 1$\sigma$ scatter in our data, with factor $\sim$2 variations among values at most.
        The significant scatter present in our data limits the possibility of a more in-depth analysis of the found differences. Also, it is important to note that \citetalias{Moline21} and our work employ different sets of cosmological parameters, spatial and mass resolutions, simulation methodologies, and subhalo identification algorithms. Given all these differences, a direct comparison between concentration values is not straightforward.

        We also examine the scatter of the $c_\mathrm{V}$ distribution in further detail. The right panel of Fig.~\ref{fig:cv} shows the histograms of $c_\mathrm{V}$ for individual subhalos with $V_\mathrm{max} \geq V_\mathrm{res}$, together with geometric means as vertical lines. The 1$\sigma$ interval of the distributions is presented as the opaque stripped regions of the histograms, and these scatter values are depicted as shaded regions on the left panel of the same figure.
        An extensive study of concentration scatter was conducted by \citet{Molin_2017} (M17 hereafter) using  DMO simulations alone. \citetalias{Molin_2017} mainly employs $c_{200}$, but also provides the conversion formula to translate $c_{200}$ values into $c_\mathrm{V}$ ones by adopting a NFW density profile. The largest scatter reported in their subhalo sample is $\sigma_{200}^\mathrm{M17} = 0.15$, which translates to $\sigma_\mathrm{V}^\mathrm{M17} = 0.39$. We detail this conversion in Appendix~\ref{appendix:CvandC200}.
        This value is very similar to the scatter found in our data both for the DMO and MHD runs, as listed in Table~\ref{tab:bestParams} and shown in Fig.~\ref{fig:cv}. Interestingly, we find the MHD scatter to be significantly larger than the one in DMO (0.38 versus 0.30).

        As detailed in this section, during the analysis of the Auriga suite we have encountered significant resolution limits that prevent us from studying low-mass subhalos accurately. Since this is critical for our objectives, we now turn to subhalo repopulation as a key technique to investigate the properties and relevance of those subhalos close to or below the resolution limit of the parent simulation.

    \begin{table*}
      \centering
        \begin{tabular}{ccccccccccc}
    \toprule\midrule
          & \multirow{3}{*}{\begin{tabular}{c}Number of\\subhalos \end{tabular}} & \multicolumn{3}{c}{SHVF [Eq.~\eqref{eq:shvf}]} & \multicolumn{3}{c}{SRD fragile [Eq.~\eqref{eq:srd_original}]} & \multicolumn{3}{c}{$c_\mathrm{V}$ [Eq.~\eqref{eq:cv}]} \\
    \cmidrule(lr){3-5} \cmidrule(lr){6-8} \cmidrule(lr){9-11} 
     && \multirow{2}{*}{$\alpha$} & \multirow{2}{*}{$V_0$} & $V_\mathrm{cut}$ & \multirow{2}{*}{a$_0$} & \multirow{2}{*}{a$_1$} & $V_\mathrm{uni}$ & c$_0$ & $1\sigma$ & $V_\mathrm{res}$ \\
     &&&& [km/s] &&& [km/s]& $\left(\times10^5\right)$  &  scatter & [km/s]\\
     \midrule
    DMO & 26,717 & $-3.9 \pm 0.2$ & $5.7 \pm 0.2$ & 7.4 & $-0.15 \pm 0.01$ & $39 \pm 2$ & 8 & $2.8 \pm 0.3$ & 0.30 & 10 \\
    MHD & 13,855 & $-4.1 \pm 0.3$ & $5.3 \pm 0.3$ & 5.0 & $-0.27 \pm 0.02$ & $16 \pm 1$ & 8 & $1.9 \pm 0.5$ & 0.38 & 10 \\
    \bottomrule
    \end{tabular}
    \caption{Summary of ingredients and their best-fit parameters that characterize the Auriga subhalo population. Full details are given in Section~\ref{sect:auriga_characterization}. SHVF stands for the subhalo velocity function (Section~\ref{subsect:shvf}), SRD for the subhalo radial distribution (Section~\ref{subsect:srd}), and $c_\mathrm{V}$ is the subhalo velocity concentration (Section~\ref{subsect:cv}).}
    \label{tab:bestParams}
    \end{table*}

\section{Repopulating Auriga with low-mass subhalos} \label{sect:repopulation_theory}

    The repopulation algorithm we adopt starts from the SHVF, SRD, and $c_\mathrm{V}-V_\mathrm{max}$ recipes that characterize the subhalo population of Auriga, derived in the previous section, to draw new subhalos according to such recipes, both above and below the original resolution limit. The latter is achieved by extrapolating the found prescriptions to lower subhalo masses not covered by the original simulation. The way these extrapolations behave is usually based on theoretical expectations deeply rooted in $\Lambda$CDM, but could in principle be of any kind. Full details of our extrapolations are provided and discussed in  detail in the subsequent subsections.
    
    The original repopulation code is a Python script originally presented and detailed in Ref.~\cite{santaella23}, which had previously been used in different works~\cite{Coronado_Bl_zquez_2022, Coronado_Bl_zquez_2019, 2019JCAP...11..045C}.
    For this work, we have implemented new functions and dependencies, as well as improved its performance, as detailed below.

    \subsection{Extrapolation recipes below the Auriga resolution limit} \label{sect:subsect_extrapolations}

        After our careful characterization of Auriga subhalo data, conveniently summarized in Table~\ref{tab:bestParams}, we concluded that there is a lack of subhalos in the SHVF below $V_\mathrm{cut}=7.4 \, (5)\,\mathrm{km}/\mathrm{s}$ in DMO (MHD) (Section~\ref{subsect:shvf}); that the SRD loses universality below $V_\mathrm{uni}= 8\,\mathrm{km}/\mathrm{s}$ (Section~\ref{subsect:srd}); and that the velocity concentration relation is only reliably defined over $V_\mathrm{res}= 10\,\mathrm{km}/\mathrm{s}$ (in Section~\ref{subsect:cv}).
        The apparent discrepancy among these different velocity values can be understood by the different requirements and restrictions at work for each of the mentioned ingredients (SHVF, SRD, $c_V$).
        For example, the minimum number of bound particles required to detect and track a collection of bound particles (e.g., a subhalo) is considerably lower than the minimum required to study its internal structure~\cite{Moline21, ErraniNavarro21}. Likewise, the scarce number of subhalos with $V_\mathrm{max}>30\,\mathrm{km}/\mathrm{s}$ limits the statistical sample of the SHVF in the high $V_\mathrm{max}$ range, yet the internal structure of such massive subhalos is well-resolved.

        The extrapolation of the SHVF below $V_\mathrm{cut}$ is straightforward. 
        Previous studies have demonstrated that with increasing resolution the SHVF extends as a power law towards lower $V_\mathrm{max}$ values \cite{Diemand_2008,Springel_2008,Grand_2021}, in agreement with predictions from the Press-Schechter formalism in $\Lambda$CDM cosmology~ \cite{1974ApJ...187..425P,1999MNRAS.308..119S,1998ApJ...499...20J}. Thus, we simply adopt a power law extrapolation of the measured SHVF (Eq.~\eqref{eq:shvf} with the best-fit parameters in Table~\ref{tab:bestParams}) down to the smallest $V_\mathrm{max}$ we aim to simulate, both for DMO and MHD. These extrapolations are shown as dashed lines in Fig.~\ref{fig:SHVF_Grand}.
        
        As for the SRD, our repopulated subhalos follow the same distributions already presented for subhalos above the resolution limit, i.e.,  Eq.~\eqref{eq:srd_original} for the fragile scenario and Eq.~\eqref{eq:srd_resilient} for the resilient one, both depicted in Fig.~\ref{fig:srd_vol}. These distributions are applied to all repopulated subhalos, independently of their $V_\mathrm{max}$. This ``universality'' of the SRD is motivated by an ample number of works, e.g.~\cite{Springel_2008,10.1111/j.1365-2966.2012.21564.x,2016MNRAS.457.3492H,2016MNRAS.457.1208H}. 
        
        We note though that, in the resilient scenario, we introduce an additional modification to the input parameters of the repopulation. 
        Indeed, in this case, if we simply repopulated with the original SHVF parameters, the number of subhalos in the outskirts would be decreased in comparison. This is so because the total number of subhalos in fragile and resilient scenarios that adopt the same SHVF would be the same, but resilient subhalos would be allowed to lie within the innermost regions of the host --where fragile subhalos cannot exist-- only at the expense of decreasing their number in the outskirts. To consistently reproduce both the distribution and the number of subhalos found in the outskirts of Auriga host halos, we increase the normalization of the resilient SHVF so that the number of repopulated subhalos with $V_\mathrm{max}>8\,\mathrm{km}/\mathrm{s}$ in the outskirts matches the number of subhalos present in the Auriga suite.
        With this change in normalization, the resilient scenario boosts the probability of subhalos being located within 10\,kpc of the GC by a factor of $\sim113$ (DMO) and $\sim2550$ (MHD).
        
        Finally, for the $V_\mathrm{max} - c_\mathrm{V}$ relation of subhalos below $V_\mathrm{res}$, we extrapolate mean DMO and MHD values by simply adopting Eq.~\eqref{eq:cv} with the best-fit normalizations we found for our data. The corresponding extrapolations were already presented as dashed lines on the left panel of Fig.~\ref{fig:cv}. Repopulated $V_\mathrm{max} - c_\mathrm{V}$ data are then scattered following a Gaussian distribution in log$_{10}$ space, with DMO and MHD standard deviations as derived from Auriga above $V_\mathrm{res}$, shown as the shaded regions on both panels of Fig.~\ref{fig:cv}. We recall that normalization and scatter values are listed in Table~\ref{tab:bestParams}.

        We list all the input parameters of our repopulations in Appendix~\ref{appendix:input_params_repops}.

    \subsection{Details of the repopulation algorithm} \label{subsect:details_repop_algorithm}
        
        \begin{figure*}
            \centering
            \includegraphics[width=0.95\linewidth]{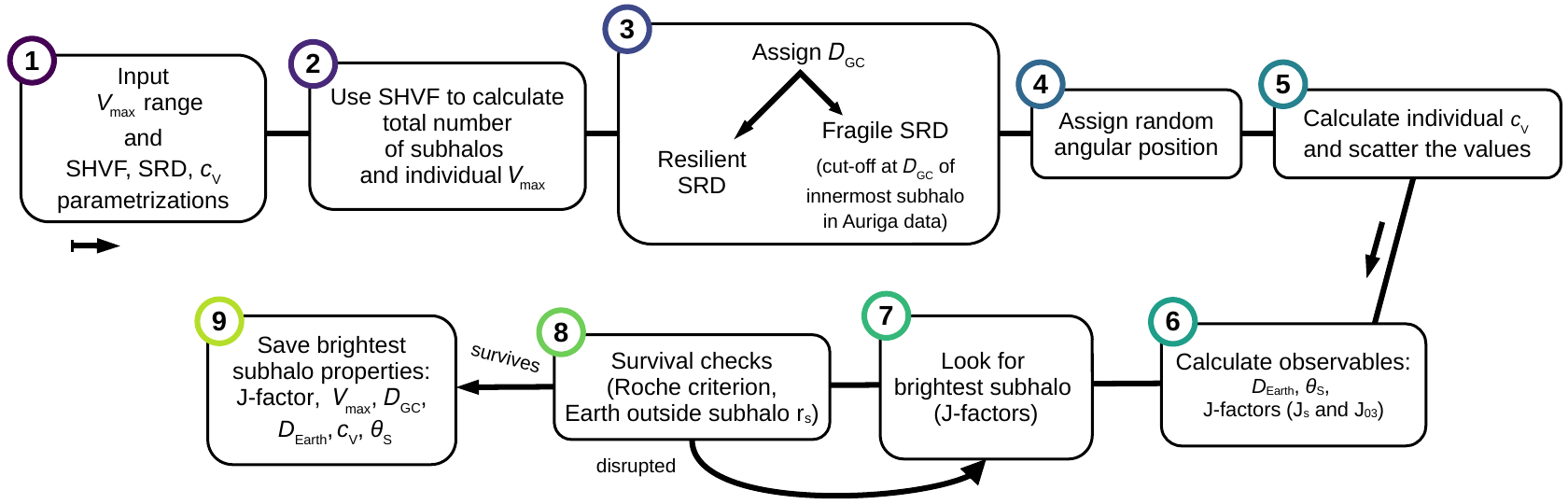}
            \caption{Flowchart of how the repopulation algorithm works; see Section~\ref{subsect:details_repop_algorithm} for full details on each of the steps.}
            \label{fig:flowchart}
        \end{figure*}
        
        The subhalo repopulation algorithm generates a new population of subhalos according to the input parametrizations described above, allowing us to create subhalos with $V_\mathrm{max}$ smaller than those present in the original Auriga runs.
        Fig.~\ref{fig:flowchart} summarizes in a flow chart how the repopulation algorithm works. Step by step, the procedure is: 
        \begin{enumerate}
            \item Input parameters: the $V_\mathrm{max}$ range for repopulation is specified, as well as the parametrizations to be used for the SHVF, SRD, and $c_\mathrm{V}$ distributions, introduced in Section~\ref{sect:auriga_characterization}, and with the parameter values listed in Table~\ref{tab:bestParams} .
            \item Calculate subhalo abundance: determine the total number of subhalos using the SHVF, and assign an individual $V_\mathrm{max}$ to each subhalo according to the SHVF in Section~\ref{subsect:shvf}. We use the inverse transform sample algorithm \cite{inverse_algorithm} to generate the individual values of $V_\mathrm{max}$.
            \item Assign subhalo $D_\mathrm{GC}$: use the SRD in Section~\ref{subsect:srd} to spread the subhalos radially throughout the host. In the fragile scenario, the SRD goes to zero at the distance of the innermost subhalo in Auriga. We adopt the same method used for $V_\mathrm{max}$ \cite{inverse_algorithm} to generate individual $D_\mathrm{GC}$ values.
            \item Angular placement: assign each subhalo a random position over the solid angle.
            \item Concentration values: assign median $c_\mathrm{V}$ using Eq.~\eqref{eq:cv}. Then, a scatter is added following the Gaussian scatter in log-space described in Section~\ref{subsect:cv}.
            \item Calculation of observables: we compute several quantities that will be later used for our purposes, i.e., the distance $D_\mathrm{Earth}$ from each subhalo to Earth (the latter placed at $D_\mathrm{GC}=8.5\,$kpc), and the angular size $\theta_S$ and the J-factor as observed from Earth for each subhalo. The latter is described in Section~\ref{sect:subs_as_gamma_targets}, and computed via Eq.~\eqref{eq:Js} and Eq.~\eqref{eq:J03}.
            \item Identification of the brightest subhalo in terms of its annihilation luminosity, i.e., the highest J-factor.
            \item Subhalo survival checks: ensure that the brightest subhalo survives the Roche criterion for MW-subhalo tidal interactions (see definition in Appendix~\ref{appendix:r_s_and_roche_criterium})\footnote{Subhalos in the original Auriga runs should ``naturally'' respect and account for the Roche criterion, as they have been realistically simulated across cosmic history. However, this may not be the case for our repopulated subhalos, as they are placed {\it ad hoc} within the host at present time, with no evolution. Although our SRD and SHVF parametrizations should implicitly carry the information on the Roche criterion, we decided to be safe and implement it explicitly as well. As expected, in the fragile scenario we found the subhalo rejection to be very low: only 0.2\% of our brightest subhalos violate the criterion (i.e., only one MHD repopulation). However, in the resilient scenario, this percentage increases to 31\% and 43\% in DMO and MHD, respectively. These numbers refer to the brightest subhalos across realizations. Instead, Appendix~\ref{appendix:repopulation_mimics_Auriga} shows results for all subhalos in a single repopulation.}, and that the Earth is placed outside the $r_\mathrm{s}$ of the subhalo.\footnote{If a subhalo is placed engulfing the Earth, the J-factor expression does not follow Eq.~\eqref{eq:Js} nor Eq.~\eqref{eq:J03} anymore. This adds a layer of complexity that will be explored elsewhere.} If these criteria are not met, the subhalo is discarded and the algorithm goes back one step, searching again.
            \item The most relevant parameters of the repopulated subhalos are saved and ready to be used for our analyses.
        \end{enumerate}
        We note that, although our main purpose is to repopulate Auriga with low-mass subhalos well beyond the original resolution limits, for each realization we utilize the repopulation algorithm to consistently create the full subhalo population of a MW-size halo, i.e., from the most massive subhalos well resolved in the parent simulation to the lightest subhalos only present in the repopulation data.    
        As a cross-check, in Appendix~\ref{appendix:repopulation_mimics_Auriga} we show that our repopulation mimics the resolved Auriga subhalo population down to its resolution limits. 

    \subsection{Creating statistically meaningful samples}
        
        At this point, we are in position to run the repopulation algorithm to generate subhalos consistently with the Auriga characterization performed in Section~\ref{sect:auriga_characterization}, and with the extrapolations detailed in Section~\ref{sect:subsect_extrapolations} for subhalos below the Auriga resolution limit.      
        We generate and analyze four scenarios, combining two levels of subhalo resilience (``fragile'' versus ``resilient'') with the presence or absence of hydrodynamics (DMO versus MHD).
        
        As discussed in Section~\ref{sect:auriga_characterization}, the original subhalo population in Auriga MHD runs exhibits a smaller number of subhalos and lower concentrations compared to its DMO counterpart. This may be due to a combined effect of an enhanced tidal stripping in the case of baryons near the Galactic center, where they reshape the host density profile, coupled with artificial numerical disruption. Since disentangling these effects is not a trivial task and represents still an open debate in the community, in our work we decide to remain agnostic on this issue and repopulate the host considering both fragile and resilient scenarios, for both DMO and MHD, following Section~\ref{subsect:srd}.

        As shown in Section~\ref{subsect:shvf} and listed in Table~\ref{tab:bestParams}, resolution issues affect the Auriga SHVF below $V_\mathrm{cut} \sim \mathcal{O}(1\,\mathrm{km/s})$. We repopulate down to $V_\mathrm{max}= 0.1\,$km/s, thus generating subhalos with $V_\mathrm{max}$ more than one order of magnitude smaller than those resolved in the original Auriga runs.
        As described in Section~\ref{sect:subsect_extrapolations}, the existence of these subhalos is not only expected by  theoretical predictions within the $\Lambda$CDM cosmological framework, but its survival down to present time is also supported by several recent works~\cite{2020MNRAS.491.4591E,2021arXiv211101148A,10.1093/mnras/stad844,Aguirre-Santaella_sheddinglight,2025AguirreSantaella}.
        
        As a final step, and for each of the four scenarios, depending on the level of subhalo resilience and the inclusion or not of hydrodynamics, we perform 500 independent repopulations to ensure a proper statistical sample. For our purposes and the remainder of this work, in each repopulation we select the brightest subhalo in terms of expected annihilation signal, i.e., the one with the highest J-factor. Thus, our results below correspond to the subsample of the 500 brightest subhalos across our realizations. Again, we recall that the set of parameters used in our repopulations can be all found in Appendix~\ref{appendix:input_params_repops}.

\section{Impact of baryons in gamma-ray subhalo searches} \label{sect:repopulation_results}

    In this section, we present two specific applications of the use of our repopulation algorithm on Auriga data, both of them in the context of DM searches. First, we derive the distribution of J-factors of the repopulated Auriga subhalo population under the different scenarios considered in our work (fragile and resilient, for both DMO and MHD runs). Then, we assess the impact of these newly derived J-factor distributions on DM constraints previously obtained from observations of gamma-ray unidentified sources.
    
    \subsection{Subhalo J-factors in the repopulated Auriga } \label{subsect:results_baryons}
            
        \begin{figure}
            \centering
            \includegraphics[width=0.95\linewidth]{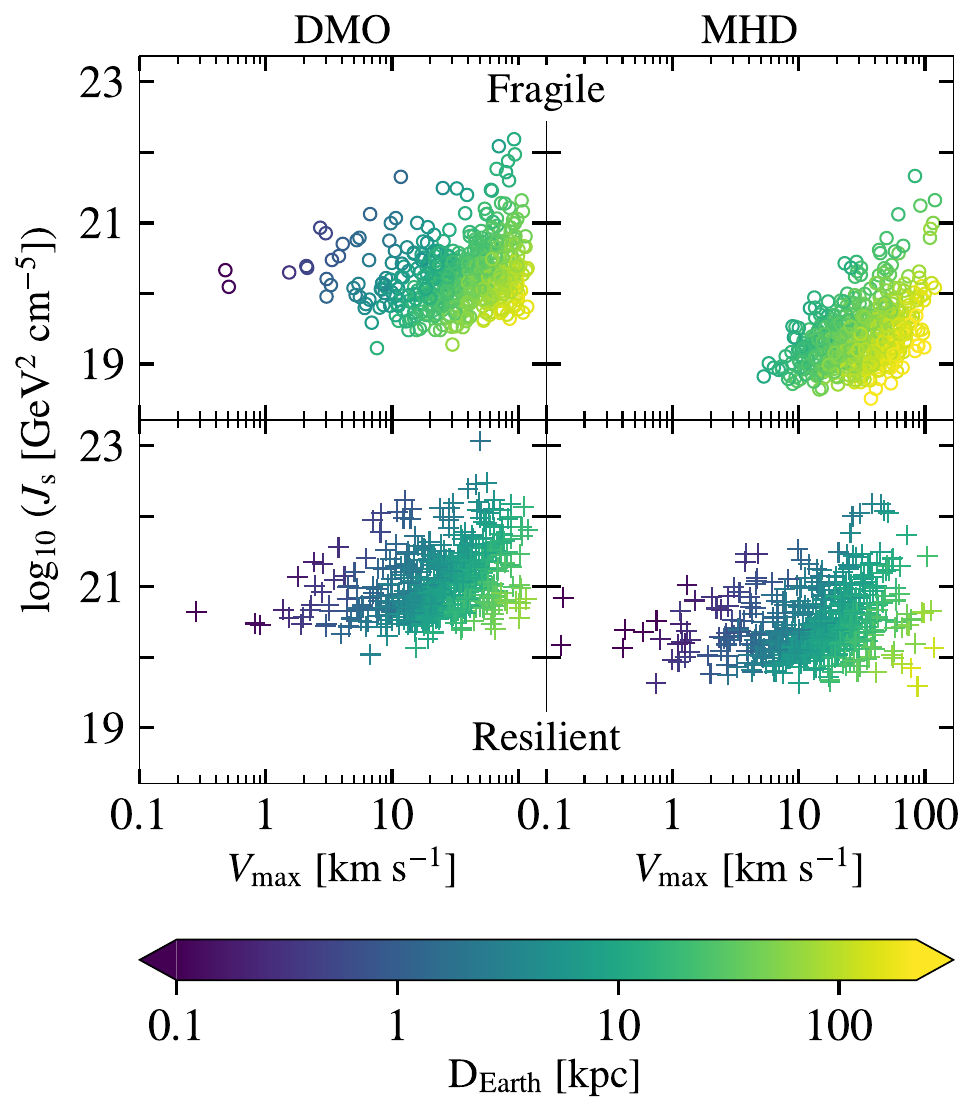}
            \caption{$J_\mathrm{S}$ (given by Eq.~\eqref{eq:Js}) versus $V_\mathrm{max}$ for the brightest subhalo in 500 repopulations with $V_\mathrm{max}\in [0.1, 120]\,$km/s, with the colored z-axis representing the distance to Earth. Different scenarios are shown depending on subhalo resilience (top and bottom panels for fragile and resilient populations, respectively) and inclusion or not of hydrodynamics (left for DMO; right for MHD).
            }
            \label{fig:VmaxJs_to120}
        \end{figure}

        Some low-mass subhalos may be potentially relevant for DM searches provided the right combination of their proximity to Earth and mass, thus we decide to adopt a wide range of $V_\mathrm{max}\in [0.1, 120]\,\mathrm{km}/\mathrm{s}$ for our Auriga repopulation, with the minimum value corresponding to subhalos a factor $\sim 10^6 - 10^7$ less massive than the resolved ones, and the maximum corresponding to the highest $V_\mathrm{max}$ value found in the Auriga ``Level 3'' suite. 

        Fig.~\ref{fig:VmaxJs_to120} shows the J-factor value (more specifically, $J_S$ given by Eq.(\ref{eq:Js}))  of the brightest subhalo in each of our 500 repopulations, as a function of $V_\mathrm{max}$. Each panel refers to one of the studied scenarios: fragile and resilient, for both DMO and MHD. Color represents $D_\mathrm{Earth}$. As it can be seen, and as expected, both the fragile and resilient MHD subhalo populations are typically less bright than their DMO counterparts, since MHD subhalos are typically less concentrated than DMO ones (Fig.~\ref{fig:cv}) and $J_S$ scales as $\sqrt{c_v}$. Also, the figure shows that the brightest fragile subhalos are generally located at larger distances from Earth than resilient ones. 
        Besides, in most repopulations, even though we generated subhalos down to $V_\mathrm{max} = 0.1$\,km/s, most of the brightest subhalos have $V_\mathrm{max} \geq 1$\,km/s, in agreement with previous studies that used DMO simulations \cite{santaella23,Coronado_Bl_zquez_2019}. The only exception is the resilient MHD scenario, for which small subhalos near Earth are the brightest in 2\% of our repopulations. 
        This is due to a combination of several factors: as said MHD subhalos are typically less concentrated than DMO ones, so for a given $V_\mathrm{max}$ value, $D_\mathrm{Earth}$ gains comparatively more relevance in the computation of the J-factor for MHD subhalos, Eq.(\ref{eq:Js}). 
        On the other hand, the latter are numerous in the innermost parts of the Galaxy --where the Sun is located-- while we expect no fragile subhalos at such small galactocentric radii, as codified in the corresponding SRDs (Fig.~\ref{fig:srd_vol}).
    
        \begin{figure}
            \centering
            \includegraphics[width=0.75\linewidth]{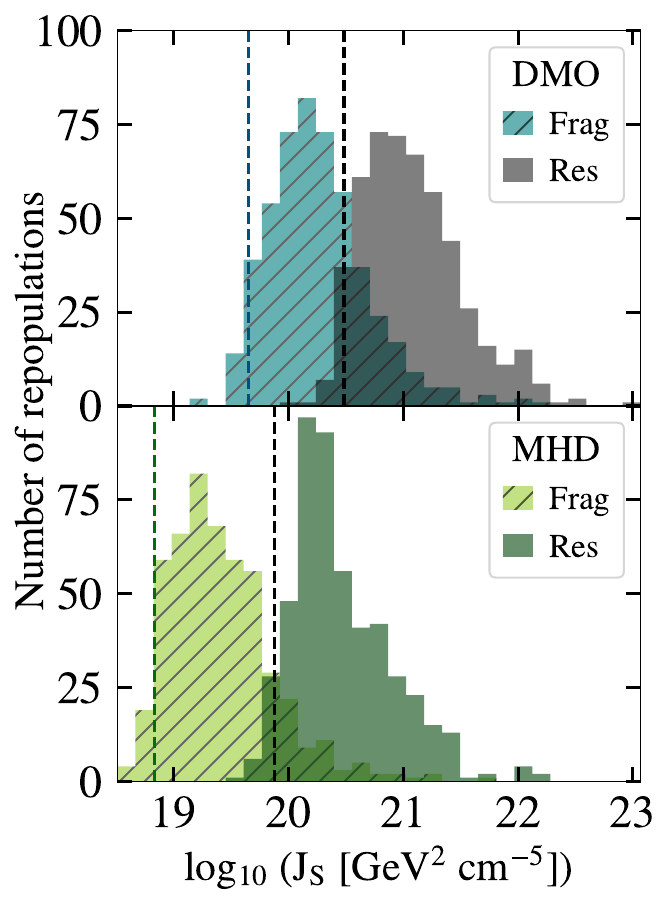}
            \caption{$J_\mathrm{S}$ (given by Eq.~\eqref{eq:Js}) histograms built from the brightest subhalo in 500 repopulations with $V_\mathrm{max}\in [0.1, 120]\,$km/s. The different histograms depict different scenarios of subhalo resilience (fragile vs.~resilient) and inclusion or not of hydrodynamics (top for DMO; bottom for MHD). The vertical dashed lines for each histogram represent $J_{95}$ values, i.e., the value of the J-factor above which 95\% of the corresponding population is contained. See text for full details.
            }
            \label{fig:J_hist_to120}
        \end{figure}

        We provide a more 'compressed' view of the subhalo J-factors in the form of histograms in Fig.~\ref{fig:J_hist_to120}. The J-factor distributions corresponding to the fragile scenario exhibit consistently lower values than the resilient ones, for both DMO and MHD runs. This indicates a direct correlation between the resilience of the population of subhalos and high J-factor values. The vertical dashed lines in the figure represent the 95$^\mathrm{th}$ percentile of the brightest subhalos, so-called $J_{95}$ in Refs.~\cite{Coronado_Bl_zquez_2022,Coronado_Bl_zquez_2019,2019JCAP...11..045C}.
        This $J_{95}$ represents the J-factor threshold above which 95\% of the distribution lies. As in the mentioned works, we here select this value to derive a benchmark, conservative J-factor indicative of the brightest subhalo in a MW-like halo.
        Furthermore, $J_{95}$ values will be particularly useful in the next subsection when discussing the impact of our findings for current DM constraints.
        Using $J_{95}$ as a proxy of the J-factors of the entire subhalo population, we conclude that DMO subhalo J-factors are higher than their MHD counterparts by 0.80~dex in fragile and 0.48~dex in resilient repopulations. Resilient subhalo J-factor values are higher than their fragile counterparts by 0.86~dex (DMO) and 1.18~dex (MHD).
        Finally, we note that the overlap between fragile and resilient histograms is more pronounced in DMO, while MHD has almost no overlap.

        \begin{figure}
            \centering
            \includegraphics[width=0.95\linewidth]{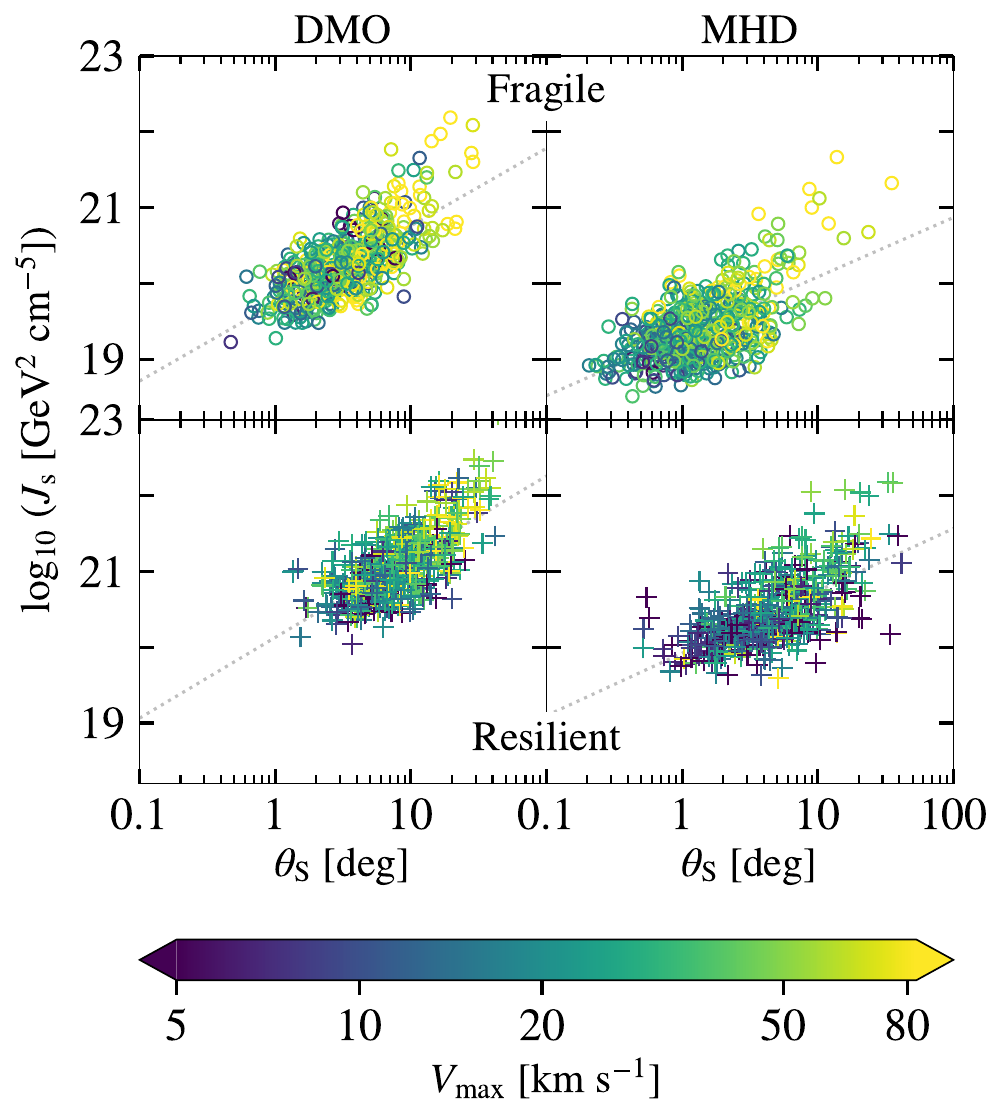}
            \caption{$J_\mathrm{S}$ (given by Eq.~\eqref{eq:Js}) versus angular radius $\theta_\mathrm{S}$ for the brightest subhalo in 500 repopulations with $V_\mathrm{max}\in [0.1, 120]\,$km/s, with the colored z-axis representing $V_\mathrm{max}$. Different scenarios are shown depending on subhalo resilience (top and bottom panels for fragile and resilient populations, respectively) and inclusion or not of hydrodynamics (left for DMO; right for MHD). Dotted gray lines show the found linear relations in $\log_{10}-\log_{10}$ scale; see text for details.}
            \label{fig:AngJs_to120}
        \end{figure}
        
        Finally, in Fig.~\ref{fig:AngJs_to120} we examine the spatial extension of the brightest subhalos. We define the angular size subtended in the sky by a subhalo as $\theta_\mathrm{S} = \arctan\left(r_\mathrm{s}/D_\mathrm{Earth}\right)$, that is, the angular extension of the scale radius as observed from Earth.
        Our results show a direct correlation between $\theta_\mathrm{S}$ and J-factor. Adopting a linear relation between both quantities in $\log_{10}-\log_{10}$ scale, we find the slopes to be $\sim1.0$ for DMO and $\sim0.8$ for MHD, both for the resilient and fragile scenarios. We show these correlations as dotted gray lines in Fig.~\ref{fig:AngJs_to120} as well. 
        No clear correlation is observed between $\theta_\mathrm{S}$ and $V_\mathrm{max}$ (the latter depicted with colors in the figure).

        Additional in-depth analyses along the lines followed in this subsection are presented in Appendix~\ref{appendix:results_to_120}: a study of the J-factor distributions as a function of $D_\mathrm{GC}$ and $V_\mathrm{max}$, as well as the corresponding $V_\mathrm{max}$ histograms.

    \subsection{Impact on current dark satellites' DM constraints}  \label{sect_sub:results_darkSatellites}
        
        We present a use case for our repopulation algorithm: the search of dark satellites, i.e., DM substructures without a baryonic counterpart.
        Specifically, we examine how the latest DM constraints derived from the lack of signals from these targets \cite{Coronado_Bl_zquez_2022} are modified, on one hand, by the inclusion of baryons and, on the other, by the level of resilience of the subhalo population. 
        
         In the context of dark satellites, the maximum value to be considered for $V_\mathrm{max}$ is constrained by the emergence of baryons in subhalos. Different works have investigated this value, finding a complex dependence with e.g.~reionization history, gas cooling, assembly bias, etc., and a minimum host halo mass for visible galaxies to be in the range $\sim10^6-10^9~M_{\odot}$~\cite{2015MNRAS.448.2941S,2016MNRAS.456...85S,2018MNRAS.473.2060J,Nadler_2020,2022ApJ...940....8M, Nadler_2025}. We simply follow \cite{Coronado_Bl_zquez_2019,Coronado_Bl_zquez_2021,2019JCAP...11..045C,Coronado_Bl_zquez_2022} and adopt as boundary between visible and invisible satellites the $V_\mathrm{max}$ that corresponds to a subhalo mass of $10^7~ \mathrm{M_{\odot}}$, i.e., $V_\mathrm{max}\approx8$\,km/s. We show this value as a dotted vertical line in Fig.~\ref{fig:RmaxVmax}.
         The adoption of a value similar to the one in these works is not only particularly useful but also key for the purposes of this section, as it will allow us to perform a simple rescaling of DM limits to show the impact of our results, as explained later below.

        Previous studies have also investigated how the brightest of these dark satellites, in terms of their expected annihilation signals, would be observed by current gamma-ray instruments~\cite{PhysRevD.102.103010,Coronado_Bl_zquez_2022}. In particular, Ref.~\cite{Coronado_Bl_zquez_2022} showed that  Fermi-LAT would always detect them as extended objects, with typical angular sizes $\sim0.3^\circ$. This information was then used by the same authors to update earlier WIMP constraints, once they verified the absence of such extended signals from potential dark satellite candidates --previously pinpointed among current gamma-ray catalogs. The updated DM constraints became a factor $2-3$ weaker, given the decrease of J-factor values used to set the constraints with respect to previous work.\footnote{Earlier works had always adopted as subhalo J-factor the one integrated up to the whole angular size of the object, which, for the brightest subhalos --the only ones of interest for DM searches-- is much larger than $0.3^\circ$, as discussed in Section~\ref{subsect:results_baryons} and shown in Fig.~\ref{fig:AngJs_to120}. Thus, adopting  a J-factor only integrated within the innermost $0.3^\circ$, instead of the whole subhalo, implies a substantial decrease of its value and, thus, of the corresponding DM constraints.} Thus, in order to make a one-to-one comparison with the J-factors used in Ref.~\cite{Coronado_Bl_zquez_2022}, we now introduce a definition of the integrated J-factor that depends on the angle of integration as seen from Earth, and use it to integrate our subhalos up to the radius that corresponds to $\theta = 0.15^\circ$ (so that they would be seen as gamma-ray blobs $0.3^\circ$ in size). We follow the notation of Ref.~\cite{Coronado_Bl_zquez_2022} and refer to this J-factor as $J_{03}$ from now on. Its computation is given by the following expression:
        \begin{equation} \label{eq:J03}
         \begin{split}
            J_{03} = &\frac{H_0}{12 \pi G^2} \frac{1}{D_\mathrm{Earth}^2} \frac{2.163^3}{f(2.163)^2}
            \sqrt{\frac{c_\mathrm{V}}{2}} \, V_\mathrm{max}^3 \\
             &\times \left(1-\frac{1}{\left(1 + \frac{D_\mathrm{Earth} \tan(0.15^\circ)} {R_\mathrm{max}/2.163}\right)^3}\right).
            \end{split}
        \end{equation}
        We detail the derivation of Eq.~(\ref{eq:J03}) in Appendix~\ref{appendix:J}.

        \begin{figure*}
            \centering
            \includegraphics[width=\linewidth]{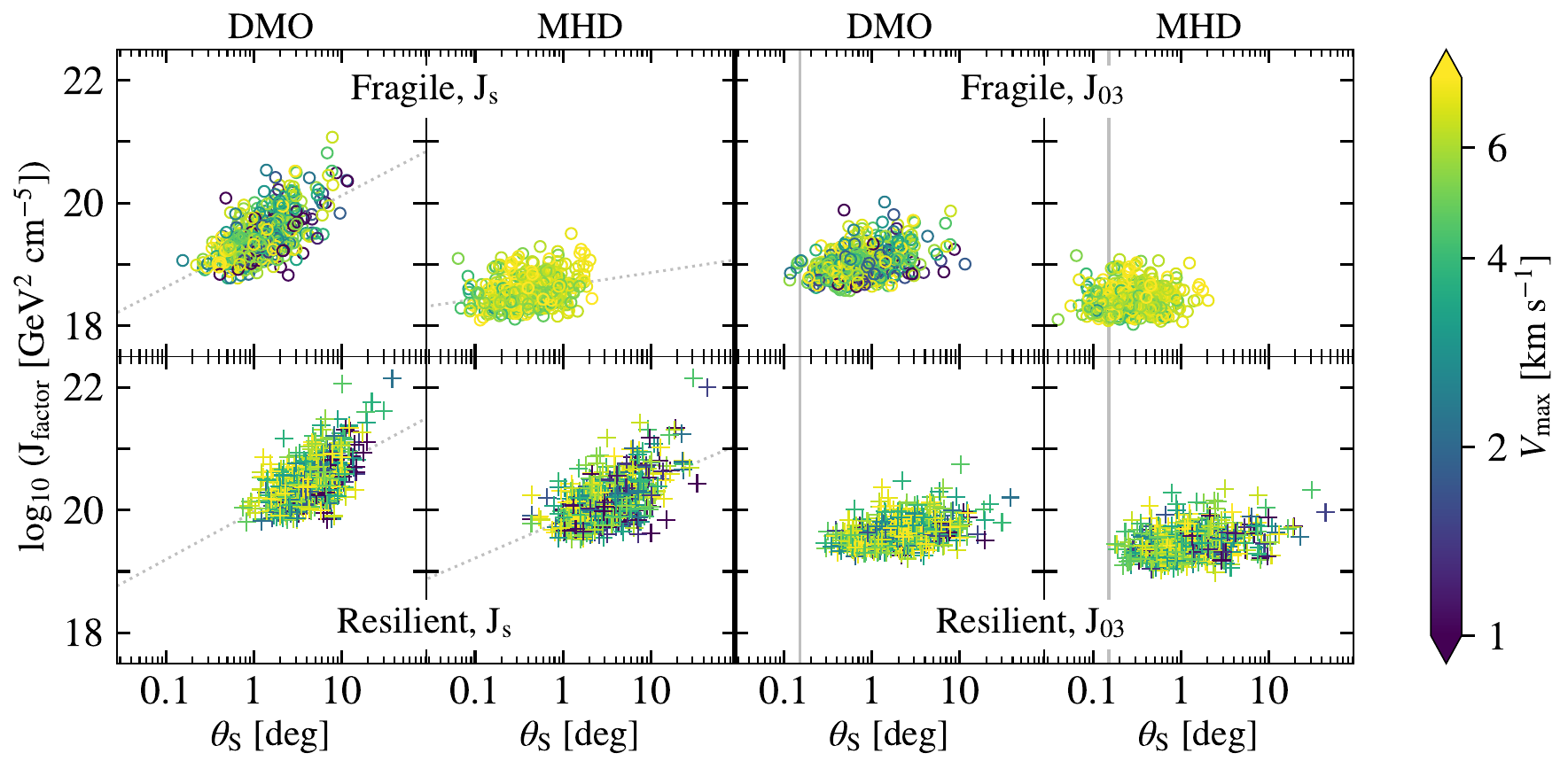}
            \caption{J-factor versus $\theta_\mathrm{S}$ for the brightest subhalo in 500 repopulations with $V_\mathrm{max}\in [0.1, 8]\,$km/s, with the colored z-axis representing the $V_\mathrm{max}$. We study different scenarios depending on subhalo resilience (fragile vs.~resilient), definition of the J-factor used ($J_\mathrm{S}$ and $J_{03}$, given by Eqs.~\eqref{eq:Js} and~\eqref{eq:J03}, respectively), and inclusion of hydrodynamics (DMO vs.~MHD).
            Vertical gray lines in the right $J_{03}$ plots are placed at $\theta=0.15^{\circ}$, i.e., the Fermi-LAT angular resolution at 1 GeV. Dotted gray lines on the left panels show log-log fits to the data. See Section~\ref{sect_sub:results_darkSatellites} for further details. 
            }
            \label{fig:AngJs_to8}
        \end{figure*}

         We follow our subhalo repopulation methodology, summarized in Fig.~\ref{fig:flowchart}, to perform 500 repopulations within the range $V_\mathrm{max}\in [0.1, 8]\,\mathrm{km}/\mathrm{s}$, and select the subhalo with the highest J-factor in each repopulation. Since we now have two different J-factor definitions, $J_\mathrm{S}$ and $J_{03}$, we collect the properties of the brightest subhalo according to both definitions. In most repopulations the same subhalo is the brightest under both definitions, but this is not a requirement. We proceed to investigate the statistical properties of the brightest subhalo over the repopulations.

        In Fig.~\ref{fig:AngJs_to8} we examine the relation between $\theta_\mathrm{S}$ and the J-factor of the brightest subhalos, considering different J-factor definitions, inclusion of hydrodynamics, and resilience.
        The leftmost four panels show $J_\mathrm{S}$, where we observe an almost linear increase in the $\log_{10}-\log_{10}$ scale between its value and $\theta_\mathrm{S}$. Fitting a power law to these data, the slopes are $\sim0.7$ for the DMO fragile scenario, $\sim0.8$ for DMO resilient and $\sim0.6$ for MHD resilient. The fragile MHD data exhibits a less pronounced relation with a slope of $\sim0.2$, but the trend persists.
        On the right panels, where we use $J_{03}$, this relation between J-factor and $\theta_\mathrm{S}$ gets flattened with slopes $\leq0.28$ as the integration is constrained to a fixed angular radius.
        In the same figure, a vertical gray line is shown on the four rightmost panels at $\theta_\mathrm{S}=0.15^\circ$, i.e., the typical angular resolution of Fermi-LAT at 1 GeV\footnote{\url{https://www.slac.stanford.edu/exp/glast/groups/canda/lat_Performance.htm}} and, indeed, most subhalos subtend $\theta_\mathrm{S}$ values larger than this. Thus, we expect most of the brightest subhalos to be spatially extended gamma-ray sources, in agreement with previous work. 
        We note that in the case of fragile MHD data,  15\% of the sample has $\theta_\mathrm{S}<0.15^\circ$. Thus, these subhalos are integrated up to radii larger than their $r_\mathrm{s}$ when computing $J_{03}$ and, consequently, their $J_{03} \geq J_\mathrm{S}$.
        Finally, we also note that there is no significant correlation between $V_\mathrm{max}$ and $\theta_\mathrm{S}$ for either of the J-factor definitions.

        \begin{figure}
            \centering
            \includegraphics[width=0.95\linewidth]{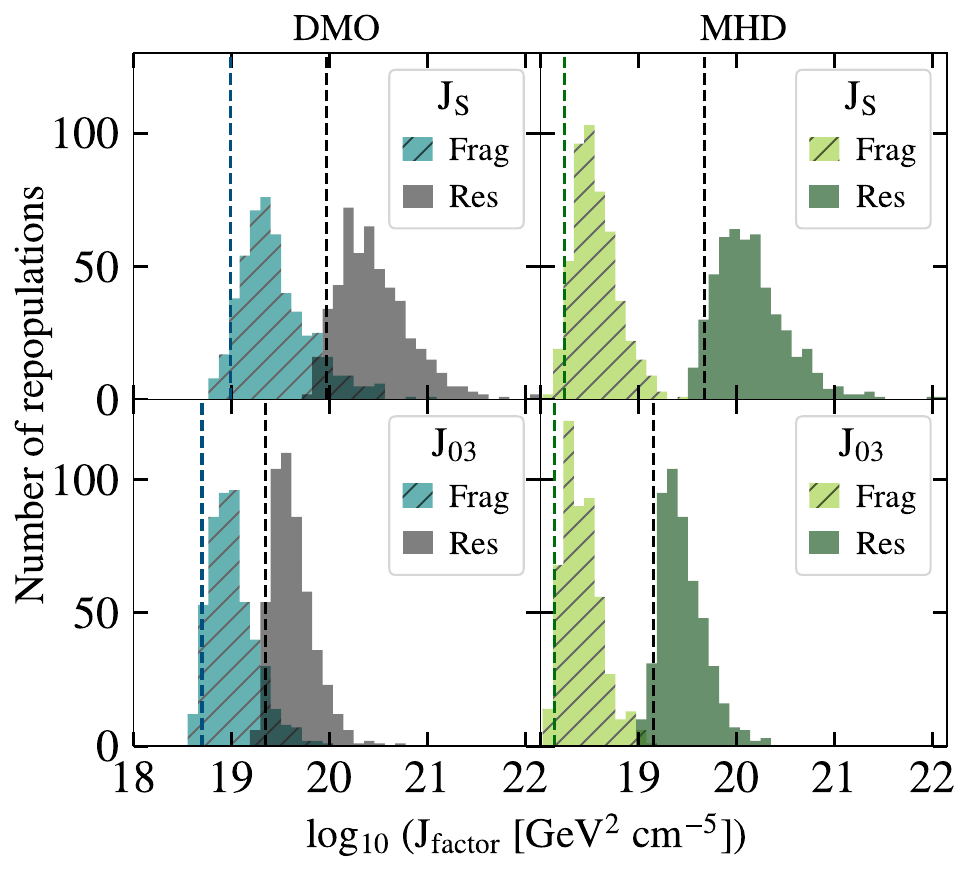}
            \caption{Histogram of J-factors for the brightest subhalo in 500 repopulations with $V_\mathrm{max}\in [0.1, 8]\,$km/s. We study different scenarios depending on subhalo resilience (fragile vs.~resilient), definition of the J-factor used ($J_\mathrm{S}$ and $J_{03}$, given by Eqs.~\eqref{eq:Js} and~\eqref{eq:J03}, respectively), and inclusion of hydrodynamics (DMO vs.~MHD).
            The vertical dashed lines represent the J-factor value above which 95\% of the corresponding distribution is contained.
            }
            \label{fig:J_hist_to8}
        \end{figure}
        
        In Fig.~\ref{fig:J_hist_to8} we show J-factor histograms for the eight scenarios under scrutiny and the 500 repopulations. Here, the fragile (stripped histogram) and resilient (non-stripped histograms) cases are included on the same panel.
        The upper row shows the $J_\mathrm{S}$ histograms, while the bottom row shows $J_{03}$. 
        The 95$^\mathrm{th}$ percentiles of the distributions are shown as vertical dashed lines for each histogram.
        Such values of the J-factor will become key when setting DM constraints from dark satellites, as explained below and introduced in Ref.~\cite{Coronado_Bl_zquez_2019}. 
        Resilient populations provide higher J-factors than the fragile ones, with differences in the 95$^\mathrm{th}$ percentile between 0.7~dex ($J_{03}$ DMO) and 1.5~dex ($J_\mathrm{S}$ MHD). 
        The sample of fragile MHD subhalos exhibits the lowest J-factors among all scenarios. 
        Furthermore, resilient DMO and MHD populations have similar 95$^\mathrm{th}$ percentiles, with variations of 0.2~dex in both $J_\mathrm{S}$ and $J_\mathrm{03}$, while fragile populations have differences of 0.8~dex in $J_\mathrm{S}$ and 0.6~dex in $J_\mathrm{03}$. 
        DMO histograms show some overlap between the fragile and resilient cases, lower resilient J-factor values being similar to those in the high tail of the fragile distributions. This is not the case for MHD, where there is no overlap between fragile and resilient.  
        Further characterization of these 500 repopulations is presented in Appendix~\ref{appendix:results_to_8}, with supplementary analysis complementing the one discussed here.
        
        Using Eq.~\eqref{Flux_eq}, which expresses the DM annihilation flux as the product of the J-factor and the $f_\mathrm{PP}$, we can set constraints on $\langle \sigma v\rangle$ at 95\% confidence level via the following expression~\cite{Coronado_Bl_zquez_2022}:
        \begin{equation}
            \langle \sigma v \rangle = \frac{8\pi\, \phi^\mathrm{min}_\mathrm{det}}{J_\mathrm{sub}\,N_\gamma}\,m_\chi^2,
        \end{equation}
        where $\phi^\mathrm{min}_\mathrm{det}$ is the minimum detectable flux by our detector, $N_\gamma$ the integrated DM spectrum per annihilation, and $J_\mathrm{sub}$ is informed by the value of the J-factor corresponding to the 95th percentile of the J-factor distribution. 
        Assuming that a number $n$ of unIDs in the Fermi-LAT catalog are dark satellites -- i.e., their observed gamma ray flux is originated by DM annihilation -- the value of $J_\mathrm{sub}$ is given by the 95th percentile of the J-factor distribution built from the n$^\mathrm{th}$ brightest subhalo across all repopulations, as detailed in Ref.~\cite{Coronado_Bl_zquez_2019}.
        Note that the case $n=1$ is the most constraining one, since the J-factors used to build the distribution are the highest of each repopulation. 
        
\begin{figure}
            \centering
            \includegraphics[width=0.99\linewidth]{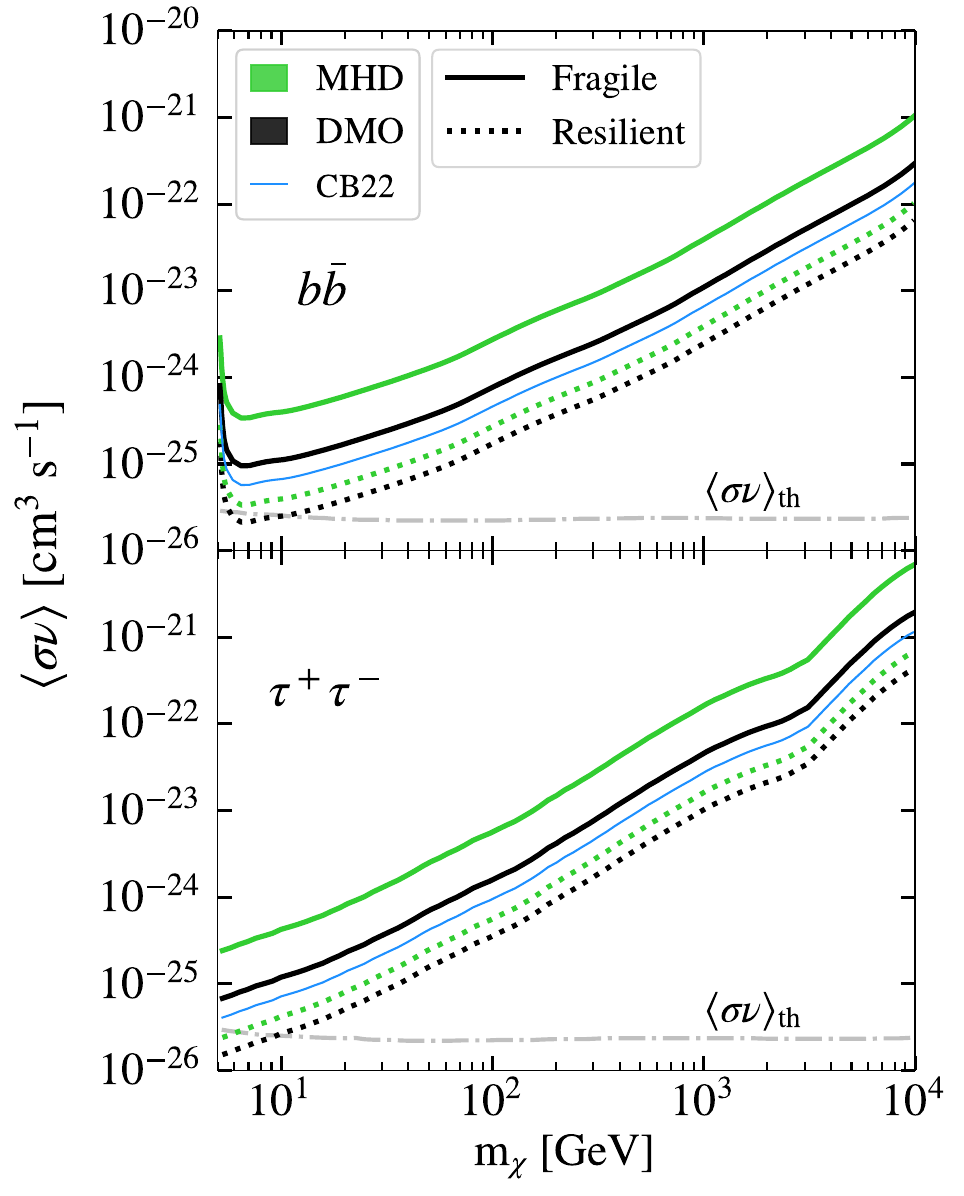}
            \caption{Example of the impact of our work on DM annihilation cross-section constraints, in this case the ones derived for dark satellites in \citetalias{Coronado_Bl_zquez_2022} (solid thin blue line) for two different annihilation channels, $b\Bar{b}$ (top) and $\tau^+\tau^-$ (bottom). Our limits have been obtained simply by rescaling those of \citetalias{Coronado_Bl_zquez_2022} accordingly to the different considered scenarios: DMO and MHD Auriga repopulations, and fragile/resilient. See Section~\ref{sect_sub:results_darkSatellites} for details on the rescaling. The dash-dotted gray line is the thermal relic cross section value \cite{Bringmann2021PreciseDM}.}
            \label{fig:cross}
        \end{figure}

        In order to illustrate how our results impact current DM constraints obtained from dark satellites via the above described methodology, we focus here only on the constraints that were derived for the $n=1$ case in Ref.~\cite{Coronado_Bl_zquez_2022} (see their Fig.12), from now on \citetalias{Coronado_Bl_zquez_2022}. Like them, we also simplify our analysis by assuming that all DM annihilates into a single channel ($B_f=1$ in Eq.~\eqref{Flux_eq}) and selecting two representative ones, $b\Bar{b}$ (quarks) and $\tau^+\tau^-$ (leptons).
        This allows for a simple rescaling of the constraint in \citetalias{Coronado_Bl_zquez_2022}, by replacing their $J^\mathrm{sub}$  with our 95$^\mathrm{th}$ percentiles of $J_{03}$.
        The results are shown in Fig.~\ref{fig:cross}, where the constraint from \citetalias{Coronado_Bl_zquez_2022} is shown together with ours as a thin blue solid line.
        In the figure, the thermal relic cross section, $\langle \sigma v \rangle _\mathrm{th}$, shown as a gray dash-dotted line, refers to the DM annihilation cross section during the early Universe, when DM was in thermal equilibrium with Standard Model particles. Should our DM limits on the cross section lie below this $\langle \sigma v\rangle _\mathrm{th}$ for a certain WIMP model, the latter cannot constitute all of the DM.
        On the contrary, DM limits above $\langle \sigma v \rangle _\mathrm{th}$ lack the sensitivity to test such WIMP model as being responsible for the whole DM content of the Universe.
        
        Fig.~\ref{fig:cross} shows that including baryons worsens the DM limits with respect to DMO by a factor of $\sim3.6$ (fragile) and $\sim1.6$ (resilient).
        Besides, the scenario of resilient subhalos provides better constraints over the fragile one for both DMO and MHD by factors of $\sim4.5$ and $\sim10$, respectively. 
        Our tightest bounds come from the DMO resilient scenario, which has the highest set of J-factor values (see Fig.~\ref{fig:J_hist_to8}). 
        We emphasize that J-factors derived from MHD simulations should be now considered as more realistic than DMO-derived J-factors, while we stay agnostic in this work about the level of resilience of the subhalo population.
        The resilient DMO constraint reaches $\langle\sigma v\rangle_\mathrm{th}$ at $m_\chi \sim 10$\,GeV for both annihilation channels, while the resilient MHD constraint does so at $m_\chi \sim 6$\,GeV only for the $\tau^+\tau^-$ channel. 
        Finally, a direct comparison with \citetalias{Coronado_Bl_zquez_2022} is only possible for the case of our DMO fragile values, given the cosmological simulations they implicitly used to set constraints in their work. We find our DMO limits to be less constraining than theirs by a factor of $\sim1.7$. 
        This mismatch is likely due to the different DMO simulations used in each case: VL-II \cite{Diemand_2008} in \citetalias{Coronado_Bl_zquez_2022} and Auriga \cite{Grand_2017} in this work. Indeed, they use a different set of cosmological parameters and initial conditions, such as $\sigma_8$, which are expected to cause significant differences in halo substructure. Additionally, both the SRD and $c_\mathrm{V}$ models differ between these suites. However, SHVF parameters are consistent across simulations. We used a bootstrapping procedure to calculate the best-fit parameters of the cumulative SHVF of the Auriga runs assuming a power law parametrization, analogous to Section~\ref{subsect:shvf}. The best-fit parameters are slope $\alpha^\mathrm{Auriga}_\mathrm{cum} = -3.0 \pm 0.1$ and normalization $V^\mathrm{Auriga}_\mathrm{0, cum} = 5.38 \pm 0.11$. The best-fit parameters from VL-II are $\alpha_\mathrm{cum}^\mathrm{VLII} = -2.97 \pm 0.08$ and $V_\mathrm{0, cum}^\mathrm{VLII} = 5.42\pm0.06$ \cite{santaella23}.

\section{Conclusions} \label{sect:discussion_conclusions}

    In this study, we have investigated how the inclusion of baryonic physics impacts the population of subhalos in Milky-Way-like halos using publicly available data from the Auriga simulation suite. Since the numerical resolution of these simulations is limited, we employed a repopulation algorithm to generate substructures orders of magnitude less massive than those well resolved in the original runs. The methodology of characterizing the original simulations and repopulating them with low-mass subhalos has previously been performed using DMO suites \cite{santaella23}. In the current work, we extended the analysis to MHD suites, this way incorporating baryonic processes in our repopulation pipeline for the first time. More precisely, we analyzed and compared four different repopulation scenarios, according to the inclusion of baryons or not, and their level of resilience.
    Additionally, we investigated the impact of our results on the latest DM annihilation cross-section constraints obtained from unID gamma-ray sources interpreted within the context of dark satellites.
    
    The Auriga suite \cite{2024Grand_releasedataauriga} is comprised of zoom-in, high-resolution cosmological simulations of Milky Way-size halos performed incorporating either DMO or MHD physics, with identical initial conditions. We focused on the ``Level 3'' resolution and assess the impact of baryonic physics on subhalo abundance, distribution, and structural properties.     
    The presence of baryons leads to a decrease of subhalos across all $V_\mathrm{max}$ scales (see the SHVF shown in Fig.~\ref{fig:SHVF_Grand}), the normalization of the DMO scenario being a factor $\sim 2.5$ higher than the MHD.
    The radial distribution of subhalos remains similar between DMO and MHD runs, as shown in the SRDs of Fig.~\ref{fig:srd_vol}. However, in the MHD scenario we find the closest subhalo to the GC to be three times further away than in DMO. 
    The internal structure of subhalos, codified in this work via the so-called velocity concentration, is shown in Fig.~\ref{fig:cv}. DMO subhalos are more concentrated than MHD ones, the corresponding DMO normalization being a factor $\sim1.5$ higher than in MHD. Table~\ref{tab:bestParams} summarizes the best-fit parameters of all these characterizations for Auriga, as well as the corresponding values of $V_\mathrm{max}$ below which resolution effects cannot be ignored.
    
    To overcome the resolution limits imposed by numerical effects, and with the intention to investigate the properties of the subhalo population down to much smaller scales, we employ the algorithm presented in Ref.~\cite{santaella23} to repopulate the original Auriga runs with millions of low-mass subhalos. The pipeline is described in Section~\ref{sect:repopulation_theory} and comprised in the flowchart of Fig.~\ref{fig:flowchart}. We extrapolate the SHVF down to $V_\mathrm{max}=0.1\,$km/s, which in terms of subhalo mass translates in approximately six to seven orders of magnitude below the resolution of the original simulations.
    We generate and analyze four different repopulation scenarios, depending on the inclusion or not of hydrodynamics (DMO versus MHD), and on the level of subhalo resilience to tidal forces ('fragile' versus 'resilient'). The latter is modeled by means of adopting different SRDs, shown in Fig.~\ref{fig:srd_vol}. The sets of input parameters used for each of these four repopulation scenarios are summarized in Appendix~\ref{appendix:input_params_repops}.
    For each of them, we generate 500 repopulations in order to ensure a proper statistical sample.

    In the context of indirect DM searches, our results in Fig.~\ref{fig:VmaxJs_to120} show that small subhalos can be the brightest of their corresponding repopulation provided that they are close enough to Earth. This is specially relevant when considering the MHD resilient scenario, where 2\% of the brightest subhalos have $V_\mathrm{max} < 1\,$km/s. This is so despite the fact that the MHD subhalo population exhibits lower $c_\mathrm{V}$ values compared to its DMO counterpart: in the resilient case there is now a non-negligible chance for subhalos to exist at the Solar galactocentric radius, thus being close enough to Earth to compete with most massive ones in terms of their J-factor values. 
    We note that the very existence of these small subhalos contradicts previous results from MHD simulations, which report a severe lack of substructure in the inner Galaxy \cite{Grand_2021, 10.1093PhatELVIS,2020MNRAS.492.5780R,2023MNRASFire}. Yet, if the disruption of such dark satellites observed in simulations is primarily due to resolution limits and such subhalos can indeed survive tidal forces, as pointed out by other authors~\cite{2018MNRAS.474.3043V, van_den_Bosch_2018, Stref_2019, 2020MNRAS.491.4591E, 2021arXiv211101148A, Aguirre-Santaella_sheddinglight}, their existence would become highly relevant for DM searches according to our results (see also \cite{santaella23} for the DMO case). In fact, these tiny structures would constitute dense, nearby structures of solely DM, whose signals could potentially be detected using indirect DM detection techniques.
    Overall, and beyond this important result, we find that the inclusion of baryons shifts the whole J-factor distribution of the brightest subhalo down to lower values with respect to the DMO scenario, and that resilient subhalos provide higher J-factors than their fragile counterparts, both for DMO and MHD (see Fig.~\ref{fig:J_hist_to120}).
    In Fig.~\ref{fig:AngJs_to120} we show that the brightest subhalo appears extended as observed from Earth in most repopulations, with the MHD fragile scenario being the one with less extended subhalos. We observe a direct correlation between the subhalo angular extension and their J-factor, with power law slopes of $\sim1.0$ for DMO and $\sim0.8$ for MHD, both for the resilient and fragile scenarios. We find no correlation between $V_\mathrm{max}$ and $\theta_\mathrm{S}$.

    In the last part of our work, we introduced a specific use case for the repopulation algorithm, just as an example of potential applications. More precisely, in Section~\ref{sect_sub:results_darkSatellites} we investigated the impact of our results within the context of DM searches using dark satellites as observational probes. 
    Once more, we performed 500 repopulations of the four considered scenarios based on the Auriga characterization, yet this time adopting for each repopulation an upper limit of $V_\mathrm{max} = 8\,$km/s for a direct comparison with previous work \cite{PhysRevD.102.103010,Coronado_Bl_zquez_2022}.
    This is the $V_\mathrm{max}$ value corresponding to subhalos with a mass $M\approx10^7 \mathrm{M_{\odot}}$, below which no subhalos are expected to host baryons according to those works (thus remaining as dark satellites).
    We also introduced an alternative definition of the J-factor, $J_\mathrm{03}$, that integrates the DM-induced gamma-ray signals only within the innermost $0.3^\circ$ of each subhalo, i.e., the expected angular extension of the brightest subhalos in the Fermi-LAT gamma-ray sky given its instrumental sensitivity (\citetalias{Coronado_Bl_zquez_2022}).
    For all cases --but marginally in part of the fragile MHD data-- $J_{03} < J_\mathrm{S}$, as expected from a smaller integration angle of the DM signal; see Figs.~\ref{fig:AngJs_to8} and~\ref{fig:J_hist_to8}.
    Finally, also following the methodology in \cite{Coronado_Bl_zquez_2019,Coronado_Bl_zquez_2021,Coronado_Bl_zquez_2022}, we calculated the 95$^\mathrm{th}$ percentile of all our J-factor distributions, i.e., the value above which the 95\% of the corresponding population is contained, and used it to rescale the DM annihilation cross section constraints derived in \citetalias{Coronado_Bl_zquez_2022}. A one-to-one comparison with the latter work --only possible for the DMO fragile case-- provides DM limits that are nearly a factor two less constraining in our case, likely due to differences in the original simulations used in either case. A more realistic case including baryons worsens the constraints by a factor of $\sim3.6$ in case of considering fragile subhalos with respect to DMO fragile results. Yet, a stronger resilience of subhalos to tidal stripping improves the DM limits by a factor $\sim4.5$ and $\sim10$ compared to their respective DMO and MHD fragile cases, respectively.

    The results found in this work show the importance of including baryons in order to achieve an accurate, more realistic characterization of the Galactic subhalo population, as well as to propose the most optimal subhalo search strategies. The latter not only via its potential DM annihilation products --for which we even included a specific example here--, but also through their gravitational signatures.
    One of these signatures are gaps in stellar streams.
    Stellar streams are the remnants of subhalos that have been heavily stripped by tidal disruption in the Galaxy~\cite{1994Natur.370..194I,2003ApJ...599.1082M,2022ApJ...926..107M}.
    The passage of a subhalo, regardless of its baryonic content, can induce a discernible gap in the stream. The properties and occurrence rate of these gaps are informed by the properties of the underlying subhalo population. An improved characterization of the subhalo population is thus crucial to correctly interpret the observed features in stellar streams \cite{2012ApJ...748...20C,2015ApJ...808...15C,2015MNRAS.450.1136E,2015MNRAS.454.3542E,2019ApJ...880...38B,2025JCAP...09..003F}.
    Strong gravitational lensing is another subhalo detection method that can largely benefit from our work.
    Indeed, halo substructure can cause irregularities such as localized distortions or anomalies in the observed lensing pattern, unlike smooth lens models.
    Again, a proper characterization of the subhalo population in the object that acts as the lens is key should we want to properly correlate the observed anomalies with the underlying DM distribution~\cite{2010MNRAS.408.1969V,2016JCAP...11..048H,2014MNRAS.442.2434N,2020Sci...369.1347M}.

    Our study represents an important step ahead in these directions, yet future simulation work will necessarily need to address in further detail the precise level of resilience of subhalos to tidal stripping, as well as to overcome current numerical limitations to directly characterize with simulation data the whole range of expected $\Lambda$CDM subhalo masses. Both are undoubtedly our main sources of uncertainty. For the time being, and by means of the different repopulation recipes and scenarios considered, in this work we provided results hopefully encapsulating such uncertainties.

\section*{Acknowledgments}
SPB acknowledges support from the European Research Council (ERC) under the European Union’s Horizon2020 research and innovation program Grant Agreement No. 948689 (AxionDM) and funding from the Deutsche Forschungsgemeinschaft (DFG, German Research Foundation) under Germany’s Excellence Strategy – EXC 2121 “Quantum Universe”– 390833306. This article is based upon work from COST Action COSMIC WISPers CA21106, supported by COST (European Cooperation in Science and Technology).

MASC and AAS were supported by the grants PID2024-155874NB-C21, PID2021-125331NB-I00 and CEX2020-001007-S, both funded by MCIN/AEI/10.13039/501100011033 and by ``ERDF A way of making Europe''. They also acknowledge the MultiDark Network, ref. RED2022-134411-T. AAS acknowledges support from the Science and Technology Facilities Council funding grant ST/X001075/1 and from the Agencia Estatal de Investigación Española (AEI; grant PID2022-138855NB-C33).

This research has made use of the Astrophysics Data System, funded by NASA under Cooperative Agreement 80NSSC21M00561.
It has also made use of \textsc{python}, along with community-developed or maintained software packages, including \textsc{matplotlib} \cite{2007CSE.....9...90H}, \textsc{numpy} \cite{harris2020array}, \textsc{scipy} \cite{2020NatMe..17..261V}, \textsc{astropy} \cite{2013A&A...558A..33A,2018AJ....156..123A,2022ApJ...935..167A} and \textsc{iminuit} \cite{iminuit}.

\section*{Data availability}
    The data underlying this article are publicly available in the website of the DAMASCO group at \url{https://projects.ift.uam-csic.es/damasco/?page_id=831}.

\appendix

\section{J-factor definitions}\label{appendix:J}

The NFW DM density profile \cite{1997ApJ490493N} of a DM halo (or subhalo) is given by
\begin{equation} \label{eq_app:NFW_density}
    \rho_\mathrm{DM}^\mathrm{NFW} (r;\, \rho_0,\,r_\mathrm{s}) = \frac{\rho_0}{\left(\frac{r}{r_\mathrm{s}}\right) \left(1 + \frac{r}{r_\mathrm{s}}\right)^2}
\end{equation}
where $\rho_0$ is the normalization density of the halo and $r_\mathrm{s}$ the scale radius.
The mass of a halo is calculated by integrating the density profile, and using the NFW expression, we obtain
\begin{equation} \label{eq_app:mass_from_halo}
    M(<r) = \int_0^rdr^{\prime} 4\pi (r')^2 \rho(r') = 4\pi \rho_0 r_\mathrm{s}^3 f(r/r_\mathrm{s}).
\end{equation}
where $f(x) = \ln(1 + x) - x/(1+x)$.

In a gravitationally bound system with dynamical equilibrium and spherical symmetry, $M(<r) = r V_\mathrm{circ}^2 / G$. If we apply this expression to the NFW density profile, the expression for the circular velocity is given by
\begin{equation} \label{eq_app:mass_general}
    V_\mathrm{circ} (r) = \sqrt{\frac{GM(<r)}{r}} = \sqrt{4\pi G \rho_0 r_\mathrm{s}^3\frac{f(r/r_\mathrm{s})}{r}},
\end{equation}
The maximum value of this expression is $V_\mathrm{max}$, which occurs at a radius $R_\mathrm{max} = 2.163\, r_\mathrm{s}$.
Moreover, we can express $\rho_0$ as a function of $V_\mathrm{max}$ and $R_\mathrm{max}$ using the same expression, as
\begin{equation} \label{eq_app:rho0}
    \rho_0 = \frac{1}{4\pi G} \frac{2.163^3}{f(2.163)} \frac{V_\mathrm{max}^2}{R_\mathrm{max}^2}.
\end{equation}

Now, we focus on the J-factor calculations. We use Eq.~\eqref{Flux_eq} to calculate the J-factor of the NFW density profile (Eq.~\eqref{eq_app:NFW_density}), and apply variable changes following Eq.~\eqref{eq_app:rho0}. The integration gives us
\begin{equation} \label{eq_app:J_general}
    \begin{split}
    J_\mathrm{NFW}(r) &= 4\pi \frac{1}{3 D_\mathrm{Earth}^2} \rho_\mathrm{0}^2 r_\mathrm{s}^3 \left(1-\frac{1}{\left(1 + \frac{r} {r_\mathrm{s}}\right)^3}\right) \\
    =& \frac{H_0}{12 \pi G^2}\frac{1}{D_\mathrm{Earth}^2} \frac{2.163^3}{f(2.163)^2}
    \sqrt{\frac{c_\mathrm{V}}{2}}V_\mathrm{max}^3 \left(1-\frac{1}{\left(1 + \frac{r} {r_\mathrm{s}}\right)^3}\right).
    \end{split}
\end{equation}
When $\frac{r} {r_\mathrm{s}} \gg 1$,  the factor $\left(1-\left(1 + r/r_\mathrm{s}\right)^{-3}\right) \rightarrow 1$. In other works this term has been omitted, e.g.~Refs.~\cite{santaella23, Moline21}, and the J-factors presented there include the integration of the whole subhalo. We work with this explicit dependence.

The J-factor integration defined in Section~\ref{sect:subs_as_gamma_targets} is performed up to $r=r_\mathrm{s}$, so
\begin{equation}
    J_\mathrm{S}= \frac{H_0}{12 \pi G^2}\frac{1}{D_\mathrm{Earth}^2} \frac{2.163^3}{f(2.163)^2}
    \sqrt{\frac{c_\mathrm{V}}{2}}V_\mathrm{max}^3 \cdot \frac{7}{8},
\end{equation}
where $H_0$ is the Hubble constant at  present time, which we take as $H_0 =67.7 \, \mathrm{km}/\mathrm{s}/\mathrm{Mpc}$ to be consistent with the Auriga simulations \cite{2024Grand_releasedataauriga}. The gravitational constant is $G = 6.67 \times 10^{-11} \, \mathrm{m}^3/\mathrm{kg}/\mathrm{s}^2$.
We arrive at Eq.\eqref{eq:Js} by combining all the numerical factors into one.

In Section~\ref{sect_sub:results_darkSatellites}, the radius of integration depends on the angle at which $r_\mathrm{s}$ subtends in the sky, $r=D_\mathrm{Earth} \cdot \tan(0.15^\circ)$, from Eq.~\eqref{eq:J03}. Then, we have
\begin{equation}
\begin{split}
    J_{03}= \frac{H_0}{12 \pi G^2}\frac{1}{D_\mathrm{Earth}^2} \frac{2.163^3}{f(2.163)^2}
    \sqrt{\frac{c_\mathrm{V}}{2}}V_\mathrm{max}^3 \\
    \times \left(1-\frac{1}{\left(1 + \frac{D_\mathrm{Earth} \cdot \tan(0.15^\circ)} {R_\mathrm{max}/2.163}\right)^3}\right).
\end{split}
    \label{Jv_parts}
\end{equation}

\section{Relation between mass and velocity concentration scatters} \label{appendix:CvandC200}

We base the relation $c_\mathrm{\Delta} - c_\mathrm{V}$ on the formalism described in \citetalias{Molin_2017}.
In their Eq.~(4), this relations is expressed as
\begin{equation} \label{eq:cvandc200}
    c_\mathrm{V} = \left(\frac{c_\Delta}{2.163}\right)^3 \frac{f(R_\mathrm{max}/r_\mathrm{s})}{f(c_\mathrm{\Delta})}  \Delta
\end{equation}
where $\Delta$ denotes the overdensity factor,
which represents a spherical overdense region that has undergone gravitational collapse and reached virial equilibrium.
We adopt the upper value $\Delta=200$.

We assume a log-gaussian distribution, given by
\begin{equation} \label{eq:log-gaussian}
    P(x) = \frac{a}{x \sigma} \exp \left[- \frac{1}{2} \left(\frac{\log_{10}(x) - \mu}{\sigma}\right)^2\right]
\end{equation}
where $a$ is a normalization constant, $\mu$ is the median point of the distribution, and $\sigma$ is the standard deviation in log space.

\begin{figure*}
    \centering
    \includegraphics[width=0.99\linewidth]{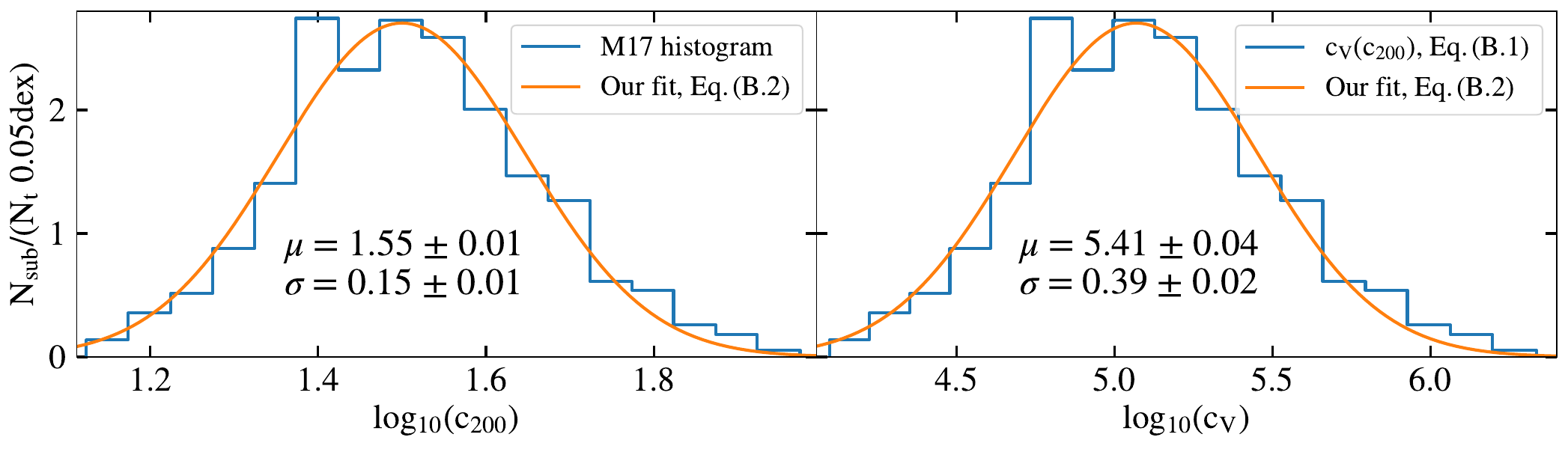}
    \caption{
    \textbf{Left:} histogram of $c_\mathrm{200}$ taken from Fig.~4 of \citetalias{Molin_2017} shown with a blue line. Fit to the histogram with a log-gaussian distribution in orange line, given in Eq.~\eqref{eq:log-gaussian}. We also list the best-fit values of $\mu$ and $\sigma$. 
    \textbf{Right: } histogram of $c_\mathrm{V}$ shown with a blue line, calculated from the $c_\mathrm{200}$ distribution using Eq.~\eqref{eq:cvandc200}. Fit to the histogram with a log-gaussian distribution in orange line, given in Eq.~\eqref{eq:log-gaussian}. We also list the best-fit values of $\mu$ and $\sigma$.}
    \label{fig:scatters_cv}
\end{figure*}

On the left panel of Fig.~\ref{fig:scatters_cv} we show one of the histograms from Fig.~4 of \citetalias{Molin_2017}. We fit the histogram to the log-gaussian distribution, and list the best-fit values of $\mu$ and $\sigma$ on the same panel. Our best-fit value for $\sigma$ agrees with that from \citetalias{Molin_2017}, $\sigma_\mathrm{M17}=0.15$. On the right side of the panel we show the distribution of $c_\mathrm{V}$, which we have calculated from the $c_\mathrm{200}$ distribution using Eq.~\eqref{eq:cvandc200}. Once again, we fit the distribution using Eq.~\eqref{eq:log-gaussian} and list the relevant parameters. Comparing the scatters, we can see that $\sigma_\mathrm{V}>2\sigma_{200}$.

\section{Input parameters for the repopulations} \label{appendix:input_params_repops}

We compile all the input parameters used in our repopulation code in Table~\ref{tab:parameters_repopulation}.
The values of the cosmological constants are taken from Ref.~\cite{PhysRevD.110.030001}. 
The values of $r_\mathrm{s}$ and $\rho_0$ of the host are the geometrical means of those from the Auriga original DMO ``Level 3'' runs, calculated using Eq.~\eqref{eq_app:rho0}. The value of $\rho_0$ is consistent with that of recent MW observations \cite{deSalas2020}. The value for $R_\mathrm{vir}$ has been taken as similar to the lower limit of the Auriga runs, since recent observations of the MW set its value to $R_\mathrm{vir} < 200\,$kpc \cite{10.1093/mnras/stae034}.

\begin{table*}
\centering
\begin{tabular}{cccccc}
\toprule
\toprule
\multicolumn{6}{c}{Cosmological constants} \\ 
\midrule
&\multicolumn{1}{r}{$\rho_\mathrm{crit}$} & 135.73 & \multicolumn{1}{l}{$\mathrm{M}_\odot / \mathrm{kpc}^3$}\\
&\multicolumn{1}{r}{$H_0$} & 67.7 & \multicolumn{1}{l}{km / s / Mpc}\\
&\multicolumn{1}{r}{$G$} & $4.297\times10^{-6}$ & \multicolumn{1}{l}{kpc / M$_\odot$ (km/s)$^2$}\\
\midrule
\midrule
\multicolumn{6}{c}{Host parameters (NFW DM density profile)} \\
\midrule
&\multicolumn{1}{r}{$R_\mathrm{vir}$} & \multicolumn{1}{r}{220} & \multicolumn{1}{l}{kpc} \\
&\multicolumn{1}{r}{$r_\mathrm{s}$} & \multicolumn{1}{r}{20} & \multicolumn{1}{l}{kpc} \\
&\multicolumn{1}{r}{$\rho_0$} & \multicolumn{1}{r}{$9.04\times10^{6}$} & \multicolumn{1}{l}{M$_\odot$ / kpc$^3$}\\
\midrule
\midrule
\multicolumn{6}{c}{Subhalo population} \\
\midrule
 &  & \multicolumn{2}{c}{DMO} & \multicolumn{2}{c}{MHD} \\
 \cmidrule(lr){3-4} \cmidrule(lr){5-6}
 &  & Fragile & Resilient & Fragile & Resilient \\
 \cmidrule(lr){3-3} \cmidrule(lr){4-4} \cmidrule(lr){5-5} \cmidrule(lr){6-6}
SHVF & $\alpha$ & -3.92 & -3.92 & -4.08 & -4.08 \\
{[Eq.~\eqref{eq:shvf}]} & $V_0$ & 5.68 & 5.78 & 5.3 & 5.51 \\
\midrule
\multirow{3}{*}{\begin{tabular}{c}SRD\\{[Eq.~\eqref{eq:srd_original}]} \end{tabular}} & $a_0$ & -0.151 & 0 & -0.27 & 0 \\
 & $a_1$ & 39.3 & 1 & 16 & 1 \\
 & Minimum $D_\mathrm{GC}$ [kpc] & 5.57 & 0 & 15.71 & 0 \\
\midrule
$c_\mathrm{V}$ & $c_0$ & 283,862 & 283,862 & 192,879 & 192,879 \\
{[Eq.~\eqref{eq:cv}]} & $1\sigma$ scatter & 0.297 & 0.297 & 0.375 & 0.375 \\
\end{tabular}
\caption{List of parameters used as input for the repopulation algorithm.}
\label{tab:parameters_repopulation}
\end{table*}

\section{Roche criterion} \label{appendix:r_s_and_roche_criterium}

The Roche criterion is the most commonly used criterion which defines when a subhalo gets fully disrupted. According to it, a subhalo survives inside a host while this expression holds: 
\begin{equation}\label{eq:roche}
    r_\mathrm{t} \ge r_\mathrm{s}
\end{equation}
where $r_\mathrm{t}$ is the tidal radius and $r_\mathrm{s}$ is the scale radius.

The tidal radius is an estimation of the size of a subhalo taking into account tidal forces. It is defined as the radius where the differential tidal force of the host is equal to the gravitational force due to the mass of the subhalo. Beyond it, we expect matter to be subject to tidal stripping, while the interior should remain bound \cite{10.1046/j.1365-8711.1998.01775.x,zavala2019dark}.
\begin{equation}
    r_\mathrm{t} (r) = D_\mathrm{GC} \left(\frac{M_\mathrm{subh}}{3\, M_\mathrm{host}(<r)}\right)^{1/3}
\end{equation}
It depends on $D_\mathrm{GC}$, and the masses of both the subhalo and the host.

The $r_\mathrm{s}$ is located in the innermost parts of a subhalo, so we define the subhalo whole disruption as the moment in which the stripping occurs up to this point, as described in Eq.~\eqref{eq:roche}. We note that other works have adopted alternatives to this definition, e.g.,~\cite{Stref_2019}.

\section{Repopulations mimicking Auriga} \label{appendix:repopulation_mimics_Auriga}

In the main text, e.g.~in Section~\ref{subsect:details_repop_algorithm}, we claim that our repopulations accurately reproduce the characterization of the subhalo population of Auriga, which allows us to extrapolate to lower values of $V_\mathrm{max}$. To substantiate this claim, we show the characterization of one repopulation based on the fragile SRD, for both DMO and MHD. 
We follow the methodology described in Section~\ref{sect:repopulation_theory} and the parameters listed in Table~\ref{tab:parameters_repopulation}, but we repopulate a range $V_\mathrm{max} \in [1, 120]\,$km/s. This range provides sufficient information for this check, while creating a manageable dataset.

\begin{figure}
    \centering
    \includegraphics[width=0.95\linewidth]{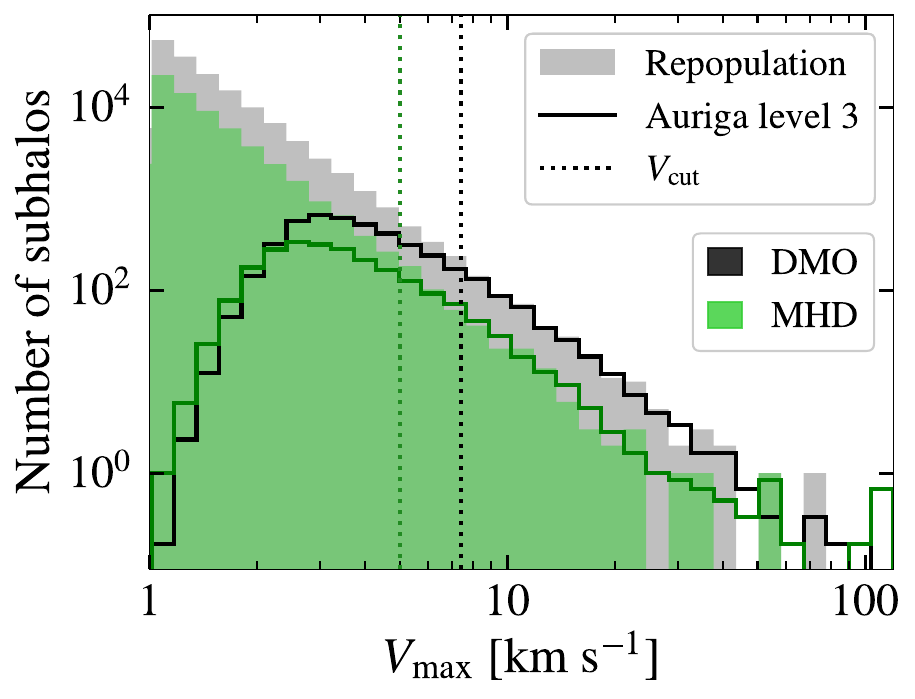}
    \caption{SHVF of one repopulation following Section~\ref{sect:repopulation_theory} and the parameters listed in Table~\ref{tab:parameters_repopulation} shown as the filled histograms, compared to the SHVF from the original Auriga runs, as calculated in Section~\ref{subsect:shvf} and shown with the empty histograms, both for DMO and MHD.}
    \label{fig:mimic_auriga_SHVF}
\end{figure}

In Fig.~\ref{fig:mimic_auriga_SHVF} we show how the SHVF of our repopulation mimics the SHVF of the Auriga runs down to $V_\mathrm{cut}$, both for the DMO and MHD characterizations. 
Below these values, the repopulation follows a power law distribution consistent with theoretical expectations.

\begin{figure}
    \centering
    \includegraphics[width=0.95\linewidth]{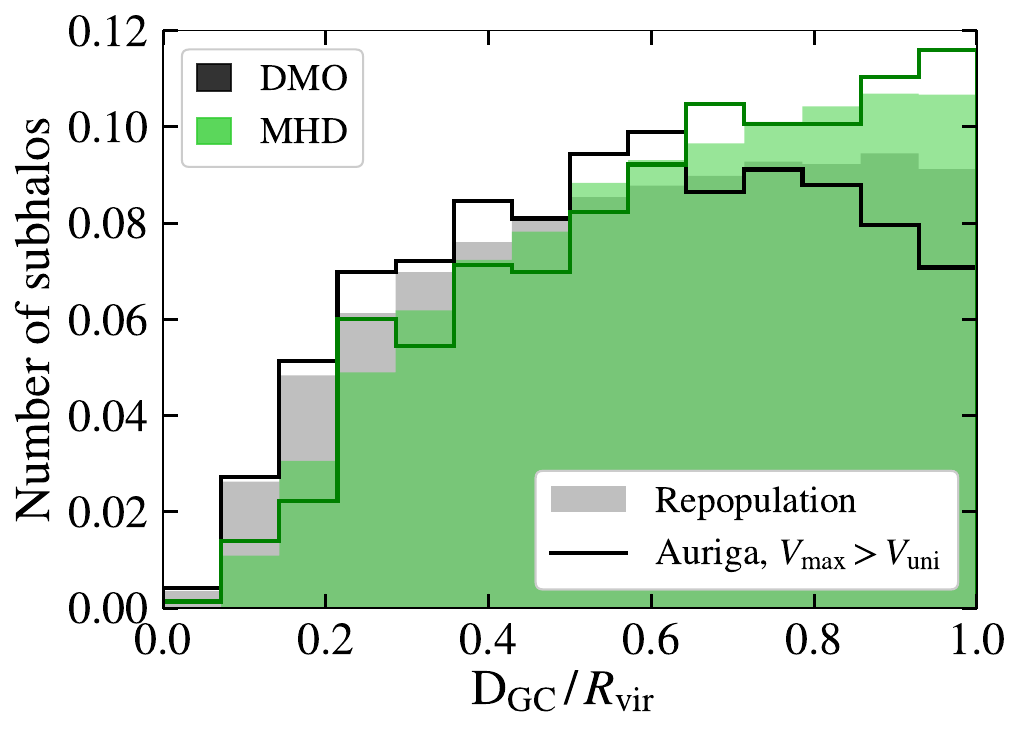}
    \caption{Fragile SRD of one repopulation following Section~\ref{sect:repopulation_theory} and the parameters listed in Table~\ref{tab:parameters_repopulation} shown as the filled histograms, compared to the SRD from the Auriga runs, as calculated in Section~\ref{subsect:srd} and shown with the empty histograms, both for DMO and MHD.}
    \label{fig:mimic_auriga_SRD}
\end{figure}

In Fig.~\ref{fig:mimic_auriga_SRD} we show the comparison between the SRD of the Auriga runs at $V_\mathrm{max} > V_\mathrm{uni}$, which was presented in Fig.~\ref{fig:srd_vol}, and that of our repopulation using all subhalos, without cuts in $V_\mathrm{max}$. For both parametrizations, DMO and MHD, the repopulation follows the analytical expression for the fragile SRD from Eq.~\eqref{eq:srd_original}, confirming the validity of our modeling approach.

None of the subhalos in this repopulation are disrupted by the Roche criterion (described in Appendix~\ref{appendix:r_s_and_roche_criterium}), either in the DMO or MHD scenario.
Although individual subhalos might disrupt in other repopulations with the same inputs, we expect similar survival numbers.
Conversely, repopulating with the resilient characterization and the same $V_\mathrm{max}$ range, some subhalos fail to survive the Roche criterion. In the DMO scenario, 99.7\% of the resilient subhalos survive ($\sim570$ subhalos disrupted), and 99.5\% in MHD ($\sim475$ subhalos disrupted). These numbers reinforce that the injected characterization is reproduced by the repopulations, and that subhalo disruption is exceptional. In Section~\ref{subsect:details_repop_algorithm} we stated that 30\%-45\% of the brightest subhalos break under the Roche criterion. The over-representation of disruption in the subsample of brightest subhalos reinforces the notion that the brightest subhalos tend to be atypical in their repopulation.

\section{Full characterization up to 120 km/s} \label{appendix:results_to_120}

    Here, we expand on the analysis of the repopulation results for the subsample of the brightest subhalo in the repopulations from Section~\ref{subsect:results_baryons}, with $V_\mathrm{max}\in [0.1, 120]\,$km/s. 

    \begin{figure}
        \centering
        \includegraphics[width=0.95\linewidth]{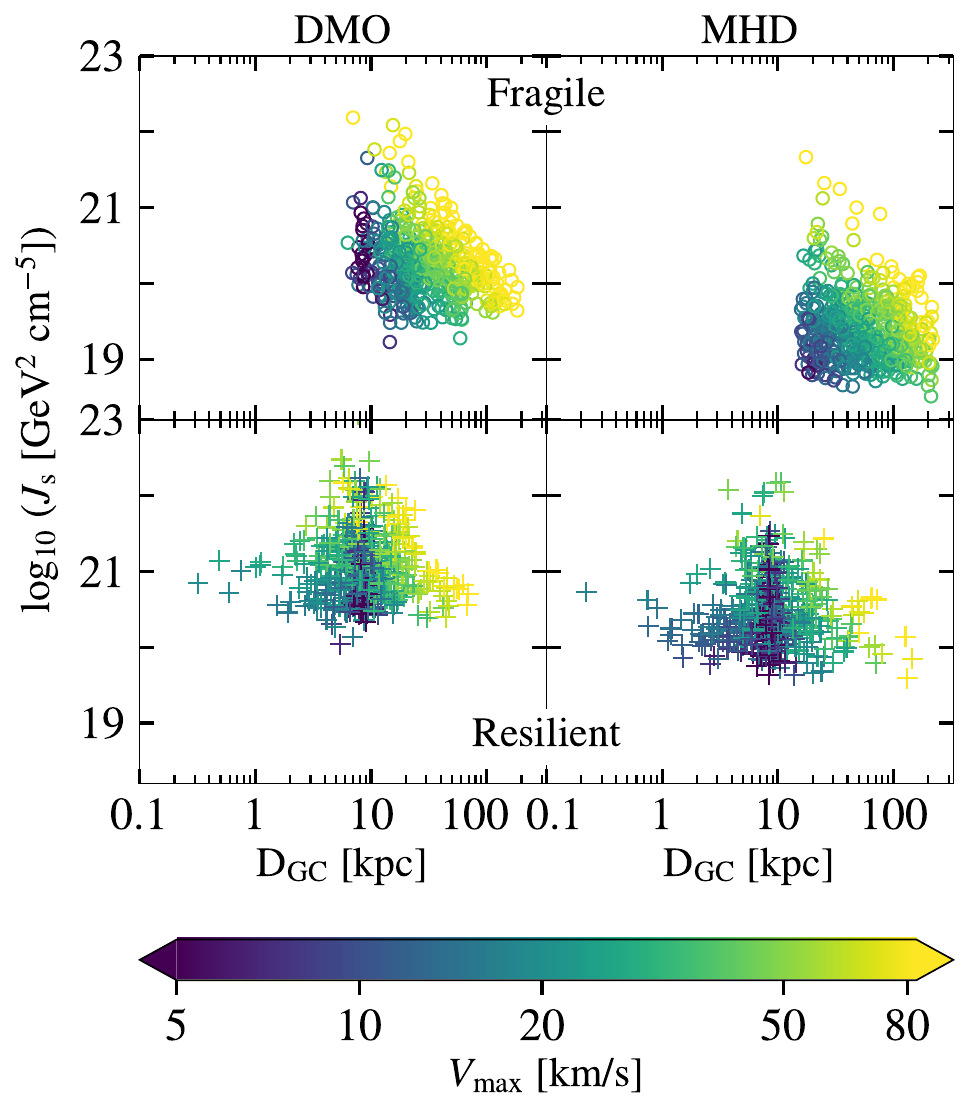}
        \caption{
        $J_\mathrm{S}$ (given by Eq.~\eqref{eq:Js}) versus $D_\mathrm{GC}$ for the brightest subhalo in 500 repopulations with $V_\mathrm{max}\in [0.1, 120]\,$km/s, with the colored z-axis representing $V_\mathrm{max}$.
        Different scenarios are shown depending on subhalo resilience (top and bottom panels for fragile and resilient populations, respectively) and inclusion or not of hydrodynamics (left for DMO; right for MHD).
        }
        \label{fig:DgcJs_to120}
    \end{figure}
    
    In Fig.~\ref{fig:DgcJs_to120} we show the J-factor as a function of the $D_\mathrm{GC}$ for all the scenarios we considered. We can clearly see the radial distance cut-offs that we enforce in the fragile SRD scenario for both DMO and MHD, as there are no subhalos below these limits. The majority of brightest subhalos with $V_\mathrm{max} \leq 10$\,km/s are located at a distance from the GC similar to Earth, $D_\odot = 8.5$ kpc. Additionally, for a certain J-factor, larger subhalos tend to be located at greater distances from the observer.
    
    We also find some caveats of the repopulation algorithm regarding large subhalos. In the four scenarios presented there are subhalos with $V_\mathrm{max} \geq 50$\,km/s appearing at $D_\mathrm{GC}\leq20$\,kpc, from 1 (fragile MHD) to 53 (resilient DMO) of the brightest subhalos. These structures clash with our observation of the Milky Way, where structures that massive are not found at distances $D_\mathrm{GC} \leq 20\,$kpc \cite{2024PhRvD.109f3024M}. These objects might be present next to the GC in other galaxies, but not in ours.
    Since there is no cosmic evolution in our repopulations, large structures can be positioned close to Earth as long as they do not violate the Roche criterion (or encompass the Earth; this latter condition is relevant for our work because we focus on studying subhalos as an ``outside'' signal, yet it might be useful to retain these subhalos for other applications).

\begin{figure}
    \centering
    \includegraphics[width=\linewidth]{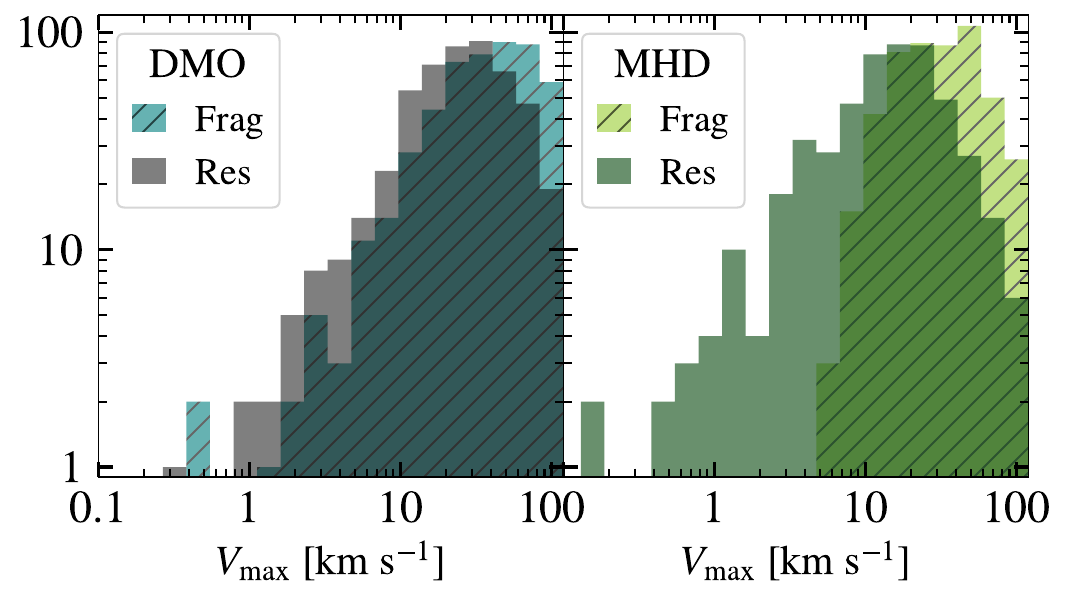}
    \caption{
        Histogram of $V_\mathrm{max}$ for the brightest subhalo in 500 repopulations with $V_\mathrm{max}\in [0.1, 120]\,$km/s. We study different scenarios depending on subhalo resilience (fragile vs.~resilient) and inclusion of hydrodynamics (left for DMO and right for MHD).
    }
    \label{fig:Vmax_hist_geom_to120}
\end{figure}

    In Fig.~\ref{fig:Vmax_hist_geom_to120} we show the $V_\mathrm{max}$ histograms of the brightest subhalo in the repopulations.
    The MHD resilient scenario is the only one where we consistently find the brightest subhalo with $V_\mathrm{max} \leq 1$\,km/s, and it happens in 2\% of the repopulations.

\section{Extra characterization of the dark satellites repopulations} \label{appendix:results_to_8}
    
    Here, we expand on the analysis of the repopulation results for the subsample of the brightest subhalo in the repopulations from Section~\ref{sect_sub:results_darkSatellites}, with $V_\mathrm{max}\in [0.1, 8]\,$km/s. 
    
    \begin{figure*}
        \centering
        \includegraphics[width=\linewidth]{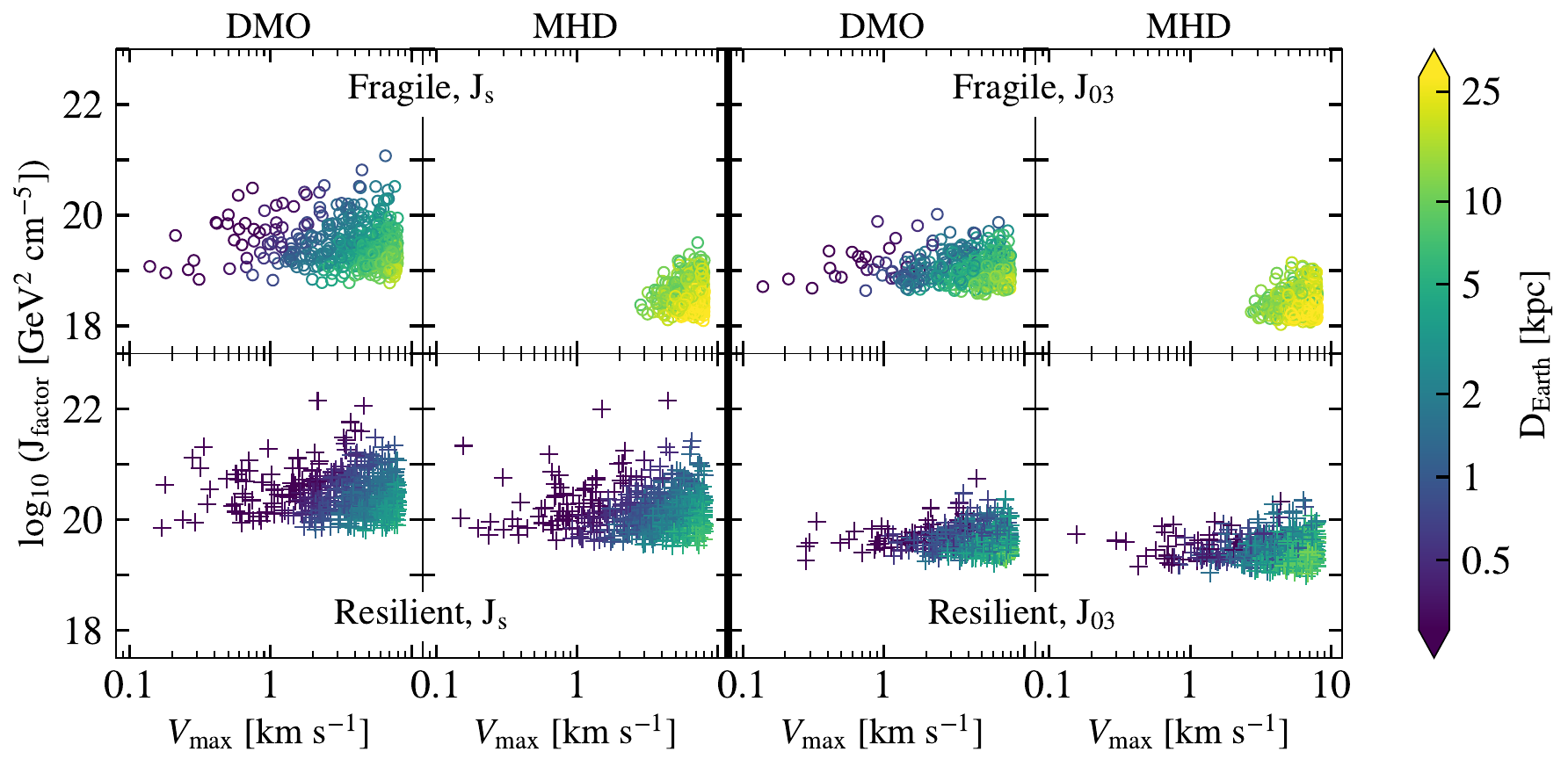}
        \caption{J-factor versus $V_\mathrm{max}$ for the brightest subhalo in 500 repopulations with $V_\mathrm{max}\in [0.1, 8]\,$km/s, with the colored z-axis representing the distance to Earth. 
        Different scenarios are shown depending on subhalo resilience (top and bottom panels for fragile and resilient populations, respectively), definition of the J-factor used (left for $J_\mathrm{S}$ (Eq.~\eqref{eq:Js}) and right for $J_{03}$ (Eq.~\eqref{eq:J03})) and inclusion or not of hydrodynamics (left for DMO; right for MHD for each J-factor definition).
        }
        \label{fig:VmaxJs_full_to8}
    \end{figure*}

    In Fig.~\ref{fig:VmaxJs_full_to8} we show the J-factors as a function of $V_\mathrm{max}$. We explicitly observe that the fragile MHD scenario does not have subhalos $V_\mathrm{max} \leq 2$\,km/s for any definition of the J-factor. Between a 2.6\% (fragile DMO, $J_\mathrm{03}$) and a 9\% (resilient MHD, $J_\mathrm{S}$) of subhalos in all other scenarios have $V_\mathrm{max} = 0.1$\,km/s, which only happened in the resilient MHD scenario in the repopulation up to $V_\mathrm{max}=120\,$km/s.
    
    \begin{figure}
        \centering
        \includegraphics[width=\linewidth]{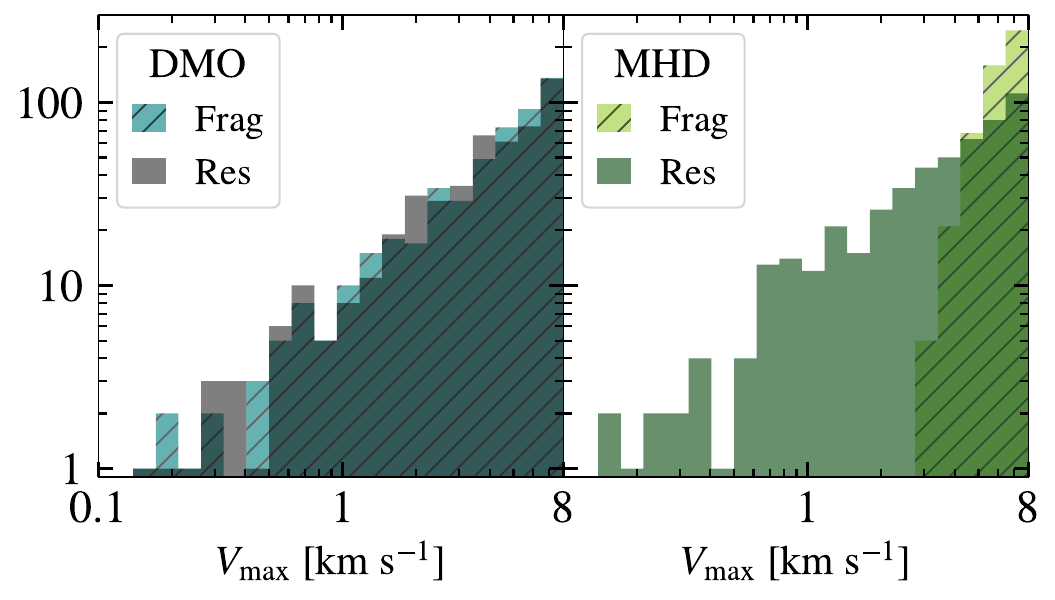}
        \caption{
        Histogram of $V_\mathrm{max}$ for the brightest subhalo using $J_\mathrm{S}$ in 500 repopulations with $V_\mathrm{max}\in [0.1, 8]\,$km/s. We study different scenarios depending on subhalo resilience (fragile vs.~resilient) and inclusion of hydrodynamics (left for DMO and right for MHD).
        }
        \label{fig:Vmax_hist_geom_to8}
    \end{figure}

    In Fig.~\ref{fig:Vmax_hist_geom_to8} we show the $V_\mathrm{max}$ histograms, using the $J_\mathrm{S}$ definition. Both resilient and fragile scenarios of DMO present subhalos with $V_\mathrm{max}\leq1\,$km/s as the brightest of their repopulations, in contrast to the MHD, where this only occurs in the resilient repopulations.

\bibliographystyle{elsarticle-num-names} 
\bibliography{biblio}

@ARTICLE{santaella23, 
       author = {{Aguirre-Santaella}, Alejandra and {S{\'a}nchez-Conde}, Miguel A.},
        title = "{The viability of low-mass subhaloes as targets for gamma-ray dark matter searches}",
      journal = {\mnras},
         year = 2024,
        month = may,
       volume = {530},
       number = {3},
        pages = {2496-2511},
          doi = {10.1093/mnras/stae940},
archivePrefix = {arXiv},
       eprint = {2309.02330},
 primaryClass = {astro-ph.GA},
       adsurl = {https://ui.adsabs.harvard.edu/abs/2024MNRAS.530.2496A}
}

@ARTICLE{2018MNRAS.474.3043V,
       author = {{van den Bosch}, Frank C. and {Ogiya}, Go and {Hahn}, Oliver and {Burkert}, Andreas},
        title = "{Disruption of dark matter substructure: fact or fiction?}",
      journal = {\mnras},
         year = 2018,
        month = mar,
       volume = {474},
       number = {3},
        pages = {3043-3066},
          doi = {10.1093/mnras/stx2956},
archivePrefix = {arXiv},
       eprint = {1711.05276},
 primaryClass = {astro-ph.GA},
       adsurl = {https://ui.adsabs.harvard.edu/abs/2018MNRAS.474.3043V}
}

@ARTICLE{2022ApJ...926..107M,
       author = {{Malhan}, Khyati and {Ibata}, Rodrigo A. and {Sharma}, Sanjib and {Famaey}, Benoit and {Bellazzini}, Michele and {Carlberg}, Raymond G. and {D'Souza}, Richard and {Yuan}, Zhen and {Martin}, Nicolas F. and {Thomas}, Guillaume F.},
        title = "{The Global Dynamical Atlas of the Milky Way Mergers: Constraints from Gaia EDR3-based Orbits of Globular Clusters, Stellar Streams, and Satellite Galaxies}",
      journal = {\apj},
         year = 2022,
        month = feb,
       volume = {926},
       number = {2},
          eid = {107},
        pages = {107},
          doi = {10.3847/1538-4357/ac4d2a},
archivePrefix = {arXiv},
       eprint = {2202.07660},
 primaryClass = {astro-ph.GA},
       adsurl = {https://ui.adsabs.harvard.edu/abs/2022ApJ...926..107M}
}

@ARTICLE{2003ApJ...599.1082M,
       author = {{Majewski}, Steven R. and {Skrutskie}, M.~F. and {Weinberg}, Martin D. and {Ostheimer}, James C.},
        title = "{A Two Micron All Sky Survey View of the Sagittarius Dwarf Galaxy. I. Morphology of the Sagittarius Core and Tidal Arms}",
      journal = {\apj},
         year = 2003,
        month = dec,
       volume = {599},
       number = {2},
        pages = {1082-1115},
          doi = {10.1086/379504},
archivePrefix = {arXiv},
       eprint = {astro-ph/0304198},
 primaryClass = {astro-ph},
       adsurl = {https://ui.adsabs.harvard.edu/abs/2003ApJ...599.1082M}
}

@ARTICLE{2020MNRAS.491.4591E,
       author = {{Errani}, Rapha{\"e}l and {Pe{\~n}arrubia}, Jorge},
        title = "{Can tides disrupt cold dark matter subhaloes?}",
      journal = {\mnras},
         year = 2020,
        month = feb,
       volume = {491},
       number = {4},
        pages = {4591-4601},
          doi = {10.1093/mnras/stz3349},
archivePrefix = {arXiv},
       eprint = {1906.01642},
 primaryClass = {astro-ph.GA},
       adsurl = {https://ui.adsabs.harvard.edu/abs/2020MNRAS.491.4591E}
}

@ARTICLE{2021arXiv211101148A,
       author = {{Amorisco}, Nicola C.},
        title = "{Cold dark matter subhaloes at arbitrarily low masses}",
      journal = {arXiv e-prints},
         year = 2021,
        month = nov,
          eid = {arXiv:2111.01148},
        pages = {arXiv:2111.01148},
archivePrefix = {arXiv},
       eprint = {2111.01148},
 primaryClass = {astro-ph.CO},
       adsurl = {https://ui.adsabs.harvard.edu/abs/2021arXiv211101148A}
}

@ARTICLE{2017MNRAS.471.1709G,
       author = {Garrison-Kimmel, Shea and {Wetzel}, Andrew and {Bullock}, James S. and {Hopkins}, Philip F. and {Boylan-Kolchin}, Michael and {Faucher-Gigu{\`e}re}, Claude-Andr{\'e} and {Kere{\v{s}}}, Du{\v{s}}an and {Quataert}, Eliot and {Sanderson}, Robyn E. and {Graus}, Andrew S. and {Kelley}, Tyler},
        title = "{Not so lumpy after all: modelling the depletion of dark matter subhaloes by Milky Way-like galaxies}",
      journal = {\mnras},
         year = 2017,
        month = oct,
       volume = {471},
       number = {2},
        pages = {1709-1727},
          doi = {10.1093/mnras/stx1710},
archivePrefix = {arXiv},
       eprint = {1701.03792},
 primaryClass = {astro-ph.GA},
       adsurl = {https://ui.adsabs.harvard.edu/abs/2017MNRAS.471.1709G}
}

@ARTICLE{2005PhR...405..279B,
       author = {{Bertone}, Gianfranco and {Hooper}, Dan and {Silk}, Joseph},
        title = "{Particle dark matter: evidence, candidates and constraints}",
      journal = {\physrep},
         year = 2005,
        month = jan,
       volume = {405},
       number = {5-6},
        pages = {279-390},
          doi = {10.1016/j.physrep.2004.08.031},
archivePrefix = {arXiv},
       eprint = {hep-ph/0404175},
 primaryClass = {hep-ph},
       adsurl = {https://ui.adsabs.harvard.edu/abs/2005PhR...405..279B}
}

@ARTICLE{1997ApJ490493N,
       author = {{Navarro}, Julio F. and {Frenk}, Carlos S. and {White}, Simon D.~M.},
        title = "{A Universal Density Profile from Hierarchical Clustering}",
      journal = {APJ},
         year = 1997,
        month = dec,
       volume = {490},
       number = {2},
        pages = {493-508},
          doi = {10.1086/304888},
archivePrefix = {arXiv},
       eprint = {astro-ph/9611107},
 primaryClass = {astro-ph},
       adsurl = {https://ui.adsabs.harvard.edu/abs/1997ApJ...490..493N}
}

@article{S_nchez_Conde_2014,
   title={The flattening of the concentration–mass relation towards low halo masses and its implications for the annihilation signal boost},
   volume={442},
   ISSN={0035-8711},
   url={http://dx.doi.org/10.1093/mnras/stu1014},
   DOI={10.1093/mnras/stu1014},
   number={3},
   journal={Monthly Notices of the Royal Astronomical Society},
   publisher={Oxford University Press (OUP)},
   author={Sánchez-Conde, Miguel A. and Prada, Francisco},
   year={2014},
   month={Jun},
   pages={2271–2277}
}

@ARTICLE{2016MNRAS.457.1208H,
       author = {{Han}, Jiaxin and {Cole}, Shaun and {Frenk}, Carlos S. and {Jing}, Yipeng},
        title = "{A unified model for the spatial and mass distribution of subhaloes}",
      journal = {\mnras},
         year = 2016,
        month = apr,
       volume = {457},
       number = {2},
        pages = {1208-1223},
          doi = {10.1093/mnras/stv2900},
archivePrefix = {arXiv},
       eprint = {1509.02175},
 primaryClass = {astro-ph.CO},
       adsurl = {https://ui.adsabs.harvard.edu/abs/2016MNRAS.457.1208H}
}

@article{Coco16,
    author = {Hellwing, Wojciech A. and Frenk, Carlos S. and Cautun, Marius and Bose, Sownak and Helly, John and Jenkins, Adrian and Sawala, Till and Cytowski, Maciej},
    title = "{The Copernicus Complexio: a high-resolution view of the small-scale Universe}",
    journal = {Monthly Notices of the Royal Astronomical Society},
    volume = {457},
    number = {4},
    pages = {3492-3509},
    year = {2016},
    month = {02},
    issn = {0035-8711},
    doi = {10.1093/mnras/stw214},
    url = {https://doi.org/10.1093/mnras/stw214},
}

@ARTICLE{2007ApJ...667..859D,
       author = {{Diemand}, J{\"u}rg and {Kuhlen}, Michael and {Madau}, Piero},
        title = "{Formation and Evolution of Galaxy Dark Matter Halos and Their Substructure}",
      journal = {\apj},
         year = 2007,
        month = oct,
       volume = {667},
       number = {2},
        pages = {859-877},
          doi = {10.1086/520573},
archivePrefix = {arXiv},
       eprint = {astro-ph/0703337},
 primaryClass = {astro-ph},
       adsurl = {https://ui.adsabs.harvard.edu/abs/2007ApJ...667..859D}
}

@ARTICLE{zavala2019dark,
       author = {{Zavala}, Jes{\'u}s and {Frenk}, Carlos S.},
        title = "{Dark Matter Haloes and Subhaloes}",
      journal = {Galaxies},
         year = 2019,
        month = sep,
       volume = {7},
       number = {4},
        pages = {81},
          doi = {10.3390/galaxies7040081},
archivePrefix = {arXiv},
       eprint = {1907.11775},
 primaryClass = {astro-ph.CO},
       adsurl = {https://ui.adsabs.harvard.edu/abs/2019Galax...7...81Z}
}

@article{van_den_Bosch_2018,
   title={Dark matter substructure in numerical simulations: a tale of discreteness noise, runaway instabilities, and artificial disruption},
   volume={475},
   ISSN={1365-2966},
   url={http://dx.doi.org/10.1093/mnras/sty084},
   DOI={10.1093/mnras/sty084},
   number={3},
   journal={Monthly Notices of the Royal Astronomical Society},
   publisher={Oxford University Press (OUP)},
   author={van den Bosch, Frank C and Ogiya, Go},
   year={2018},
   month={Jan},
   pages={4066–4087}
}

@article{Atwood_2009,
   title={THE LARGE AREA TELESCOPE ON THE FERMI GAMMA-RAY SPACE TELESCOPE MISSION},
   volume={697},
   ISSN={1538-4357},
   url={http://dx.doi.org/10.1088/0004-637X/697/2/1071},
   DOI={10.1088/0004-637x/697/2/1071},
   number={2},
   journal={The Astrophysical Journal},
   publisher={American Astronomical Society},
   author={Atwood, W. B. and Abdo, A. A. and Ackermann, M. and Althouse, W. and Anderson, B. and Axelsson, M. and Baldini, L. and Ballet, J. and Band, D. L. and Barbiellini, G. and et al.},
   year={2009},
   month={May},
   pages={1071–1102}
}

@article{Bringmann_2009,
   title={Particle models and the small-scale structure of dark matter},
   volume={11},
   ISSN={1367-2630},
   url={http://dx.doi.org/10.1088/1367-2630/11/10/105027},
   DOI={10.1088/1367-2630/11/10/105027},
   number={10},
   journal={New Journal of Physics},
   publisher={IOP Publishing},
   author={Bringmann, Torsten},
   year={2009},
   month={Oct},
   pages={105027}
}

@article{Sawala_2015,
   title={The chosen few: the low-mass haloes that host faint galaxies},
   volume={456},
   ISSN={1365-2966},
   url={http://dx.doi.org/10.1093/mnras/stv2597},
   DOI={10.1093/mnras/stv2597},
   number={1},
   journal={Monthly Notices of the Royal Astronomical Society},
   publisher={Oxford University Press (OUP)},
   author={Sawala, Till and Frenk, Carlos S. and Fattahi, Azadeh and Navarro, Julio F. and Theuns, Tom and Bower, Richard G. and Crain, Robert A. and Furlong, Michelle and Jenkins, Adrian and Schaller, Matthieu and et al.},
   year={2015},
   month={Dec},
   pages={85–97}
}

@ARTICLE{2015MNRAS.448.2941S,
       author = {{Sawala}, Till and {Frenk}, Carlos S. and {Fattahi}, Azadeh and {Navarro}, Julio F. and {Bower}, Richard G. and {Crain}, Robert A. and {Dalla Vecchia}, Claudio and {Furlong}, Michelle and {Jenkins}, Adrian and {McCarthy}, Ian G. and {Qu}, Yan and {Schaller}, Matthieu and {Schaye}, Joop and {Theuns}, Tom},
        title = "{Bent by baryons: the low-mass galaxy-halo relation}",
      journal = {\mnras},
         year = 2015,
        month = apr,
       volume = {448},
       number = {3},
        pages = {2941-2947},
          doi = {10.1093/mnras/stu2753},
archivePrefix = {arXiv},
       eprint = {1404.3724},
 primaryClass = {astro-ph.GA},
       adsurl = {https://ui.adsabs.harvard.edu/abs/2015MNRAS.448.2941S}
}

@article{Molin_2017,
    author = {Moliné, {\'A}ngeles and Sánchez-Conde, Miguel A. and Palomares-Ruiz, Sergio and Prada, Francisco},
    title = {Characterization of subhalo structural properties and implications for dark matter annihilation signals},
    journal = {Monthly Notices of the Royal Astronomical Society},
    volume = {466},
    number = {4},
    pages = {4974-4990},
    year = {2017},
    month = {01},
    issn = {0035-8711},
    doi = {10.1093/mnras/stx026},
    url = {https://doi.org/10.1093/mnras/stx026},
}

@ARTICLE{Bringmann2021PreciseDM,
       author = {{Bringmann}, Torsten and {Depta}, Paul Frederik and {Hufnagel}, Marco and {Schmidt-Hoberg}, Kai},
        title = "{Precise dark matter relic abundance in decoupled sectors}",
      journal = {Physics Letters B},
         year = 2021,
        month = jun,
       volume = {817},
          eid = {136341},
        pages = {136341},
          doi = {10.1016/j.physletb.2021.136341},
archivePrefix = {arXiv},
       eprint = {2007.03696},
 primaryClass = {hep-ph},
       adsurl = {https://ui.adsabs.harvard.edu/abs/2021PhLB..81736341B}
}

@article{Springel_2008,
author = {{Springel}, V. and {Wang}, J. and {Vogelsberger}, M. and {Ludlow}, A. and {Jenkins}, A. and {Helmi}, A. and {Navarro}, J.~F. and {Frenk}, C.~S. and {White}, S.~D.~M.},
        title = "{The Aquarius Project: the subhaloes of galactic haloes}",
      journal = {\mnras},
         year = 2008,
        month = dec,
       volume = {391},
       number = {4},
        pages = {1685-1711},
          doi = {10.1111/j.1365-2966.2008.14066.x},
archivePrefix = {arXiv},
       eprint = {0809.0898},
 primaryClass = {astro-ph},
       adsurl = {https://ui.adsabs.harvard.edu/abs/2008MNRAS.391.1685S}
}

@article{Grand_2017,
    author = {Grand, Robert J. J. and Gómez, Facundo A. and Marinacci, Federico and Pakmor, Rüdiger and Springel, Volker and Campbell, David J. R. and Frenk, Carlos S. and Jenkins, Adrian and White, Simon D. M.},
    title = {The Auriga Project: the properties and formation mechanisms of disc galaxies across cosmic time},
    journal = {Monthly Notices of the Royal Astronomical Society},
    volume = {467},
    number = {1},
    pages = {179-207},
    year = {2017},
    month = {01},
    issn = {0035-8711},
    doi = {10.1093/mnras/stx071},
    url = {https://doi.org/10.1093/mnras/stx071},
}

@article{Planck2018results,
   title={Planck2018 results},
   volume={641},
   ISSN={1432-0746},
   url={http://dx.doi.org/10.1051/0004-6361/201833880},
   DOI={10.1051/0004-6361/201833880},
   journal={Astronomy \& Astrophysics},
   publisher={EDP Sciences},
   author={Aghanim, N. and Akrami, Y. and Arroja, F. and Ashdown, M. and Aumont, J. and Baccigalupi, C. and Ballardini, M. and Banday, A. J. and Barreiro, R. B. and et al.},
   year={2020},
   month={Sep},
   pages={A1}
}

@article{Grand_2018,
	doi = {10.1093/mnras/sty2403},
  
	url = {https://doi.org/10.1093\%2Fmnras\%2Fsty2403},
  
	year = 2018,
	month = {sep},
  
	publisher = {Oxford University Press ({OUP})},
  
	volume = {481},
  
	number = {2},
  
	pages = {1726--1743},
  
	author = {Robert J J Grand and John Helly and Azadeh Fattahi and Marius Cautun and Shaun Cole and Andrew P Cooper and Alis J Deason and Carlos Frenk and Facundo A G{\'{o}}mez and Jason A S Hunt and Federico Marinacci and Rüdiger Pakmor and Christine M Simpson and Volker Springel and Dandan Xu},
  
	title = {Aurigaia: mock Gaia {DR}2 stellar catalogues from the auriga cosmological simulations},
  
	journal = {Monthly Notices of the Royal Astronomical Society}
}

@article{Coronado_Bl_zquez_2022,
	doi = {10.1103/physrevd.105.083006},
  
	url = {https://doi.org/10.1103\%2Fphysrevd.105.083006},
  
	year = 2022,
	month = {apr},
  
	publisher = {American Physical Society ({APS})},
  
	volume = {105},
  
	number = {8},
  
	author = {Javier Coronado-Bl{\'{a}}zquez and Miguel A. S{\'{a}}nchez-Conde and Judit P{\'{e}}rez-Romero and Alejandra Aguirre-Santaella and},
  
	title = {Spatial extension of dark subhalos as seen by 
		Fermi-LAT and the implications for {WIMP} constraints},
  
	journal = {Physical Review D}
}

@article{Coronado_Bl_zquez_2021,
	doi = {10.1016/j.dark.2021.100845},
  
	url = {https://doi.org/10.1016\%2Fj.dark.2021.100845},
  
	year = 2021,
	month = {may},
  
	publisher = {Elsevier {BV}
},
  
	volume = {32},
  
	pages = {100845},
  
	author = {Javier Coronado-Bl{\'{a}}zquez and Michele Doro and Miguel A. S{\'{a}}nchez-Conde and Alejandra Aguirre-Santaella},
  
	title = {Sensitivity of the Cherenkov Telescope Array to dark subhalos},
  
	journal = {Physics of the Dark Universe}
}

@article{Coronado_Bl_zquez_2019,
   title={Unidentified gamma-ray sources as targets for indirect dark matter detection with theFermi-Large Area Telescope},
   volume={2019},
   ISSN={1475-7516},
   url={http://dx.doi.org/10.1088/1475-7516/2019/07/020},
   DOI={10.1088/1475-7516/2019/07/020},
   number={07},
   journal={Journal of Cosmology and Astroparticle Physics},
   publisher={IOP Publishing},
   author={Coronado-Blázquez, Javier and Sánchez-Conde, Miguel A. and Domínguez, Alberto and Aguirre-Santaella, Alejandra and Mauro, Mattia Di and Mirabal, Néstor and Nieto, Daniel and Charles, Eric},
   year={2019},
   month={Jul},
   pages={020–020}
}

@article{Grand_2020,
	doi = {10.1093/mnras/staa3993},
  
	url = {https://doi.org/10.1093\%2Fmnras\%2Fstaa3993},
  
	year = 2020,
	month = {dec},
  
	publisher = {Oxford University Press ({OUP})},
  
	author = {Robert J J Grand and Simon D M White},
  
	title = {Baryonic effects on the detectability of annihilation radiation from dark matter subhaloes around the Milky Way},
  
	journal = {Monthly Notices of the Royal Astronomical Society}
}

@article{Grand_2021,
	doi = {10.1093/mnras/stab2492},
  
	url = {https://doi.org/10.1093\%2Fmnras\%2Fstab2492},
  
	year = 2021,
	month = {sep},
  
	publisher = {Oxford University Press ({OUP})},
  
	volume = {507},
  
	number = {4},
  
	pages = {4953--4967},
  
	author = {Robert J J Grand and Federico Marinacci and Rüdiger Pakmor and Christine M Simpson and Ashley J Kelly and Facundo A G{\'{o}}mez and Adrian Jenkins and Volker Springel and Carlos S Frenk and Simon D M White},
  
	title = {Determining the full satellite population of a Milky Way-mass halo in a highly resolved cosmological hydrodynamic simulation},
  
	journal = {Monthly Notices of the Royal Astronomical Society}
}

@ARTICLE{Moline21,
       author = {{Molin{\'e}}, {\'A}ngeles and {S{\'a}nchez-Conde}, Miguel A. and {Aguirre-Santaella}, Alejandra and {Ishiyama}, Tomoaki and {Prada}, Francisco and {Cora}, Sof{\'\i}a A. and {Croton}, Darren and {Jullo}, Eric and {Metcalf}, R. Benton and {Oogi}, Taira and {Ruedas}, Jos{\'e}},
        title = "{{\ensuremath{\Lambda}}CDM halo substructure properties revealed with high-resolution and large-volume cosmological simulations}",
      journal = {\mnras},
         year = 2023,
        month = jan,
       volume = {518},
       number = {1},
        pages = {157-173},
          doi = {10.1093/mnras/stac2930},
archivePrefix = {arXiv},
       eprint = {2110.02097},
 primaryClass = {astro-ph.CO},
       adsurl = {https://ui.adsabs.harvard.edu/abs/2023MNRAS.518..157M}
}

@article{Stref_2019,
	doi = {10.3390/galaxies7020065},
  
	url = {https://doi.org/10.3390\%2Fgalaxies7020065},
  
	year = 2019,
	month = {jun},
  
	publisher = {{MDPI} {AG}
},
  
	volume = {7},
  
	number = {2},
  
	pages = {65},
  
	author = {Martin Stref and Thomas Lacroix and Julien Lavalle},
  
	title = {Remnants of Galactic Subhalos and Their Impact on Indirect Dark-Matter Searches},
  
	journal = {Galaxies}
}

@article{Klypin_2002,
	doi = {10.1086/340656},
  
	url = {https://doi.org/10.1086\%2F340656},
  
	year = 2002,
	month = {jul},
  
	publisher = {American Astronomical Society},
  
	volume = {573},
  
	number = {2},
  
	pages = {597--613},
  
	author = {Anatoly Klypin and HongSheng Zhao and Rachel S. Somerville},
  
	title = {{LCDM}-based Models for the Milky Way and M31. I. Dynamical Models},
  
	journal = {The Astrophysical Journal}
}

@ARTICLE{Diemand_2008,
       author = {{Diemand}, J. and {Kuhlen}, M. and {Madau}, P. and {Zemp}, M. and {Moore}, B. and {Potter}, D. and {Stadel}, J.},
        title = "{Clumps and streams in the local dark matter distribution}",
      journal = {\nat},
         year = 2008,
        month = aug,
       volume = {454},
       number = {7205},
        pages = {735-738},
          doi = {10.1038/nature07153},
archivePrefix = {arXiv},
       eprint = {0805.1244},
 primaryClass = {astro-ph},
       adsurl = {https://ui.adsabs.harvard.edu/abs/2008Natur.454..735D}
}

@ARTICLE{2024Grand_releasedataauriga,
       author = {{Grand}, Robert J.~J. and {Fragkoudi}, Francesca and {G{\'o}mez}, Facundo A. and {Jenkins}, Adrian and {Marinacci}, Federico and {Pakmor}, R{\"u}diger and {Springel}, Volker},
        title = "{Overview and public data release of the augmented Auriga Project: cosmological simulations of dwarf and Milky Way-mass galaxies}",
      journal = {\mnras},
         year = 2024,
        month = aug,
       volume = {532},
       number = {2},
        pages = {1814-1831},
          doi = {10.1093/mnras/stae1598},
archivePrefix = {arXiv},
       eprint = {2401.08750},
 primaryClass = {astro-ph.GA},
       adsurl = {https://ui.adsabs.harvard.edu/abs/2024MNRAS.532.1814G}
}

@article{Aguirre-Santaella_sheddinglight,
    author = {Aguirre-Santaella, Alejandra and Sánchez-Conde, Miguel A and Ogiya, Go and Stücker, Jens and Angulo, Raul E},
    title = {Shedding light on low-mass subhalo survival and annihilation luminosity with numerical simulations},
    journal = {Monthly Notices of the Royal Astronomical Society},
    volume = {518},
    number = {1},
    pages = {93-110},
    year = {2022},
    month = {10},
    issn = {0035-8711},
    doi = {10.1093/mnras/stac2921},
    url = {https://doi.org/10.1093/mnras/stac2921}
}

@article{WEEKES2002221,
title = {VERITAS: the Very Energetic Radiation Imaging Telescope Array System},
journal = {Astroparticle Physics},
volume = {17},
number = {2},
pages = {221-243},
year = {2002},
issn = {0927-6505},
doi = {https://doi.org/10.1016/S0927-6505(01)00152-9},
url = {https://www.sciencedirect.com/science/article/pii/S0927650501001529},
author = {T.C Weekes and H Badran and S.D Biller and I Bond and S Bradbury and J Buckley and D Carter-Lewis and M Catanese and S Criswell and W Cui and P Dowkontt and C Duke and D.J Fegan and J Finley and L Fortson and J Gaidos and G.H Gillanders and J Grindlay and T.A Hall and K Harris and A.M Hillas and P Kaaret and M Kertzman and D Kieda and F Krennrich and M.J Lang and S LeBohec and R Lessard and J Lloyd-Evans and J Knapp and B McKernan and J McEnery and P Moriarty and D Muller and P Ogden and R Ong and D Petry and J Quinn and N.W Reay and P.T Reynolds and J Rose and M Salamon and G Sembroski and R Sidwell and P Slane and N Stanton and S.P Swordy and V.V Vassiliev and S.P Wakely}
}

@ARTICLE{2004NewAR..48..331H,
       author = {{Hinton}, J.~A. and {HESS Collaboration}},
        title = "{The status of the HESS project}",
      journal = {\nar},
         year = 2004,
        month = apr,
       volume = {48},
       number = {5-6},
        pages = {331-337},
          doi = {10.1016/j.newar.2003.12.004},
archivePrefix = {arXiv},
       eprint = {astro-ph/0403052},
 primaryClass = {astro-ph},
       adsurl = {https://ui.adsabs.harvard.edu/abs/2004NewAR..48..331H}
}

@ARTICLE{2005NIMPA.553..274F,
       author = {{Ferenc}, D. and {MAGIC Collaboration}},
        title = "{The MAGIC gamma-ray observatory}",
      journal = {Nuclear Instruments and Methods in Physics Research A},
         year = 2005,
        month = nov,
       volume = {553},
       number = {1-2},
        pages = {274-281},
          doi = {10.1016/j.nima.2005.08.085},
       adsurl = {https://ui.adsabs.harvard.edu/abs/2005NIMPA.553..274F}
}

@article{Ackermann_2012,
doi = {10.1088/0004-637X/747/2/121},
url = {https://dx.doi.org/10.1088/0004-637X/747/2/121},
year = {2012},
month = {feb},
publisher = {The American Astronomical Society},
volume = {747},
number = {2},
pages = {121},
author = {Ackermann, M. and Albert, A. and Baldini, L. and Ballet, J. and Barbiellini, G. and Bastieri, D. and Bechtol, K. and Bellazzini, R. and Blandford, R. D. and Bloom, E. D. and Bonamente, E. and Borgland, A. W. and Bottacini, E. and Brandt, T. J. and Bregeon, J. and Brigida, M. and Bruel, P. and Buehler, R. and Burnett, T. H. and Caliandro, G. A. and Cameron, R. A. and Caraveo, P. A. and Casandjian, J. M. and Cecchi, C. and Charles, E. and Chiang, J. and Ciprini, S. and Claus, R. and Cohen-Tanugi, J. and Conrad, J. and Cutini, S. and de Palma, F. and Dermer, C. D. and Digel, S. W. and do Couto e Silva, E. and Drell, P. S. and Drlica-Wagner, A. and Essig, R. and Falletti, L. and Favuzzi, C. and Fegan, S. J. and Focke, W. B. and Fukazawa, Y. and Funk, S. and Fusco, P. and Gargano, F. and Germani, S. and Giglietto, N. and Giordano, F. and Giroletti, M. and Glanzman, T. and Godfrey, G. and Grenier, I. A. and Guiriec, S. and Gustafsson, M. and Hadasch, D. and Hayashida, M. and Hou, X. and Hughes, R. E. and Johnson, R. P. and Johnson, A. S. and Kamae, T. and Katagiri, H. and Kataoka, J. and Knödlseder, J. and Kuss, M. and Lande, J. and Latronico, L. and Lee, S.-H. and Lionetto, A. M. and Garde, M. Llena and Longo, F. and Loparco, F. and Lovellette, M. N. and Lubrano, P. and Mazziotta, M. N. and McEnery, J. E. and Michelson, P. F. and Mitthumsiri, W. and Mizuno, T. and Moiseev, A. A. and Monte, C. and Monzani, M. E. and Morselli, A. and Moskalenko, I. V. and Murgia, S. and Naumann-Godo, M. and Norris, J. P. and Nuss, E. and Ohsugi, T. and Okumura, A. and Orlando, E. and Ormes, J. F. and Ozaki, M. and Paneque, D. and Pelassa, V. and Pierbattista, M. and Piron, F. and Pivato, G. and Porter, T. A. and Rainò, S. and Rando, R. and Razzano, M. and Reimer, A. and Reimer, O. and Ritz, S. and Sadrozinski, H. F.-W. and Sehgal, N. and Sgrò, C. and Siskind, E. J. and Spinelli, P. and Strigari, L. and Suson, D. J. and Tajima, H. and Takahashi, H. and Tanaka, T. and Thayer, J. G. and Thayer, J. B. and Tibaldo, L. and Tinivella, M. and Torres, D. F. and Troja, E. and Uchiyama, Y. and Usher, T. L. and Vandenbroucke, J. and Vasileiou, V. and Vianello, G. and Vitale, V. and Waite, A. P. and Wang, P. and Winer, B. L. and Wood, K. S. and Yang, Z. and Zalewski, S. and Zimmer, S.},
title = {SEARCH FOR DARK MATTER SATELLITES USING FERMI-LAT},
journal = {The Astrophysical Journal}
}

@ARTICLE{2019JCAP...11..045C,
       author = {{Coronado-Bl{\'a}zquez}, Javier and {S{\'a}nchez-Conde}, Miguel A. and {Di Mauro}, Mattia and {Aguirre-Santaella}, Alejandra and {Ciuc{\u{a}}}, Ioana and {Dom{\'\i}nguez}, Alberto and {Kawata}, Daisuke and {Mirabal}, N{\'e}stor},
        title = "{Spectral and spatial analysis of the dark matter subhalo candidates among Fermi Large Area Telescope unidentified sources}",
      journal = {\jcap},
         year = 2019,
        month = nov,
       volume = {2019},
       number = {11},
          eid = {045},
        pages = {045},
          doi = {10.1088/1475-7516/2019/11/045},
archivePrefix = {arXiv},
       eprint = {1910.14429},
 primaryClass = {astro-ph.HE},
       adsurl = {https://ui.adsabs.harvard.edu/abs/2019JCAP...11..045C}
}

@article{ErraniNavarro21,
    author = {Errani, Raphaël and Navarro, Julio F},
    title = {The asymptotic tidal remnants of cold dark matter subhaloes},
    journal = {Monthly Notices of the Royal Astronomical Society},
    volume = {505},
    number = {1},
    pages = {18-32},
    year = {2021},
    month = {05},
    issn = {0035-8711},
    doi = {10.1093/mnras/stab1215},
    url = {https://doi.org/10.1093/mnras/stab1215},
}

@article{LHAASOsciencebook,
  TITLE = {{The Large High Altitude Air Shower Observatory (LHAASO) Science Book (2021 Edition)}},
  AUTHOR = {Bai, X. and Bi, B.Y. and Bi, X.J. and Cao, Z. and Chen, S.Z. and Chen, Y. and Chiavassa, A. and Cui, X.H. and Dai, Z.G. and Della Volpe, D. and Di Girolamo, T. and Di Sciascio, Giuseppe and Fan, Y.Z. and Giacalone, J. and Guo, Y.Q. and He, H.H. and He, T.L. and Heller, M. and Huang, D. and Huang, Y.F. and Jia, H. and Ksenofontov, L.T. and Leahy, D. and Li, F. and Li, Z. and Liang, E.W. and Lipari, P. and Liu, R.Y. and Liu, Y. and Liu, S. and Ma, X. and Martineau-Huynh, O. and Martraire, D. and Montaruli, T. and Ruffolo, D. and Stenkin, Y.V. and Su, H.Q. and Tam, T. and Tang, Q.W. and Tian, W.W. and Vallania, P. and Vernetto, S. and Vigorito, C. and Wang, J..C. and Wang, L.Z. and Wang, X.Y. and Wang, X.J. and Wang, Z.X. and Wei, D.M. and Wei, J.J. and Wu, D. and Wu, H.R. and Wu, X.F. and Yan, D.H. and Yang, A.Y. and Yang, R.Z. and Yao, Z.G. and Yin, L.Q. and Yuan, Q. and Zhang, Bing and Zhang, L. and Zhang, M.F. and Zhang, S.S. and Zhang, X. and Zhao, Yi and Zhou, X.X. and Zhu, F.R. and Zhu, H.},
  URL = {https://hal.science/hal-02153252},
  JOURNAL = {{Chinese Physics C}},
  PUBLISHER = {{IOP Publishing}},
  VOLUME = {46},
  PAGES = {035001-035007},
  YEAR = {2022},
  HAL_ID = {hal-02153252},
  HAL_VERSION = {v1},
}

@ARTICLE{2012Hawc,
       author = {{Abeysekara}, A.~U. and {Aguilar}, J.~A. and {Aguilar}, S. and {Alfaro}, R. and {Almaraz}, E. and {{\'A}lvarez}, C. and others},
        title = "{On the sensitivity of the HAWC observatory to gamma-ray bursts}",
      journal = {Astroparticle Physics},
         year = 2012,
        month = may,
       volume = {35},
       number = {10},
        pages = {641-650},
          doi = {10.1016/j.astropartphys.2012.02.001},
archivePrefix = {arXiv},
       eprint = {1108.6034},
 primaryClass = {astro-ph.HE},
       adsurl = {https://ui.adsabs.harvard.edu/abs/2012APh....35..641A}
}

@article{BERGSTROM1998137,
title = {Observability of gamma rays from dark matter neutralino annihilations in the Milky Way halo},
journal = {Astroparticle Physics},
volume = {9},
number = {2},
pages = {137-162},
year = {1998},
issn = {0927-6505},
doi = {https://doi.org/10.1016/S0927-6505(98)00015-2},
url = {https://www.sciencedirect.com/science/article/pii/S0927650598000152},
author = {Lars {Bergstr{\"o}m} and Piero Ullio and James H. Buckley},
}

@ARTICLE{1997Bregstroem,
       author = {{Bergstr{\"o}m}, Lars and {Ullio}, Piero},
        title = "{Full one-loop calculation of neutralino annihilation into two photons}",
      journal = {Nuclear Physics B},
         year = 1997,
        month = feb,
       volume = {504},
        pages = {27-44},
          doi = {10.1016/S0550-3213(97)00530-0},
archivePrefix = {arXiv},
       eprint = {hep-ph/9706232},
 primaryClass = {hep-ph},
       adsurl = {https://ui.adsabs.harvard.edu/abs/1997NuPhB.504...27B}
}

@article{PhysRevDEvans2004,
  title = {A travel guide to the dark matter annihilation signal},
  author = {Evans, N. W. and Ferrer, F. and Sarkar, S.},
  journal = {Phys. Rev. D},
  volume = {69},
  issue = {12},
  pages = {123501},
  numpages = {10},
  year = {2004},
  month = {Jun},
  publisher = {American Physical Society},
  doi = {10.1103/PhysRevD.69.123501},
  url = {https://link.aps.org/doi/10.1103/PhysRevD.69.123501}
}

@ARTICLE{2018B_history,
       author = {{Bertone}, Gianfranco and {Hooper}, Dan},
        title = "{History of dark matter}",
      journal = {Reviews of Modern Physics},
         year = 2018,
        month = oct,
       volume = {90},
       number = {4},
          eid = {045002},
        pages = {045002},
          doi = {10.1103/RevModPhys.90.045002},
archivePrefix = {arXiv},
       eprint = {1605.04909},
 primaryClass = {astro-ph.CO},
       adsurl = {https://ui.adsabs.harvard.edu/abs/2018RvMP...90d5002B}
}

@ARTICLE{2018RPPh...81f6201R,
       author = {{Roszkowski}, Leszek and {Sessolo}, Enrico Maria and {Trojanowski}, Sebastian},
        title = "{WIMP dark matter candidates and searches{\textemdash}current status and future prospects}",
      journal = {Reports on Progress in Physics},
         year = 2018,
        month = jun,
       volume = {81},
       number = {6},
          eid = {066201},
        pages = {066201},
          doi = {10.1088/1361-6633/aab913},
archivePrefix = {arXiv},
       eprint = {1707.06277},
 primaryClass = {hep-ph},
       adsurl = {https://ui.adsabs.harvard.edu/abs/2018RPPh...81f6201R}
}

@ARTICLE{2018Natur.562...51B,
       author = {{Bertone}, Gianfranco and {Tait}, Tim M.~P.},
        title = "{A new era in the search for dark matter}",
      journal = {\nat},
         year = 2018,
        month = oct,
       volume = {562},
       number = {7725},
        pages = {51-56},
          doi = {10.1038/s41586-018-0542-z},
archivePrefix = {arXiv},
       eprint = {1810.01668},
 primaryClass = {astro-ph.CO},
       adsurl = {https://ui.adsabs.harvard.edu/abs/2018Natur.562...51B}
}

@article{garret2011,
author = {Garrett, Katherine and Dūda, Gintaras},
title = {Dark Matter: A Primer},
journal = {Advances in Astronomy},
volume = {2011},
number = {1},
pages = {968283},
doi = {https://doi.org/10.1155/2011/968283},
url = {https://onlinelibrary.wiley.com/doi/abs/10.1155/2011/968283},
year = {2011}
}

@article{JUNGMAN1996195,
title = {Supersymmetric dark matter},
journal = {Physics Reports},
volume = {267},
number = {5},
pages = {195-373},
year = {1996},
issn = {0370-1573},
doi = {https://doi.org/10.1016/0370-1573(95)00058-5},
url = {https://www.sciencedirect.com/science/article/pii/0370157395000585},
author = {Gerard Jungman and Marc Kamionkowski and Kim Griest}
}

@ARTICLE{2025AguirreSantaella,
    author = {Aguirre-Santaella, Alejandra and Sánchez-Conde, Miguel A and Ogiya, Go},
    title = {New insights on low-mass dark matter subhalo tidal tracks via numerical simulations},
    journal = {Monthly Notices of the Royal Astronomical Society},
    volume = {545},
    number = {2},
    pages = {staf2058},
    year = {2025},
    month = {11},
    issn = {0035-8711},
    doi = {10.1093/mnras/staf2058},
    url = {https://doi.org/10.1093/mnras/staf2058},
}

@article{10.1093PhatELVIS,
    author = {Kelley, Tyler and Bullock, James S and Garrison-Kimmel, Shea and Boylan-Kolchin, Michael and Pawlowski, Marcel S and Graus, Andrew S},
    title = {Phat ELVIS: The inevitable effect of the Milky Way’s disc on its dark matter subhaloes},
    journal = {Monthly Notices of the Royal Astronomical Society},
    volume = {487},
    number = {3},
    pages = {4409-4423},
    year = {2019},
    month = {06},
    issn = {0035-8711},
    doi = {10.1093/mnras/stz1553},
    url = {https://doi.org/10.1093/mnras/stz1553},
}

@ARTICLE{2023MNRASFire,
       author = {{Barry}, Megan and {Wetzel}, Andrew and {Chapman}, Sierra and {Samuel}, Jenna and {Sanderson}, Robyn and {Arora}, Arpit},
        title = "{The dark side of FIRE: predicting the population of dark matter subhaloes around Milky Way-mass galaxies}",
      journal = {\mnras},
         year = 2023,
        month = jul,
       volume = {523},
       number = {1},
        pages = {428-440},
          doi = {10.1093/mnras/stad1395},
archivePrefix = {arXiv},
       eprint = {2303.05527},
 primaryClass = {astro-ph.GA},
       adsurl = {https://ui.adsabs.harvard.edu/abs/2023MNRAS.523..428B}
}

@article{10.1093/mnras/fire2017,
    author = {Garrison-Kimmel, Shea and Wetzel, Andrew and Bullock, James S. and Hopkins, Philip F. and Boylan-Kolchin, Michael and Faucher-Giguère, Claude-André and Kereš, Dušan and Quataert, Eliot and Sanderson, Robyn E. and Graus, Andrew S. and Kelley, Tyler},
    title = {Not so lumpy after all: modelling the depletion of dark matter subhaloes by Milky Way-like galaxies },
    journal = {Monthly Notices of the Royal Astronomical Society},
    volume = {471},
    number = {2},
    pages = {1709-1727},
    year = {2017},
    month = {07},
    issn = {0035-8711},
    doi = {10.1093/mnras/stx1710},
    url = {https://doi.org/10.1093/mnras/stx1710},
}

@article{10.1093Uchuu_release,
    author = {Ishiyama, Tomoaki and Prada, Francisco and Klypin, Anatoly A and Sinha, Manodeep and Metcalf, R Benton and Jullo, Eric and Altieri, Bruno and Cora, Sofía A and Croton, Darren and de la Torre, Sylvain and Millán-Calero, David E and Oogi, Taira and Ruedas, José and Vega-Martínez, Cristian A},
    title = {The Uchuu simulations: Data Release 1 and dark matter halo concentrations},
    journal = {Monthly Notices of the Royal Astronomical Society},
    volume = {506},
    number = {3},
    pages = {4210-4231},
    year = {2021},
    month = {06},
    issn = {0035-8711},
    doi = {10.1093/mnras/stab1755},
    url = {https://doi.org/10.1093/mnras/stab1755},
}

@ARTICLE{1974ApJ...187..425P,
       author = {{Press}, William H. and {Schechter}, Paul},
        title = "{Formation of Galaxies and Clusters of Galaxies by Self-Similar Gravitational Condensation}",
      journal = {\apj},
         year = 1974,
        month = feb,
       volume = {187},
        pages = {425-438},
          doi = {10.1086/152650},
       adsurl = {https://ui.adsabs.harvard.edu/abs/1974ApJ...187..425P}
}

@ARTICLE{1999MNRAS.308..119S,
       author = {{Sheth}, Ravi K. and {Tormen}, Giuseppe},
        title = "{Large-scale bias and the peak background split}",
      journal = {\mnras},
         year = 1999,
        month = sep,
       volume = {308},
       number = {1},
        pages = {119-126},
          doi = {10.1046/j.1365-8711.1999.02692.x},
archivePrefix = {arXiv},
       eprint = {astro-ph/9901122},
 primaryClass = {astro-ph},
       adsurl = {https://ui.adsabs.harvard.edu/abs/1999MNRAS.308..119S}
}

@ARTICLE{1998ApJ...499...20J,
       author = {{Jenkins}, A. and {Frenk}, C.~S. and {Pearce}, F.~R. and {Thomas}, P.~A. and {Colberg}, J.~M. and {White}, S.~D.~M. and {Couchman}, H.~M.~P. and {Peacock}, J.~A. and {Efstathiou}, G. and {Nelson}, A.~H.},
        title = "{Evolution of Structure in Cold Dark Matter Universes}",
      journal = {\apj},
         year = 1998,
        month = may,
       volume = {499},
       number = {1},
        pages = {20-40},
          doi = {10.1086/305615},
archivePrefix = {arXiv},
       eprint = {astro-ph/9709010},
 primaryClass = {astro-ph},
       adsurl = {https://ui.adsabs.harvard.edu/abs/1998ApJ...499...20J}
}

@article{PhysRevLett.115.231301,
  title = {Searching for Dark Matter Annihilation from Milky Way Dwarf Spheroidal Galaxies with Six Years of Fermi Large Area Telescope Data},
  author = {Ackermann, M. and Albert, A. and Anderson, B. and Atwood, W. B. and Baldini, L. and Barbiellini, G. and Bastieri, D. and Bechtol, K. and Bellazzini, R. and Bissaldi, E. and Blandford, R. D. and Bloom, E. D. and Bonino, R. and Bottacini, E. and Brandt, T. J. and Bregeon, J. and Bruel, P. and Buehler, R. and Caliandro, G. A. and Cameron, R. A. and Caputo, R. and Caragiulo, M. and Caraveo, P. A. and Cecchi, C. and Charles, E. and Chekhtman, A. and Chiang, J. and Chiaro, G. and Ciprini, S. and Claus, R. and Cohen-Tanugi, J. and Conrad, J. and Cuoco, A. and Cutini, S. and D'Ammando, F. and de Angelis, A. and de Palma, F. and Desiante, R. and Digel, S. W. and Di Venere, L. and Drell, P. S. and Drlica-Wagner, A. and Essig, R. and Favuzzi, C. and Fegan, S. J. and Ferrara, E. C. and Focke, W. B. and Franckowiak, A. and Fukazawa, Y. and Funk, S. and Fusco, P. and Gargano, F. and Gasparrini, D. and Giglietto, N. and Giordano, F. and Giroletti, M. and Glanzman, T. and Godfrey, G. and Gomez-Vargas, G. A. and Grenier, I. A. and Guiriec, S. and Gustafsson, M. and Hays, E. and Hewitt, J. W. and Horan, D. and Jogler, T. and J\'ohannesson, G. and Kuss, M. and Larsson, S. and Latronico, L. and Li, J. and Li, L. and Llena Garde, M. and Longo, F. and Loparco, F. and Lubrano, P. and Malyshev, D. and Mayer, M. and Mazziotta, M. N. and McEnery, J. E. and Meyer, M. and Michelson, P. F. and Mizuno, T. and Moiseev, A. A. and Monzani, M. E. and Morselli, A. and Murgia, S. and Nuss, E. and Ohsugi, T. and Orienti, M. and Orlando, E. and Ormes, J. F. and Paneque, D. and Perkins, J. S. and Pesce-Rollins, M. and Piron, F. and Pivato, G. and Porter, T. A. and Rain\`o, S. and Rando, R. and Razzano, M. and Reimer, A. and Reimer, O. and Ritz, S. and S\'anchez-Conde, M. and Schulz, A. and Sehgal, N. and Sgr\`o, C. and Siskind, E. J. and Spada, F. and Spandre, G. and Spinelli, P. and Strigari, L. and Tajima, H. and Takahashi, H. and Thayer, J. B. and Tibaldo, L. and Torres, D. F. and Troja, E. and Vianello, G. and Werner, M. and Winer, B. L. and Wood, K. S. and Wood, M. and Zaharijas, G. and Zimmer, S.},
  collaboration = {The Fermi-LAT Collaboration},
  journal = {Phys. Rev. Lett.},
  volume = {115},
  issue = {23},
  pages = {231301},
  numpages = {8},
  year = {2015},
  month = {Nov},
  publisher = {American Physical Society},
  doi = {10.1103/PhysRevLett.115.231301},
  url = {https://link.aps.org/doi/10.1103/PhysRevLett.115.231301}
}

@article{PhysRevD.82.063501,
  title = {Dark matter subhalos in the Fermi first source catalog},
  author = {Buckley, Matthew R. and Hooper, Dan},
  journal = {Phys. Rev. D},
  volume = {82},
  issue = {6},
  pages = {063501},
  numpages = {11},
  year = {2010},
  month = {Sep},
  publisher = {American Physical Society},
  doi = {10.1103/PhysRevD.82.063501},
  url = {https://link.aps.org/doi/10.1103/PhysRevD.82.063501}
}

@article{10.1111/j.1745-3933.2012.01287.x,
    author = {Mirabal, N. and Frías-Martinez, V. and Hassan, T. and Frías-Martinez, E.},
    title = {Fermi's SIBYL: mining the gamma-ray sky for dark matter subhaloes},
    journal = {Monthly Notices of the Royal Astronomical Society: Letters},
    volume = {424},
    number = {1},
    pages = {L64-L68},
    year = {2012},
    month = {07},
    issn = {1745-3925},
    doi = {10.1111/j.1745-3933.2012.01287.x},
    url = {https://doi.org/10.1111/j.1745-3933.2012.01287.x},
}

@article{HannesSZechlin_2012,
doi = {10.1088/1475-7516/2012/11/050},
url = {https://doi.org/10.1088/1475-7516/2012/11/050},
year = {2012},
month = {nov},
publisher = {},
volume = {2012},
number = {11},
pages = {050},
author = {Hannes-S. Zechlin and Dieter Horns},
title = {Unidentified sources in the Fermi-LAT second source catalog:
  the case for DM subhalos},
journal = {Journal of Cosmology and Astroparticle Physics}
}

@article{Bertoni_2016,
doi = {10.1088/1475-7516/2016/05/049},
url = {https://dx.doi.org/10.1088/1475-7516/2016/05/049},
year = {2016},
month = {may},
publisher = {},
volume = {2016},
number = {05},
pages = {049},
author = {Bertoni, Bridget and Hooper, Dan and Linden, Tim},
title = {Is the gamma-ray source 3FGL J2212.5+0703 a dark matter subhalo?},
journal = {Journal of Cosmology and Astroparticle Physics}
}

@Article{galaxies7040090,
AUTHOR = {Calore, Francesca and Hütten, Moritz and Stref, Martin},
TITLE = {Gamma-Ray Sensitivity to Dark Matter Subhalo Modelling at High Latitudes},
JOURNAL = {Galaxies},
VOLUME = {7},
YEAR = {2019},
NUMBER = {4},
ARTICLE-NUMBER = {90},
URL = {https://www.mdpi.com/2075-4434/7/4/90},
ISSN = {2075-4434},
DOI = {10.3390/galaxies7040090}
}

@article{2022MNRAS.509.2624G,
       author = {{Green}, Sheridan B. and {van den Bosch}, Frank C. and {Jiang}, Fangzhou},
        title = "{SatGen - II. Assessing the impact of a disc potential on subhalo populations}",
      journal = {\mnras},
         year = 2022,
        month = jan,
       volume = {509},
       number = {2},
        pages = {2624-2636},
          doi = {10.1093/mnras/stab3130},
archivePrefix = {arXiv},
       eprint = {2110.13044},
 primaryClass = {astro-ph.GA},
       adsurl = {https://ui.adsabs.harvard.edu/abs/2022MNRAS.509.2624G}
}

@article{10.1093/mnras/stv2970,
    author = {Fattahi, Azadeh and Navarro, Julio F. and Sawala, Till and Frenk, Carlos S. and Oman, Kyle A. and Crain, Robert A. and Furlong, Michelle and Schaller, Matthieu and Schaye, Joop and Theuns, Tom and Jenkins, Adrian},
    title = {The apostle project: Local Group kinematic mass constraints and simulation candidate selection},
    journal = {Monthly Notices of the Royal Astronomical Society},
    volume = {457},
    number = {1},
    pages = {844-856},
    year = {2016},
    month = {01},
    issn = {0035-8711},
    doi = {10.1093/mnras/stv2970},
    url = {https://doi.org/10.1093/mnras/stv2970},
}

@ARTICLE{2020MNRAS.492.5780R,
       author = {{Richings}, Jack and {Frenk}, Carlos and {Jenkins}, Adrian and {Robertson}, Andrew and {Fattahi}, Azadeh and {Grand}, Robert J.~J. and {Navarro}, Julio and {Pakmor}, R{\"u}diger and {Gomez}, Facundo A. and {Marinacci}, Federico and {Oman}, Kyle A.},
        title = "{Subhalo destruction in the APOSTLE and AURIGA simulations}",
      journal = {\mnras},
         year = 2020,
        month = mar,
       volume = {492},
       number = {4},
        pages = {5780-5793},
          doi = {10.1093/mnras/stz3448},
archivePrefix = {arXiv},
       eprint = {1811.12437},
 primaryClass = {astro-ph.GA},
       adsurl = {https://ui.adsabs.harvard.edu/abs/2020MNRAS.492.5780R}
}

@ARTICLE{deSalas2020,
       author = {{de Salas}, Pablo F. and {Widmark}, A.},
        title = "{Dark matter local density determination: recent observations and future prospects}",
      journal = {Reports on Progress in Physics},
         year = 2021,
        month = oct,
       volume = {84},
       number = {10},
          eid = {104901},
        pages = {104901},
          doi = {10.1088/1361-6633/ac24e7},
archivePrefix = {arXiv},
       eprint = {2012.11477},
 primaryClass = {astro-ph.GA},
       adsurl = {https://ui.adsabs.harvard.edu/abs/2021RPPh...84j4901D}
}

@article{PhysRevD.110.030001,
  title = {Review of Particle Physics},
  author = {Navas, S. and Amsler, C. and Gutsche, T. and Hanhart, C. and Hern\'andez-Rey, J. J. and Louren\ifmmode \mbox{\c{c}}\else \c{c}\fi{}o, C. and Masoni, A. and Mikhasenko, M. and Mitchell, R. E. and Patrignani, C. and Schwanda, C. and Spanier, S. and Venanzoni, G. and Yuan, C. Z. and Agashe, K. and Aielli, G. and Allanach, B. C. and Alvarez-Mu\~niz, J. and Antonelli, M. and Aschenauer, E. C. and Asner, D. M. and Assamagan, K. and Baer, H. and Banerjee, Sw. and Barnett, R. M. and Baudis, L. and Bauer, C. W. and Beatty, J. J. and Beringer, J. and Bettini, A. and Biebel, O. and Black, K. M. and Blucher, E. and Bonventre, R. and Briere, R. A. and Buckley, A. and Burkert, V. D. and Bychkov, M. A. and Cahn, R. N. and Cao, Z. and Carena, M. and Casarosa, G. and Ceccucci, A. and Cerri, A. and Chivukula, R. S. and Cowan, G. and Cranmer, K. and Crede, V. and Cremonesi, O. and D'Ambrosio, G. and Damour, T. and de Florian, D. and de Gouv\^ea, A. and DeGrand, T. and Demers, S. and Demiragli, Z. and Dobrescu, B. A. and D'Onofrio, M. and Doser, M. and Dreiner, H. K. and Eerola, P. and Egede, U. and Eidelman, S. and El-Khadra, A. X. and Ellis, J. and Eno, S. C. and Erler, J. and Ezhela, V. V. and Fava, A. and Fetscher, W. and Fields, B. D. and Freitas, A. and Gallagher, H. and Gershon, T. and Gershtein, Y. and Gherghetta, T. and Gonzalez-Garcia, M. C. and Goodman, M. and Grab, C. and Gritsan, A. V. and Grojean, C. and Groom, D. E. and Gr\"unewald, M. and Gurtu, A. and Haber, H. E. and Hamel, M. and Hashimoto, S. and Hayato, Y. and Hebecker, A. and Heinemeyer, S. and Hikasa, K. and Hisano, J. and H\"ocker, A. and Holder, J. and Hsu, L. and Huston, J. and Hyodo, T. and Ianni, Al. and Kado, M. and Karliner, M. and Katz, U. F. and Kenzie, M. and Khoze, V. A. and Klein, S. R. and Krauss, F. and Kreps, M. and Kri\ifmmode \check{z}\else \v{z}\fi{}an, P. and Krusche, B. and Kwon, Y. and Lahav, O. and Lellouch, L. P. and Lesgourgues, J. and Liddle, A. R. and Ligeti, Z. and Lin, C.-J. and Lippmann, C. and Liss, T. M. and Lister, A. and Littenberg, L. and Lugovsky, K. S. and Lugovsky, S. B. and Lusiani, A. and Makida, Y. and Maltoni, F. and Manohar, A. V. and Marciano, W. J. and Matthews, J. and Mei\ss{}ner, U.-G. and Melzer-Pellmann, I.-A. and Mertsch, P. and Miller, D. J. and Milstead, D. and M\"onig, K. and Molaro, P. and Moortgat, F. and Moskovic, M. and Nagata, N. and Nakamura, K. and Narain, M. and Nason, P. and Nelles, A. and Neubert, M. and Nir, Y. and O'Connell, H. B. and O'Hare, C. A. J. and Olive, K. A. and Peacock, J. A. and Pianori, E. and Pich, A. and Piepke, A. and Pietropaolo, F. and Pomarol, A. and Pordes, S. and Profumo, S. and Quadt, A. and Rabbertz, K. and Rademacker, J. and Raffelt, G. and Ramsey-Musolf, M. and Richardson, P. and Ringwald, A. and Robinson, D. J. and Roesler, S. and Rolli, S. and Romaniouk, A. and Rosenberg, L. J and Rosner, J. L. and Rybka, G. and Ryskin, M. G. and Ryutin, R. A. and Safdi, B. and Sakai, Y. and Sarkar, S. and Sauli, F. and Schneider, O. and Sch\"onert, S. and Scholberg, K. and Schwartz, A. J. and Schwiening, J. and Scott, D. and Sefkow, F. and Seljak, U. and Sharma, V. and Sharpe, S. R. and Shiltsev, V. and Signorelli, G. and Silari, M. and Simon, F. and Sj\"ostrand, T. and Skands, P. and Skwarnicki, T. and Smoot, G. F. and Soffer, A. and Sozzi, M. S. and Spiering, C. and Stahl, A. and Sumino, Y. and Takahashi, F. and Tanabashi, M. and Tanaka, J. and Ta\ifmmode \check{s}\else \v{s}\fi{}evsk\'y, M. and Terao, K. and Terashi, K. and Terning, J. and Thoma, U. and Thorne, R. S. and Tiator, L. and Titov, M. and Tovey, D. R. and Trabelsi, K. and Urquijo, P. and Valencia, G. and Van de Water, R. and Varelas, N. and Verde, L. and Vivarelli, I. and Vogel, P. and Vogelsang, W. and Vorobyev, V. and Wakely, S. P. and Walkowiak, W. and Walter, C. W. and Wands, D. and Weinberg, D. H. and Weinberg, E. J. and Wermes, N. and White, M. and Wiencke, L. R. and Willocq, S. and Woody, C. L. and Workman, R. L. and Yao, W.-M. and Yokoyama, M. and Yoshida, R. and Zanderighi, G. and Zeller, G. P. and Zhu, R.-Y. and Zhu, S.-L. and Zimmermann, F. and Zyla, P. A. and Anderson, J. and Kramer, M. and Schaffner, P. and Zheng, W.},
  collaboration = {Particle Data Group Collaboration},
  journal = {Phys. Rev. D},
  volume = {110},
  issue = {3},
  pages = {030001},
  numpages = {5},
  year = {2024},
  month = {Aug},
  publisher = {American Physical Society},
  doi = {10.1103/PhysRevD.110.030001},
  url = {https://link.aps.org/doi/10.1103/PhysRevD.110.030001}
}

@article{Nadler_2025,
doi = {10.3847/2041-8213/adbc6e},
url = {https://dx.doi.org/10.3847/2041-8213/adbc6e},
year = {2025},
month = {apr},
publisher = {The American Astronomical Society},
volume = {983},
number = {1},
pages = {L23},
author = {Nadler, Ethan O.},
title = {The Impact of Molecular Hydrogen Cooling on the Galaxy Formation Threshold},
journal = {The Astrophysical Journal Letters}
}

@ARTICLE{2016MNRAS.456...85S,
       author = {{Sawala}, Till and {Frenk}, Carlos S. and {Fattahi}, Azadeh and {Navarro}, Julio F. and {Theuns}, Tom and {Bower}, Richard G. and {Crain}, Robert A. and {Furlong}, Michelle and {Jenkins}, Adrian and {Schaller}, Matthieu and {Schaye}, Joop},
        title = "{The chosen few: the low-mass haloes that host faint galaxies}",
      journal = {\mnras},
         year = 2016,
        month = feb,
       volume = {456},
       number = {1},
        pages = {85-97},
          doi = {10.1093/mnras/stv2597},
archivePrefix = {arXiv},
       eprint = {1406.6362},
 primaryClass = {astro-ph.CO},
       adsurl = {https://ui.adsabs.harvard.edu/abs/2016MNRAS.456...85S}
}

@book{inverse_algorithm,
    author    = "Luc Devroye",
    title     = "Chapter 2.2, pages 27-39, from Non-Uniform Random Variate Generation",
    year      = "1986",
    publisher = "Springer-Verlag New York Inc.",
    address   = "175 Fifth Avenue, New York, New York 10010, U.S.A."
}

@article{PhysRevD.102.103010,
  title = {Investigating the detection of dark matter subhalos as extended sources with Fermi-LAT},
  author = {Di Mauro, Mattia and Stref, Martin and Calore, Francesca},
  journal = {Phys. Rev. D},
  volume = {102},
  issue = {10},
  pages = {103010},
  numpages = {12},
  year = {2020},
  month = {Nov},
  publisher = {American Physical Society},
  doi = {10.1103/PhysRevD.102.103010},
  url = {https://link.aps.org/doi/10.1103/PhysRevD.102.103010}
}

@article{10.1093/mnras/stae034,
    author = {Ou, Xiaowei and Eilers, Anna-Christina and Necib, Lina and Frebel, Anna},
    title = {The dark matter profile of the Milky Way inferred from its circular velocity curve},
    journal = {Monthly Notices of the Royal Astronomical Society},
    volume = {528},
    number = {1},
    pages = {693-710},
    year = {2024},
    month = {01},
    issn = {0035-8711},
    doi = {10.1093/mnras/stae034},
    url = {https://doi.org/10.1093/mnras/stae034},
}

@ARTICLE{2018MNRAS.473.2060J,
       author = {{Jethwa}, P. and {Erkal}, D. and {Belokurov}, V.},
        title = "{The upper bound on the lowest mass halo}",
      journal = {\mnras},
         year = 2018,
        month = jan,
       volume = {473},
       number = {2},
        pages = {2060-2083},
          doi = {10.1093/mnras/stx2330},
archivePrefix = {arXiv},
       eprint = {1612.07834},
 primaryClass = {astro-ph.GA},
       adsurl = {https://ui.adsabs.harvard.edu/abs/2018MNRAS.473.2060J}
}

@article{Nadler_2020,
doi = {10.3847/1538-4357/ab846a},
url = {https://dx.doi.org/10.3847/1538-4357/ab846a},
year = {2020},
month = {apr},
publisher = {The American Astronomical Society},
volume = {893},
number = {1},
pages = {48},
author = {Nadler, E. O. and Wechsler, R. H. and Bechtol, K. and Mao, Y.-Y. and Green, G. and Drlica-Wagner, A. and McNanna, M. and Mau, S. and Pace, A. B. and Simon, J. D. and Kravtsov, A. and Dodelson, S. and Li, T. S. and Riley, A. H. and Wang, M. Y. and Abbott, T. M. C. and Aguena, M. and Allam, S. and Annis, J. and Avila, S. and Bernstein, G. M. and Bertin, E. and Brooks, D. and Burke, D. L. and Rosell, A. Carnero and Kind, M. Carrasco and Carretero, J. and Costanzi, M. and da Costa, L. N. and De Vicente, J. and Desai, S. and Evrard, A. E. and Flaugher, B. and Fosalba, P. and Frieman, J. and García-Bellido, J. and Gaztanaga, E. and Gerdes, D. W. and Gruen, D. and Gschwend, J. and Gutierrez, G. and Hartley, W. G. and Hinton, S. R. and Honscheid, K. and Krause, E. and Kuehn, K. and Kuropatkin, N. and Lahav, O. and Maia, M. A. G. and Marshall, J. L. and Menanteau, F. and Miquel, R. and Palmese, A. and Paz-Chinchón, F. and Plazas, A. A. and Romer, A. K. and Sanchez, E. and Santiago, B. and Scarpine, V. and Serrano, S. and Smith, M. and Soares-Santos, M. and Suchyta, E. and Tarle, G. and Thomas, D. and Varga, T. N. and Walker, A. R. and (DES Collaboration)},
title = {Milky Way Satellite Census. II. Galaxy–Halo Connection Constraints Including the Impact of the Large Magellanic Cloud},
journal = {The Astrophysical Journal}
}

@ARTICLE{2022ApJ...940....8M,
       author = {{McQuinn}, Kristen. B.~W. and {Adams}, Elizabeth A.~K. and {Cannon}, John M. and {Fuson}, Jackson and {Skillman}, Evan D. and {Brooks}, Alyson and {Rhode}, Katherine L. and {Haynes}, Martha P. and {Inoue}, John L. and {Marine}, Joshua and {Salzer}, John. J. and {Talluri}, Anjana K.},
        title = "{The Turndown of the Baryonic Tully-Fisher Relation and Changing Baryon Fraction at Low Galaxy Masses}",
      journal = {\apj},
         year = 2022,
        month = nov,
       volume = {940},
       number = {1},
          eid = {8},
        pages = {8},
          doi = {10.3847/1538-4357/ac9285},
archivePrefix = {arXiv},
       eprint = {2203.10105},
 primaryClass = {astro-ph.GA},
       adsurl = {https://ui.adsabs.harvard.edu/abs/2022ApJ...940....8M}
}

@ARTICLE{2012ApJ...748...20C,
       author = {{Carlberg}, R.~G.},
        title = "{Dark Matter Sub-halo Counts via Star Stream Crossings}",
      journal = {\apj},
         year = 2012,
        month = mar,
       volume = {748},
       number = {1},
          eid = {20},
        pages = {20},
          doi = {10.1088/0004-637X/748/1/20},
archivePrefix = {arXiv},
       eprint = {1109.6022},
 primaryClass = {astro-ph.CO},
       adsurl = {https://ui.adsabs.harvard.edu/abs/2012ApJ...748...20C}
}

@ARTICLE{2015ApJ...808...15C,
       author = {{Carlberg}, R.~G.},
        title = "{Star Streams in Triaxial Isochrone Potentials with Sub-halos}",
      journal = {\apj},
         year = 2015,
        month = jul,
       volume = {808},
       number = {1},
          eid = {15},
        pages = {15},
          doi = {10.1088/0004-637X/808/1/15},
archivePrefix = {arXiv},
       eprint = {1506.00957},
 primaryClass = {astro-ph.GA},
       adsurl = {https://ui.adsabs.harvard.edu/abs/2015ApJ...808...15C}
}

@ARTICLE{2015MNRAS.454.3542E,
       author = {{Erkal}, Denis and {Belokurov}, Vasily},
        title = "{Properties of dark subhaloes from gaps in tidal streams}",
      journal = {\mnras},
         year = 2015,
        month = dec,
       volume = {454},
       number = {4},
        pages = {3542-3558},
          doi = {10.1093/mnras/stv2122},
archivePrefix = {arXiv},
       eprint = {1507.05625},
 primaryClass = {astro-ph.GA},
       adsurl = {https://ui.adsabs.harvard.edu/abs/2015MNRAS.454.3542E}
}

@ARTICLE{2015MNRAS.450.1136E,
       author = {{Erkal}, Denis and {Belokurov}, Vasily},
        title = "{Forensics of subhalo-stream encounters: the three phases of gap growth}",
      journal = {\mnras},
         year = 2015,
        month = jun,
       volume = {450},
       number = {1},
        pages = {1136-1149},
          doi = {10.1093/mnras/stv655},
archivePrefix = {arXiv},
       eprint = {1412.6035},
 primaryClass = {astro-ph.GA},
       adsurl = {https://ui.adsabs.harvard.edu/abs/2015MNRAS.450.1136E}
}

@ARTICLE{2019ApJ...880...38B,
       author = {{Bonaca}, Ana and {Hogg}, David W. and {Price-Whelan}, Adrian M. and {Conroy}, Charlie},
        title = "{The Spur and the Gap in GD-1: Dynamical Evidence for a Dark Substructure in the Milky Way Halo}",
      journal = {\apj},
         year = 2019,
        month = jul,
       volume = {880},
       number = {1},
          eid = {38},
        pages = {38},
          doi = {10.3847/1538-4357/ab2873},
archivePrefix = {arXiv},
       eprint = {1811.03631},
 primaryClass = {astro-ph.GA},
       adsurl = {https://ui.adsabs.harvard.edu/abs/2019ApJ...880...38B}
}

@ARTICLE{2010MNRAS.408.1969V,
       author = {{Vegetti}, S. and {Koopmans}, L.~V.~E. and {Bolton}, A. and {Treu}, T. and {Gavazzi}, R.},
        title = "{Detection of a dark substructure through gravitational imaging}",
      journal = {\mnras},
         year = 2010,
        month = nov,
       volume = {408},
       number = {4},
        pages = {1969-1981},
          doi = {10.1111/j.1365-2966.2010.16865.x},
archivePrefix = {arXiv},
       eprint = {0910.0760},
 primaryClass = {astro-ph.CO},
       adsurl = {https://ui.adsabs.harvard.edu/abs/2010MNRAS.408.1969V}
}

@ARTICLE{2016JCAP...11..048H,
       author = {{Hezaveh}, Yashar and {Dalal}, Neal and {Holder}, Gilbert and {Kisner}, Theodore and {Kuhlen}, Michael and {Perreault Levasseur}, Laurence},
        title = "{Measuring the power spectrum of dark matter substructure using strong gravitational lensing}",
      journal = {\jcap},
         year = 2016,
        month = nov,
       volume = {2016},
       number = {11},
          eid = {048},
        pages = {048},
          doi = {10.1088/1475-7516/2016/11/048},
archivePrefix = {arXiv},
       eprint = {1403.2720},
 primaryClass = {astro-ph.CO},
       adsurl = {https://ui.adsabs.harvard.edu/abs/2016JCAP...11..048H}
}

@ARTICLE{2014MNRAS.442.2434N,
       author = {{Nierenberg}, A.~M. and {Treu}, T. and {Wright}, S.~A. and {Fassnacht}, C.~D. and {Auger}, M.~W.},
        title = "{Detection of substructure with adaptive optics integral field spectroscopy of the gravitational lens B1422+231}",
      journal = {\mnras},
         year = 2014,
        month = aug,
       volume = {442},
       number = {3},
        pages = {2434-2445},
          doi = {10.1093/mnras/stu862},
archivePrefix = {arXiv},
       eprint = {1402.1496},
 primaryClass = {astro-ph.GA},
       adsurl = {https://ui.adsabs.harvard.edu/abs/2014MNRAS.442.2434N}
}

@ARTICLE{2020Sci...369.1347M,
       author = {{Meneghetti}, Massimo and {Davoli}, Guido and {Bergamini}, Pietro and {Rosati}, Piero and {Natarajan}, Priyamvada and {Giocoli}, Carlo and {Caminha}, Gabriel B. and {Metcalf}, R. Benton and {Rasia}, Elena and {Borgani}, Stefano and {Calura}, Francesco and {Grillo}, Claudio and {Mercurio}, Amata and {Vanzella}, Eros},
        title = "{An excess of small-scale gravitational lenses observed in galaxy clusters}",
      journal = {Science},
         year = 2020,
        month = sep,
       volume = {369},
       number = {6509},
        pages = {1347-1351},
          doi = {10.1126/science.aax5164},
archivePrefix = {arXiv},
       eprint = {2009.04471},
 primaryClass = {astro-ph.GA},
       adsurl = {https://ui.adsabs.harvard.edu/abs/2020Sci...369.1347M}
}

@article{Hayashi_2003,
doi = {10.1086/345788},
url = {https://doi.org/10.1086/345788},
year = {2003},
month = {feb},
publisher = {},
volume = {584},
number = {2},
pages = {541},
author = {Hayashi, Eric and Navarro, Julio F. and Taylor, James E. and Stadel, Joachim and Quinn, Thomas},
title = {The Structural Evolution of Substructure},
journal = {The Astrophysical Journal},
}

@ARTICLE{2019MNRAS.490.2091G,
       author = {{Green}, Sheridan B. and {van den Bosch}, Frank C.},
        title = "{The tidal evolution of dark matter substructure - I. subhalo density profiles}",
      journal = {\mnras},
         year = 2019,
        month = dec,
       volume = {490},
       number = {2},
        pages = {2091-2101},
          doi = {10.1093/mnras/stz2767},
archivePrefix = {arXiv},
       eprint = {1908.08537},
 primaryClass = {astro-ph.GA},
       adsurl = {https://ui.adsabs.harvard.edu/abs/2019MNRAS.490.2091G}
}

@article{201200212FrenkWhite,
author = {Frenk, C.S. and White, S.D.M.},
title = {Dark matter and cosmic structure},
journal = {Annalen der Physik},
volume = {524},
number = {9-10},
pages = {507-534},
doi = {https://doi.org/10.1002/andp.201200212},
url = {https://onlinelibrary.wiley.com/doi/abs/10.1002/andp.201200212},
year = {2012}
}

@ARTICLE{2020NatRP...2...42V,
       author = {{Vogelsberger}, Mark and {Marinacci}, Federico and {Torrey}, Paul and {Puchwein}, Ewald},
        title = "{Cosmological simulations of galaxy formation}",
      journal = {Nature Reviews Physics},
         year = 2020,
        month = jan,
       volume = {2},
       number = {1},
        pages = {42-66},
          doi = {10.1038/s42254-019-0127-2},
archivePrefix = {arXiv},
       eprint = {1909.07976},
 primaryClass = {astro-ph.GA},
       adsurl = {https://ui.adsabs.harvard.edu/abs/2020NatRP...2...42V}
}

@ARTICLE{2001MNRAS.321..559B,
       author = {{Bullock}, J.~S. and {Kolatt}, T.~S. and {Sigad}, Y. and {Somerville}, R.~S. and {Kravtsov}, A.~V. and {Klypin}, A.~A. and {Primack}, J.~R. and {Dekel}, A.},
        title = "{Profiles of dark haloes: evolution, scatter and environment}",
      journal = {\mnras},
         year = 2001,
        month = mar,
       volume = {321},
       number = {3},
        pages = {559-575},
          doi = {10.1046/j.1365-8711.2001.04068.x},
archivePrefix = {arXiv},
       eprint = {astro-ph/9908159},
 primaryClass = {astro-ph},
       adsurl = {https://ui.adsabs.harvard.edu/abs/2001MNRAS.321..559B}
}

@ARTICLE{2008MNRAS.391.1940M,
       author = {{Macci{\`o}}, Andrea V. and {Dutton}, Aaron A. and {van den Bosch}, Frank C.},
        title = "{Concentration, spin and shape of dark matter haloes as a function of the cosmological model: WMAP1, WMAP3 and WMAP5 results}",
      journal = {\mnras},
         year = 2008,
        month = dec,
       volume = {391},
       number = {4},
        pages = {1940-1954},
          doi = {10.1111/j.1365-2966.2008.14029.x},
archivePrefix = {arXiv},
       eprint = {0805.1926},
 primaryClass = {astro-ph},
       adsurl = {https://ui.adsabs.harvard.edu/abs/2008MNRAS.391.1940M}
}

@ARTICLE{2013MNRAS.432.1103L,
       author = {{Ludlow}, Aaron D. and {Navarro}, Julio F. and {Boylan-Kolchin}, Michael and {Bett}, Philip E. and {Angulo}, Ra{\'u}l E. and {Li}, Ming and {White}, Simon D.~M. and {Frenk}, Carlos and {Springel}, Volker},
        title = "{The mass profile and accretion history of cold dark matter haloes}",
      journal = {\mnras},
         year = 2013,
        month = jun,
       volume = {432},
       number = {2},
        pages = {1103-1113},
          doi = {10.1093/mnras/stt526},
archivePrefix = {arXiv},
       eprint = {1302.0288},
 primaryClass = {astro-ph.CO},
       adsurl = {https://ui.adsabs.harvard.edu/abs/2013MNRAS.432.1103L}
}

@ARTICLE{2012MNRAS.423.3018P,
       author = {{Prada}, Francisco and {Klypin}, Anatoly A. and {Cuesta}, Antonio J. and {Betancort-Rijo}, Juan E. and {Primack}, Joel},
        title = "{Halo concentrations in the standard {\ensuremath{\Lambda}} cold dark matter cosmology}",
      journal = {\mnras},
         year = 2012,
        month = jul,
       volume = {423},
       number = {4},
        pages = {3018-3030},
          doi = {10.1111/j.1365-2966.2012.21007.x},
archivePrefix = {arXiv},
       eprint = {1104.5130},
 primaryClass = {astro-ph.CO},
       adsurl = {https://ui.adsabs.harvard.edu/abs/2012MNRAS.423.3018P}
}

@ARTICLE{2017MNRAS.472.4918P,
       author = {{Pilipenko}, Sergey V. and {S{\'a}nchez-Conde}, Miguel A. and {Prada}, Francisco and {Yepes}, Gustavo},
        title = "{Pushing down the low-mass halo concentration frontier with the Lomonosov cosmological simulations}",
      journal = {\mnras},
         year = 2017,
        month = dec,
       volume = {472},
       number = {4},
        pages = {4918-4927},
          doi = {10.1093/mnras/stx2319},
archivePrefix = {arXiv},
       eprint = {1703.06012},
 primaryClass = {astro-ph.CO},
       adsurl = {https://ui.adsabs.harvard.edu/abs/2017MNRAS.472.4918P}
}

@ARTICLE{2016MNRAS.457.3492H,
       author = {{Hellwing}, Wojciech A. and {Frenk}, Carlos S. and {Cautun}, Marius and {Bose}, Sownak and {Helly}, John and {Jenkins}, Adrian and {Sawala}, Till and {Cytowski}, Maciej},
        title = "{The Copernicus Complexio: a high-resolution view of the small-scale Universe}",
      journal = {\mnras},
         year = 2016,
        month = apr,
       volume = {457},
       number = {4},
        pages = {3492-3509},
          doi = {10.1093/mnras/stw214},
archivePrefix = {arXiv},
       eprint = {1505.06436},
 primaryClass = {astro-ph.CO},
       adsurl = {https://ui.adsabs.harvard.edu/abs/2016MNRAS.457.3492H}
}

@article{10.1111/j.1365-2966.2012.21564.x,
    author = {Gao, L. and Navarro, J. F. and Frenk, C. S. and Jenkins, A. and Springel, V. and White, S. D. M.},
    title = {The Phoenix Project: the dark side of rich Galaxy clusters},
    journal = {Monthly Notices of the Royal Astronomical Society},
    volume = {425},
    number = {3},
    pages = {2169-2186},
    year = {2012},
    month = {09},
    issn = {0035-8711},
    doi = {10.1111/j.1365-2966.2012.21564.x},
    url = {https://doi.org/10.1111/j.1365-2966.2012.21564.x},
}

@article{10.1093/mnras/stad844,
    author = {Stücker, Jens and Ogiya, Go and Angulo, Raul E and Aguirre-Santaella, Alejandra and Sánchez-Conde, Miguel A},
    title = {Tidal stripping in the adiabatic limit},
    journal = {Monthly Notices of the Royal Astronomical Society},
    volume = {521},
    number = {3},
    pages = {4432-4461},
    year = {2023},
    month = {03},
    issn = {0035-8711},
    doi = {10.1093/mnras/stad844},
    url = {https://doi.org/10.1093/mnras/stad844},
}

@ARTICLE{2007CSE.....9...90H,
       author = {{Hunter}, John D.},
        title = "{Matplotlib: A 2D Graphics Environment}",
      journal = {Computing in Science and Engineering},
         year = 2007,
        month = jan,
       volume = {9},
       number = {3},
        pages = {90-95},
          doi = {10.1109/MCSE.2007.55},
       adsurl = {https://ui.adsabs.harvard.edu/abs/2007CSE.....9...90H}
}

@Article{         harris2020array,
 title         = {Array programming with {NumPy}},
 author        = {Charles R. Harris and K. Jarrod Millman and St{\'{e}}fan J.
                 van der Walt and Ralf Gommers and Pauli Virtanen and David
                 Cournapeau and Eric Wieser and Julian Taylor and Sebastian
                 Berg and Nathaniel J. Smith and Robert Kern and Matti Picus
                 and Stephan Hoyer and Marten H. van Kerkwijk and Matthew
                 Brett and Allan Haldane and Jaime Fern{\'{a}}ndez del
                 R{\'{i}}o and Mark Wiebe and Pearu Peterson and Pierre
                 G{\'{e}}rard-Marchant and Kevin Sheppard and Tyler Reddy and
                 Warren Weckesser and Hameer Abbasi and Christoph Gohlke and
                 Travis E. Oliphant},
 year          = {2020},
 month         = sep,
 journal       = {Nature},
 volume        = {585},
 number        = {7825},
 pages         = {357--362},
 doi           = {10.1038/s41586-020-2649-2},
 publisher     = {Springer Science and Business Media {LLC}},
 url           = {https://doi.org/10.1038/s41586-020-2649-2}
}

@ARTICLE{2020NatMe..17..261V,
       author = {{Virtanen}, Pauli and {Gommers}, Ralf and {Oliphant}, Travis E. and {Haberland}, Matt and {Reddy}, Tyler and {Cournapeau}, David and {Burovski}, Evgeni and {Peterson}, Pearu and {Weckesser}, Warren and {Bright}, Jonathan and {van der Walt}, St{\'e}fan J. and {Brett}, Matthew and {Wilson}, Joshua and {Millman}, K. Jarrod and {Mayorov}, Nikolay and {Nelson}, Andrew R.~J. and {Jones}, Eric and {Kern}, Robert and {Larson}, Eric and {Carey}, C.~J. and {Polat}, {\.I}lhan and {Feng}, Yu and {Moore}, Eric W. and {VanderPlas}, Jake and {Laxalde}, Denis and {Perktold}, Josef and {Cimrman}, Robert and {Henriksen}, Ian and {Quintero}, E.~A. and {Harris}, Charles R. and {Archibald}, Anne M. and {Ribeiro}, Ant{\^o}nio H. and {Pedregosa}, Fabian and {van Mulbregt}, Paul and {SciPy 1.  0 Contributors}},
        title = "{SciPy 1.0: fundamental algorithms for scientific computing in Python}",
      journal = {Nature Medicine},
         year = 2020,
        month = feb,
       volume = {17},
        pages = {261-272},
          doi = {10.1038/s41592-019-0686-2},
archivePrefix = {arXiv},
       eprint = {1907.10121},
 primaryClass = {cs.MS},
       adsurl = {https://ui.adsabs.harvard.edu/abs/2020NatMe..17..261V}
}

@ARTICLE{2013A&A...558A..33A,
       author = {{Astropy Collaboration} and {Robitaille}, Thomas P. and {Tollerud}, Erik J. and {Greenfield}, Perry and {Droettboom}, Michael and {Bray}, Erik and {Aldcroft}, Tom and {Davis}, Matt and {Ginsburg}, Adam and {Price-Whelan}, Adrian M. and {Kerzendorf}, Wolfgang E. and {Conley}, Alexander and {Crighton}, Neil and {Barbary}, Kyle and {Muna}, Demitri and {Ferguson}, Henry and {Grollier}, Fr{\'e}d{\'e}ric and {Parikh}, Madhura M. and {Nair}, Prasanth H. and {Unther}, Hans M. and {Deil}, Christoph and {Woillez}, Julien and {Conseil}, Simon and {Kramer}, Roban and {Turner}, James E.~H. and {Singer}, Leo and {Fox}, Ryan and {Weaver}, Benjamin A. and {Zabalza}, Victor and {Edwards}, Zachary I. and {Azalee Bostroem}, K. and {Burke}, D.~J. and {Casey}, Andrew R. and {Crawford}, Steven M. and {Dencheva}, Nadia and {Ely}, Justin and {Jenness}, Tim and {Labrie}, Kathleen and {Lim}, Pey Lian and {Pierfederici}, Francesco and {Pontzen}, Andrew and {Ptak}, Andy and {Refsdal}, Brian and {Servillat}, Mathieu and {Streicher}, Ole},
        title = "{Astropy: A community Python package for astronomy}",
      journal = {\aap},
         year = 2013,
        month = oct,
       volume = {558},
          eid = {A33},
        pages = {A33},
          doi = {10.1051/0004-6361/201322068},
archivePrefix = {arXiv},
       eprint = {1307.6212},
 primaryClass = {astro-ph.IM},
       adsurl = {https://ui.adsabs.harvard.edu/abs/2013A&A...558A..33A}
}

@ARTICLE{2018AJ....156..123A,
       author = {{Astropy Collaboration} and {Price-Whelan}, A.~M. and {Sip{\H{o}}cz}, B.~M. and {G{\"u}nther}, H.~M. and {Lim}, P.~L. and {Crawford}, S.~M. and {Conseil}, S. and {Shupe}, D.~L. and {Craig}, M.~W. and {Dencheva}, N. and {Ginsburg}, A. and {VanderPlas}, J.~T. and {Bradley}, L.~D. and {P{\'e}rez-Su{\'a}rez}, D. and {de Val-Borro}, M. and {Aldcroft}, T.~L. and {Cruz}, K.~L. and {Robitaille}, T.~P. and {Tollerud}, E.~J. and {Ardelean}, C. and {Babej}, T. and {Bach}, Y.~P. and {Bachetti}, M. and {Bakanov}, A.~V. and {Bamford}, S.~P. and {Barentsen}, G. and {Barmby}, P. and {Baumbach}, A. and {Berry}, K.~L. and {Biscani}, F. and {Boquien}, M. and {Bostroem}, K.~A. and {Bouma}, L.~G. and {Brammer}, G.~B. and {Bray}, E.~M. and {Breytenbach}, H. and {Buddelmeijer}, H. and {Burke}, D.~J. and {Calderone}, G. and {Cano Rodr{\'\i}guez}, J.~L. and {Cara}, M. and {Cardoso}, J.~V.~M. and {Cheedella}, S. and {Copin}, Y. and {Corrales}, L. and {Crichton}, D. and {D'Avella}, D. and {Deil}, C. and {Depagne}, {\'E}. and {Dietrich}, J.~P. and {Donath}, A. and {Droettboom}, M. and {Earl}, N. and {Erben}, T. and {Fabbro}, S. and {Ferreira}, L.~A. and {Finethy}, T. and {Fox}, R.~T. and {Garrison}, L.~H. and {Gibbons}, S.~L.~J. and {Goldstein}, D.~A. and {Gommers}, R. and {Greco}, J.~P. and {Greenfield}, P. and {Groener}, A.~M. and {Grollier}, F. and {Hagen}, A. and {Hirst}, P. and {Homeier}, D. and {Horton}, A.~J. and {Hosseinzadeh}, G. and {Hu}, L. and {Hunkeler}, J.~S. and {Ivezi{\'c}}, {\v{Z}}. and {Jain}, A. and {Jenness}, T. and {Kanarek}, G. and {Kendrew}, S. and {Kern}, N.~S. and {Kerzendorf}, W.~E. and {Khvalko}, A. and {King}, J. and {Kirkby}, D. and {Kulkarni}, A.~M. and {Kumar}, A. and {Lee}, A. and {Lenz}, D. and {Littlefair}, S.~P. and {Ma}, Z. and {Macleod}, D.~M. and {Mastropietro}, M. and {McCully}, C. and {Montagnac}, S. and {Morris}, B.~M. and {Mueller}, M. and {Mumford}, S.~J. and {Muna}, D. and {Murphy}, N.~A. and {Nelson}, S. and {Nguyen}, G.~H. and {Ninan}, J.~P. and {N{\"o}the}, M. and {Ogaz}, S. and {Oh}, S. and {Parejko}, J.~K. and {Parley}, N. and {Pascual}, S. and {Patil}, R. and {Patil}, A.~A. and {Plunkett}, A.~L. and {Prochaska}, J.~X. and {Rastogi}, T. and {Reddy Janga}, V. and {Sabater}, J. and {Sakurikar}, P. and {Seifert}, M. and {Sherbert}, L.~E. and {Sherwood-Taylor}, H. and {Shih}, A.~Y. and {Sick}, J. and {Silbiger}, M.~T. and {Singanamalla}, S. and {Singer}, L.~P. and {Sladen}, P.~H. and {Sooley}, K.~A. and {Sornarajah}, S. and {Streicher}, O. and {Teuben}, P. and {Thomas}, S.~W. and {Tremblay}, G.~R. and {Turner}, J.~E.~H. and {Terr{\'o}n}, V. and {van Kerkwijk}, M.~H. and {de la Vega}, A. and {Watkins}, L.~L. and {Weaver}, B.~A. and {Whitmore}, J.~B. and {Woillez}, J. and {Zabalza}, V. and {Astropy Contributors}},
        title = "{The Astropy Project: Building an Open-science Project and Status of the v2.0 Core Package}",
      journal = {\aj},
         year = 2018,
        month = sep,
       volume = {156},
       number = {3},
          eid = {123},
        pages = {123},
          doi = {10.3847/1538-3881/aabc4f},
archivePrefix = {arXiv},
       eprint = {1801.02634},
 primaryClass = {astro-ph.IM},
       adsurl = {https://ui.adsabs.harvard.edu/abs/2018AJ....156..123A}
}

@ARTICLE{2022ApJ...935..167A,
       author = {{Astropy Collaboration} and {Price-Whelan}, Adrian M. and {Lim}, Pey Lian and {Earl}, Nicholas and {Starkman}, Nathaniel and {Bradley}, Larry and {Shupe}, David L. and {Patil}, Aarya A. and {Corrales}, Lia and {Brasseur}, C.~E. and {N{\"o}the}, Maximilian and {Donath}, Axel and {Tollerud}, Erik and {Morris}, Brett M. and {Ginsburg}, Adam and {Vaher}, Eero and {Weaver}, Benjamin A. and {Tocknell}, James and {Jamieson}, William and {van Kerkwijk}, Marten H. and {Robitaille}, Thomas P. and {Merry}, Bruce and {Bachetti}, Matteo and {G{\"u}nther}, H. Moritz and {Aldcroft}, Thomas L. and {Alvarado-Montes}, Jaime A. and {Archibald}, Anne M. and {B{\'o}di}, Attila and {Bapat}, Shreyas and {Barentsen}, Geert and {Baz{\'a}n}, Juanjo and {Biswas}, Manish and {Boquien}, M{\'e}d{\'e}ric and {Burke}, D.~J. and {Cara}, Daria and {Cara}, Mihai and {Conroy}, Kyle E. and {Conseil}, Simon and {Craig}, Matthew W. and {Cross}, Robert M. and {Cruz}, Kelle L. and {D'Eugenio}, Francesco and {Dencheva}, Nadia and {Devillepoix}, Hadrien A.~R. and {Dietrich}, J{\"o}rg P. and {Eigenbrot}, Arthur Davis and {Erben}, Thomas and {Ferreira}, Leonardo and {Foreman-Mackey}, Daniel and {Fox}, Ryan and {Freij}, Nabil and {Garg}, Suyog and {Geda}, Robel and {Glattly}, Lauren and {Gondhalekar}, Yash and {Gordon}, Karl D. and {Grant}, David and {Greenfield}, Perry and {Groener}, Austen M. and {Guest}, Steve and {Gurovich}, Sebastian and {Handberg}, Rasmus and {Hart}, Akeem and {Hatfield-Dodds}, Zac and {Homeier}, Derek and {Hosseinzadeh}, Griffin and {Jenness}, Tim and {Jones}, Craig K. and {Joseph}, Prajwel and {Kalmbach}, J. Bryce and {Karamehmetoglu}, Emir and {Ka{\l}uszy{\'n}ski}, Miko{\l}aj and {Kelley}, Michael S.~P. and {Kern}, Nicholas and {Kerzendorf}, Wolfgang E. and {Koch}, Eric W. and {Kulumani}, Shankar and {Lee}, Antony and {Ly}, Chun and {Ma}, Zhiyuan and {MacBride}, Conor and {Maljaars}, Jakob M. and {Muna}, Demitri and {Murphy}, N.~A. and {Norman}, Henrik and {O'Steen}, Richard and {Oman}, Kyle A. and {Pacifici}, Camilla and {Pascual}, Sergio and {Pascual-Granado}, J. and {Patil}, Rohit R. and {Perren}, Gabriel I. and {Pickering}, Timothy E. and {Rastogi}, Tanuj and {Roulston}, Benjamin R. and {Ryan}, Daniel F. and {Rykoff}, Eli S. and {Sabater}, Jose and {Sakurikar}, Parikshit and {Salgado}, Jes{\'u}s and {Sanghi}, Aniket and {Saunders}, Nicholas and {Savchenko}, Volodymyr and {Schwardt}, Ludwig and {Seifert-Eckert}, Michael and {Shih}, Albert Y. and {Jain}, Anany Shrey and {Shukla}, Gyanendra and {Sick}, Jonathan and {Simpson}, Chris and {Singanamalla}, Sudheesh and {Singer}, Leo P. and {Singhal}, Jaladh and {Sinha}, Manodeep and {Sip{\H{o}}cz}, Brigitta M. and {Spitler}, Lee R. and {Stansby}, David and {Streicher}, Ole and {{\v{S}}umak}, Jani and {Swinbank}, John D. and {Taranu}, Dan S. and {Tewary}, Nikita and {Tremblay}, Grant R. and {de Val-Borro}, Miguel and {Van Kooten}, Samuel J. and {Vasovi{\'c}}, Zlatan and {Verma}, Shresth and {de Miranda Cardoso}, Jos{\'e} Vin{\'\i}cius and {Williams}, Peter K.~G. and {Wilson}, Tom J. and {Winkel}, Benjamin and {Wood-Vasey}, W.~M. and {Xue}, Rui and {Yoachim}, Peter and {Zhang}, Chen and {Zonca}, Andrea and {Astropy Project Contributors}},
        title = "{The Astropy Project: Sustaining and Growing a Community-oriented Open-source Project and the Latest Major Release (v5.0) of the Core Package}",
      journal = {\apj},
         year = 2022,
        month = aug,
       volume = {935},
       number = {2},
          eid = {167},
        pages = {167},
          doi = {10.3847/1538-4357/ac7c74},
archivePrefix = {arXiv},
       eprint = {2206.14220},
 primaryClass = {astro-ph.IM},
       adsurl = {https://ui.adsabs.harvard.edu/abs/2022ApJ...935..167A}
}

@article{iminuit,
  author={Hans Dembinski and Piti Ongmongkolkul et al.},
  title={scikit-hep/iminuit},
  DOI={10.5281/zenodo.3949207},
  publisher={Zenodo},
  year={2020},
  month={Dec},
  url={https://doi.org/10.5281/zenodo.3949207}
}

@article{10.1046/j.1365-8711.1998.01775.x,
    author = {Tormen, Giuseppe and Diaferio, Antonaldo and Syer, D.},
    title = {Survival of substructure within dark matter haloes},
    journal = {Monthly Notices of the Royal Astronomical Society},
    volume = {299},
    number = {3},
    pages = {728-742},
    year = {1998},
    month = {09},
    issn = {0035-8711},
    doi = {10.1046/j.1365-8711.1998.01775.x},
    url = {https://doi.org/10.1046/j.1365-8711.1998.01775.x},
}

@ARTICLE{2024PhRvD.109f3024M,
       author = {{McDaniel}, Alex and {Ajello}, Marco and {Karwin}, Christopher M. and {Di Mauro}, Mattia and {Drlica-Wagner}, Alex and {S{\'a}nchez-Conde}, Miguel A.},
        title = "{Legacy analysis of dark matter annihilation from the Milky Way dwarf spheroidal galaxies with 14 years of Fermi -LAT data}",
      journal = {\prd},
         year = 2024,
        month = mar,
       volume = {109},
       number = {6},
          eid = {063024},
        pages = {063024},
          doi = {10.1103/PhysRevD.109.063024},
archivePrefix = {arXiv},
       eprint = {2311.04982},
 primaryClass = {astro-ph.HE},
       adsurl = {https://ui.adsabs.harvard.edu/abs/2024PhRvD.109f3024M}
}

@ARTICLE{1994Natur.370..194I,
       author = {{Ibata}, R.~A. and {Gilmore}, G. and {Irwin}, M.~J.},
        title = "{A dwarf satellite galaxy in Sagittarius}",
      journal = {\nat},
         year = 1994,
        month = jul,
       volume = {370},
       number = {6486},
        pages = {194-196},
          doi = {10.1038/370194a0},
       adsurl = {https://ui.adsabs.harvard.edu/abs/1994Natur.370..194I}
}

@ARTICLE{2025JCAP...09..003F,
       author = {{Fern{\'a}ndez-Su{\'a}rez}, Cristina and {S{\'a}nchez-Conde}, Miguel {\'A}.},
        title = "{A search for dark matter annihilation in stellar streams with the Fermi-LAT}",
      journal = {\jcap},
         year = 2025,
        month = sep,
       volume = {2025},
       number = {9},
          eid = {003},
        pages = {003},
          doi = {10.1088/1475-7516/2025/09/003},
archivePrefix = {arXiv},
       eprint = {2502.15656},
 primaryClass = {astro-ph.HE},
       adsurl = {https://ui.adsabs.harvard.edu/abs/2025JCAP...09..003F}
}

\end{document}